%% file: main.tex
\documentclass[a4paper,11pt]{article}
\pdfoutput=1
\usepackage{jheppub}
\usepackage[T1]{fontenc}
\usepackage{diagbox}
\usepackage{adjustbox}
\usepackage{multirow}
\usepackage{pifont}
\usepackage{array}
\usepackage{mathtools}
\usepackage{booktabs}
\usepackage[dvipsnames]{xcolor}
\usepackage{tcolorbox}
\usepackage{float}
\usepackage{placeins}
\newcommand{\eqal}[1]{\begin{align}#1\end{align}}
\newcommand{\xmark}{\color{red}\ding{55}}
\newcommand{\cmark}{\color{blue}\checkmark}

\usepackage{verbatim}

\title{\boldmath Neutrino seesaw models at one-loop matching: Discrimination by effective operators}

\author[a]{Yong Du,}
\author[b]{Xu-Xiang Li,}
\author[a,c,d,e,f]{Jiang-Hao Yu}

\affiliation[a]{CAS Key Laboratory of Theoretical Physics, Institute of Theoretical Physics, Chinese Academy of Sciences, Beijing 100190, China}
\affiliation[b]{Department of Physics and State Key Laboratory of Nuclear Physics and Technology, Peking University, Beijing 100871, China}
\affiliation[c]{School of Physical Sciences, University of Chinese Academy of Sciences, Beijing 100049, P.\ R.\ China}
\affiliation[d]{Center for High Energy Physics, Peking University, Beijing 100871, China}
\affiliation[e]{School of Fundamental Physics and Mathematical Sciences, Hangzhou Institute for Advanced
Study, UCAS, Hangzhou 310024, China}
\affiliation[f]{International Centre for Theoretical Physics Asia-Pacific, Beijing/Hangzhou, China}

\emailAdd{yongdu@itp.ac.cn}
\emailAdd{xuxiangli@pku.edu.cn}
\emailAdd{jhyu@itp.ac.cn}

\abstract{Using the functional method, one-loop matching of the type-I, -II and -III seesaw models are investigated and the results are presented in both the Green's and the Warsaw bases.
Although these models generate the same dimension-5 Weinberg operator, they could induce quite different types of dimension-6 effective operators that can be utilized for model discrimination. 
We also find the threshold effects from one-loop matching could be significant, which turn out to allow triggering electroweak symmetry breaking radiatively in type-II seesaw while forbid that in type-I/-III models. An analytical criterion for such radiative symmetry breaking is also derived in type-II seesaw.
Finally, we investigate the indirect signatures from different types of dimension-6 operators at high-energy colliders, low-energy precision experiments and forward physics facilities for model discrimination.

}

\begin{document} 
\maketitle
\flushbottom

\input{intro}


\input{func_match}
\input{type123seesaw}


\input{EWSB}


\input{pheno}
\input{conclusions}

\paragraph{Note added.} During the preparation of this article, Refs.\,\cite{Ohlsson:2022hfl,Li:2022ipc} on a similar topic appeared. The authors of Ref.\,\cite{Ohlsson:2022hfl} only considered the type-I seesaw model using the amplitude/diagram approach for one-loop matching and the authors of Ref\,\cite{Li:2022ipc} only focused on the type-II seesaw model, while we considered all the type-I, -II and -III seesaw models using the functional method. In addition, we also studied the RGEs of the three seesaw models in detail and qualitatively discussed the phenomenological aspects of the three seesaw models.

\acknowledgments{We thank Michael Robert Trott, Rupert Coy, Shun Zhou, Di Zhang for their kind discussion and correspondence. YD wishes to thank Zhengkang Zhang and Lohan Sartore for helpful communication.
J.\ H.\ Y.\ and Y.\ D.\ are  supported by the National Science Foundation of China under Grants No. 12022514, No. 11875003 and No. 12047503, and National Key Research and Development Program of China Grant No. 2020YFC2201501, No. 2021YFA0718304, and CAS Project for Young Scientists in Basic Research YSBR-006, the Key Research Program of the CAS Grant No. XDPB15.}

\input{appendix}

\bibliographystyle{JHEP}
\bibliography{ref}
\end{document}

%% file: intro.tex
\section{Introduction}
Neutrino oscillations, predicted by Pontecorvo in 1957\,\cite{Pontecorvo:1957cp, Pontecorvo:1967fh} and answering the solar neutrino problem from Homestake\,\cite{Bahcall:1976zz}, were firstly observed by the Super-Kamiokande experiment in 1998 with atmospheric neutrinos\,\cite{Fukuda:1998mi} and by the Sudbury Neutrino Observatory in 2001 from solar neutrinos\,\cite{Ahmad:2001an}. The oscillations immediately suggest the fact that neutrinos are massive particles, contradicting the prediction of the Standard Model (SM) and directly signaling new physics beyond the SM.

Model independently, neutrino masses can be naturally generated through the dimension-5 Weinberg operator\,\cite{Weinberg:1979sa}. At tree level, these operators can be realized by the seesaw models, known as the type-I, -II and -III models\,\cite{Minkowski:1977sc,Ramond:1979py,GellMann:1980vs,Yanagida:1979as,Mohapatra:1979ia,Schechter:1980gr,Schechter:1981cv,Konetschny:1977bn,Cheng:1980qt,Lazarides:1980nt,Magg:1980ut,Foot:1988aq,Witten:1985bz,Mohapatra:1986aw,Mohapatra:1986bd,Val86,Barr:2003nn,Mohapatra:1980yp}. One the one hand, fulfilling the requirement of naturalness, it is known that the type-I and -III seesaw models would live above the GUT scale\,\cite{Broncano:2002rw,Abada:2007ux,Du:2020qwi}, making them impossible to be produced or tested directly at current experiments. The type-II seesaw model, however, stands out in the sense that the neutrino Yukawa couplings could be of $\mathcal{O}(1)$ with a relatively light triplet at the $\mathcal{O}(1\rm\,TeV)$ scale as long as the triplet vacuum expectation value (vev) is agnostically small. On the other hand, putting naturalness aside, the three seesaw models around the $\mathcal{O}(1\,\rm TeV)$ scale, commonly referred to as low-scale seesaw scenarios, would be allowed, making them also testable at colliders, see, for example, Refs.\,\cite{Han:2006ip,Atre:2009rg,FileviezPerez:2009hdc,Alva:2014gxa,Cai:2017mow,Dev:2018sel,Du:2018eaw,Mandal:2022zmy,Ashanujjaman:2021txz,Ashanujjaman:2021jhi,Ashanujjaman:2021zrh}.

In the low-scale seesaw scenarios, however, one immediate question arises: How could one distinguish one seesaw model from the rest if some signals, especially those common in two or three of the seesaw models, are observed? One such example is Higgs pair production at lepton and/or hadron colliders, which receives extra contributions from all the three seesaw models at one loop. To answer this question, one direction would be to perform a detailed simulation at different colliders, apply dedicated cuts to separate the signals from various backgrounds, and then look for ideal benchmark points in hoping to distinguish different seesaw models in the end.

However, in the mass region not accessible at colliders, 10\,TeV to $10^{12}$\,GeV for example, one can instead focus on low-energy observables, where various experiments have been carried out and very high precision has been achieved in the past. Precision measurements of the lepton magnetic dipole moments\,\cite{Muong-2:2006rrc,Muong-2:2021ojo}, for example. More importantly, there will be even more precision experiments in the near future, examples are precision measurements of the effective number of relativistic species, $N_{\rm eff}$, in the early universe at the Cosmic Microwave Background-Stage 4 (CMB-S4),\footnote{Recently, this has been studied model independently in the low-energy effective field theory framework up to dimension 7\,\cite{Du:2021idh,Du:2021nyb}.} lepton mixing parameters from neutrino oscillation experiments such as JUNO\,\cite{JUNO:2015zny}, T2HK\,\cite{Hyper-KamiokandeProto-:2015xww} and DUNE\,\cite{DUNE:2016hlj}, and precision measurements of the weak mixing angle from parity-violating electron scatterings\,\cite{MOLLER:2014iki,Berger:2015aaa}. All these precision observables will receive extra corrections from all the three seesaw models, making them possible identifiers for the seesaw models.

On the other hand, the big energy gap between low-energy precision observables and the high-energy seesaw models renders studying the effects from the latter in the effective field theory (EFT) framework possible. The EFTs can be obtained by integrating out the heavy particles in the UV models, either from the amplitude/diagram matching, see for example, \cite{Skiba:2010xn}, or from the recently developed functional method\,\cite{Gaillard:1985uh,Chan:1986jq,Cheyette:1987qz,Henning:2014wua,Dedes:2021abc}. While the former method is quite intuitive, the latter is perhaps convenient at one loop, especially with the recently developed {\tt Mathematica} packages {\tt STrEAM}\,\cite{Cohen:2020qvb} and {\tt SuperTracer}\,\cite{Fuentes-Martin:2020udw}.\footnote{We also note the recently released package {\tt Matchmakereft}\,\cite{Carmona:2021xtq} that makes the amplitude matching approach also straightforward at one loop.} In this work, we use the functional method to match the three seesaw models onto the SM EFT (SMEFT) up to one loop, and only keep operators up to dimensional 6 after the matching. We parameterize the SMEFT Lagrangian as follows:
\eqal{\mathcal{L}_{\rm SMEFT}^{\rm dim-6} = \mathcal{L}_{\rm SM} + C_5\mathcal{O}^{(5)} + \sum\limits_i C_6\mathcal{O}_i^{(6)},}
where the superscripts of $\mathcal{O}$ represent the mass dimensions of the operators, and $C_i$'s are the dimensional Wilson coefficients as functions of model parameters. Note that the operators obtained after the functional matching are generically over-complete and dependent. These dependent operators can be related to each other, for example, from equations of motion (EOM) and/or Fierz identities. This operator dependence is eliminated manually, after which we choose to present our final results in the Warsaw basis\,\cite{Grzadkowski:2010es}, together with the intermediate results in the Green's basis\,\cite{Jiang:2018pbd}. We comment on that the Green's basis is over-complete when all particles in the operators become on-shell, and this over-completeness can be removed through applying the EOM to reduce the Green's basis into the Warsaw basis. 

\begin{figure}
	\centering
	\includegraphics[width=\textwidth]{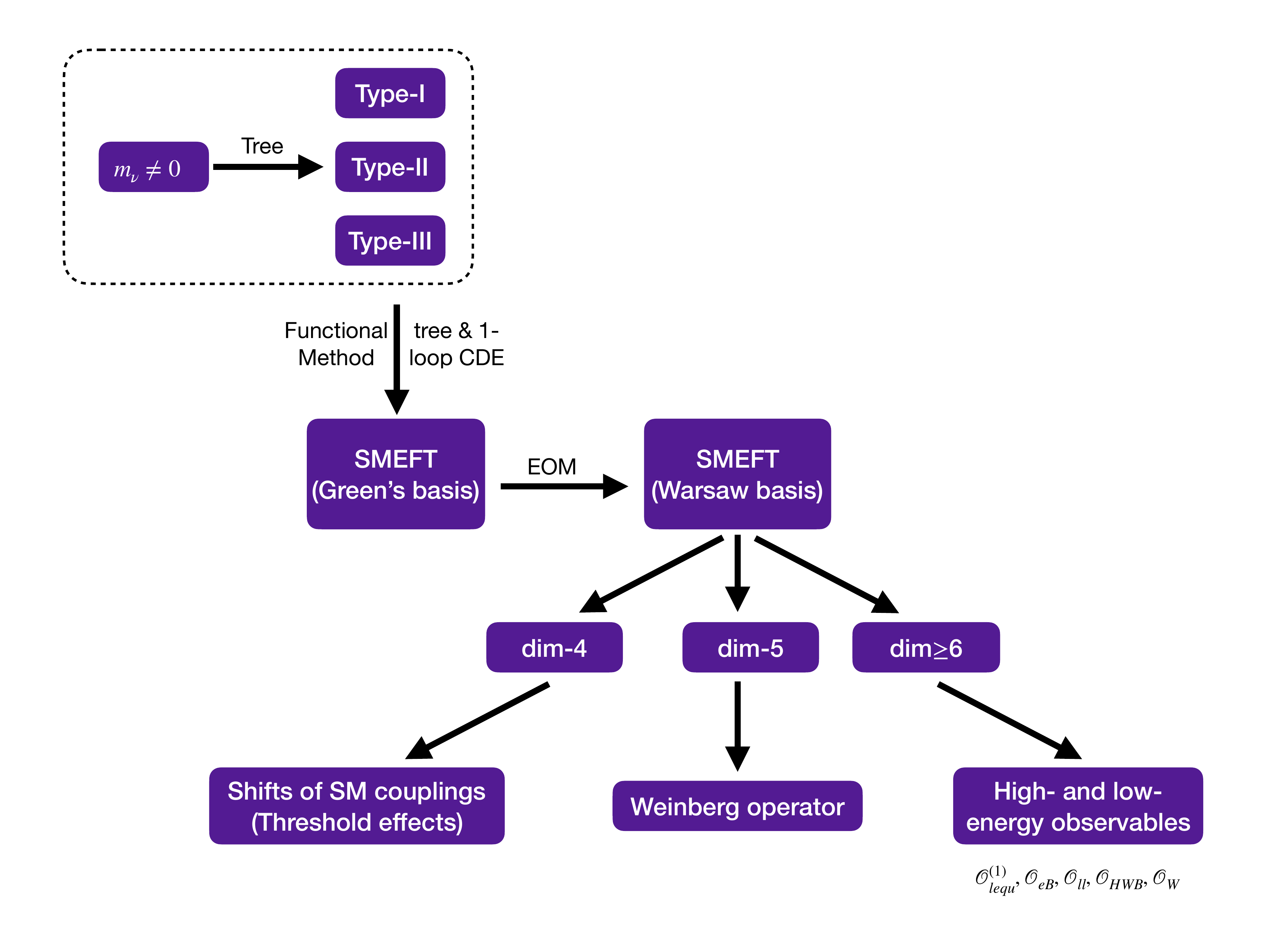}
	\caption{A schematic plot of the workflow, where the dashed box represent the three UV models that are responsible for neutrino masses at tree level. The operators after the functional matching can be reduced to the Green's basis, subject to equation of motion redundancy. Eliminating this redundancy results in the Warsaw basis. See the main text for details.}\label{fig:workflow}
\end{figure}

In the resulting Warsaw basis, both the dimension-4 and -6 operators would contribute to shifting the bare parameters in $\mathcal{L}_{\rm SM}$ at the matching scale, though the dimension-4 ones dominate when both of them are present. {\color{black}This shift at the matching scale would in turn affect the Higgs mass, Higgs couplings, and also EWSB through radiative corrections that we will investigate in detail in this work.} The dimension-5 operators are nothing but the Weinberg operators. The dimension-6 operators, which could be used for model discrimination, have very rich phenomenological effects as we will discuss later in this article.

The rest of this article is organized as follows, which is also pictorially summarized in figure\,\ref{fig:workflow}: We briefly review the functional method and the covariant derivative expansion technique in section\,\ref{sec:functionalmethod}, and also discuss how to translate the Green's basis into the Warsaw basis after removing the redundancy from EOM. We then present our tree and one-loop matching results for all the three seesaw models in section\,\ref{sec:seesaw123}. In section\,\ref{sec:ewsb}, we solve the complete one-loop running of the renormalization group equations (RGEs) using {\tt PyR@TE3}\,\cite{Sartore:2020gou}, addressing the difference of the three seesaw models in triggering electroweak symmetry breaking (EWSB). Phenomenological implications are then qualitatively discussed in section\,\ref{sec:pheno}, with a special emphasis on operators listed in figure\,\ref{fig:workflow} due to their rich phenomena. We then conclude in section\,\ref{sec:conclusions}.

%% file: func_match.tex
\section{Brief review of functional matching}\label{sec:functionalmethod}
Conventionally, there are two major methods to perform the matching between EFTs and UV theories: For amplitude/diagram matching, one firstly selects an operator basis, and then computes and compares the amplitudes in both the UV theory and the EFT to obtain the matching between the Wilson coefficients and the UV parameters. By contrast, the functional method, using the effective action to match the two theories, provides a direct way to that end. An explicit gauge-invariant method was introduced for the latter method firstly in Ref.\,\cite{Henning:2014wua}, which was then developed further and became routinizable in recent years \cite{Henning:2016lyp, Zhang:2016pja, Cohen:2020fcu}. 
For this reason, we adopt the functional method for one-loop matching in this work. Before presenting our matching results, we briefly review the functional method in this section.

\subsection{tree and one-loop level of functional matching}\label{subsec:ProcOfFuncMatch}
The basic idea of the functional method is to find the effective Lagrangian ${\cal L}_{\rm EFT}[\phi]$ such that the one-light-particle-irreducible (1LPI) effective action of light fields in the EFT matches with the 1LPI effective action induced from certain UV Lagrangian ${\cal L}_{\rm UV}[\phi,\Phi]$, where we use $\phi$ and $\Phi$ symbolically for any light and heavy degrees of freedom, respectively. This can be done with the background field method:
\eqal{\Gamma_{\rm L, EFT}[\phi_b] \overset{!}{=} \Gamma_{\rm L, UV}[\phi_b],}
where $\Gamma_{\rm L}$ is the 1LPI effective action, and the subscript $b$ means the background fields. 

The effective action can be derived from the generating functional,
\eqal{\Gamma_{\rm L, EFT}[\bar{\phi}] := -i\log Z_{\rm EFT}[J_\phi] - \int {\rm d}^4 x J_\phi \bar{\phi},}
where
\eqal{Z_{\rm EFT}[J_\phi] := \int {\cal D}\phi e^{i\int {\rm d}^4x ({\cal L}_{\rm EFT}[\phi] + J_\phi \phi)}}
and 
\eqal{\bar{\phi} := -i\frac{\delta}{\delta J_\phi}\log Z_{\rm EFT}[J_\phi].}
In the background field method, all fields are separated into classical backgrounds and quantum fluctuations (noted by prime): $\phi = \phi_b + \phi'$. The classical backgrounds are obtained as the extrema of the generating functional, which satisfy
\eqal{\frac{\delta}{\delta \phi}{\cal L}_{\rm EFT}[\phi_b] + J_\phi = 0.}
The effective action of the UV theory is derived in the same way,
\eqal{\Gamma_{\rm L, UV}[\phi_b] := -i\log Z_{\rm UV}[J_\phi, J_\Phi = 0] - \int {\rm d}^4x J_\phi \phi_b,}
except that the external source of heavy field $J_\Phi$ is set to zero and the heavy background fields $\Phi_b$ are replaced by the extremum conditions
\eqal{\frac{\delta}{\Phi}{\cal L}_{\rm UV}[\phi_b,\Phi_b] = \frac{\delta}{\phi}{\cal L}_{\rm UV}[\phi_b,\Phi_b] + J_\phi = 0 ~ \Rightarrow ~ \Phi_b = \Phi_c[\phi_b].}

${\cal L}_{\rm EFT}[\phi]$ can be derived order by order,
\eqal{{\cal L}_{\rm EFT}^{(0)}[\phi] = {\cal L}_{\rm EFT}^{(0)}[\phi] + {\cal L}_{\rm EFT}^{(1)}[\phi] + \cdots,}
where the superscript represents the Lagrangian that is retained in matching with $n$th-loop expansion of $Z_{\rm UV}$. At tree level, the matching is trivial:
\eqal{\Gamma_{\rm L,UV}^{(0)}[\phi_b] = \int {\rm d}^4x {\cal L}_{\rm UV}[\phi_b, \Phi_b]\big|_{\Phi_b = \Phi_c[\phi_b]},}
with
\eqal{\Gamma_{\rm L, EFT}^{(0)}[\phi_b] = \int {\rm d}^4x {\cal L}_{\rm EFT}^{(0)}[\phi_b].}
At one loop, the effective action can be expressed as
\eqal{\Gamma_{\rm L,UV}^{(1)}[\phi_b] & = \frac{i}{2}{\rm STr}\log \left(-\frac{\delta^2 S_{\rm UV}[\phi_b,\Phi_b]}{\delta \varphi_i \delta \varphi_j} \Bigg|_{\Phi_b = \Phi_c[\phi_b]} \right), \\
\Gamma_{\rm L, EFT}^{(1)}[\phi_b] & = \int {\rm d}^4x {\cal L}_{\rm EFT}^{(1)}[\phi_b] + \frac{i}{2}{\rm STr} \log \left(-\frac{\delta^2 S_{\rm EFT}^{(0)}[\phi_b]}{\delta \varphi_i \delta \varphi_j} \right),}
where $S$ is the action and the superscript $(0)$ represents the action from ${\cal L}_{\rm EFT}^{(0)}$. The variation is taken for all fields $\varphi_i = \phi,\Phi$, and the supertrace ${\rm STr}$ means the trace in both the internal, gauge and spin for example, and the external space, the momentum space for example. Note that the supertrace takes opposite signs for fermionic and bosonic degrees of freedoms due to spin statistics. Using the method of regions \cite{Beneke:1997zp,Smirnov:2002pj}, the one-loop level 1LPI effective action of the UV theory can be divided into contributions from the hard and the soft regions, and the latter matches onto the ${\rm STr}$ term from ${\cal L}_{\rm EFT}^{(0)}$ \cite{Zhang:2016pja}. As a result, the matching condition gives rise to
\eqal{\int {\rm d}^4x {\cal L}_{\rm EFT}^{(0)}[\phi_b] & = \int {\rm d}^4x {\cal L}_{\rm UV}[\phi_b, \Phi_b]\big|_{\Phi_b = \Phi_c[\phi_b]}, \\
\int {\rm d}^4x {\cal L}_{\rm EFT}^{(1)}[\phi_b] & = \frac{i}{2}{\rm STr} \log \left(-\frac{\delta^2 S_{\rm UV}[\phi_b,\Phi_b]}{\delta \varphi_i \delta \varphi_j} \Bigg|_{\Phi_b = \Phi_c[\phi_b]} \right)\Bigg|_{\rm hard}.}

\subsection{Supertrace evaluation}\label{subsec:EvalOfSupertrace}
For any UV theory, the variation of $S_{\rm UV}$ can be generically classified into a kinematic part $\Delta^{-1}$ and an interaction part $X$. Due to the heavy scale $M$ suppression since $\Delta X\sim1/M$, one readily has
\eqal{\int {\rm d}^4x {\cal L}_{\rm EFT}^{(1)} = \frac{i}{2}{\rm STr} \log(\Delta^{-1} - X)\big|_{\rm hard} = \frac{i}{2}{\rm STr} \log\Delta^{-1}\big|_{\rm hard} - \frac{i}{2}\sum_{n = 1}\frac{1}{n}{\rm STr}(\Delta X)^n\big|_{\rm hard},}
where the first term, called the log-type terms, can be calculated in a universal form since the kinematic part only depends on the field types. The second term, referred to as the power-type terms, specifies the expansion order from the fact that $\Delta X\sim1/M$. Therefore, the power-type terms naturally signals the truncation order up to a certain dimension for the EFTs. 

We stress that each term in the supertrace can be presented covariantly, meaning that it only depends on $\varphi$ and the covariant derivative $D_\mu$. Therefore, with partial derivative expansion, one can firstly break the covariant derivatives into normal derivatives and gauge fields, and then expand each propagator into series of the gauge fields. However, this approach ignores gauge symmetries and yields a large number of terms from both the expansion of propagators and possibly existing open derivatives in the interacting terms $X$. Fortunately, this complication can be cured significantly by covariant derivative expansion (CDE) in an explicit gauge covariant way\cite{Gaillard:1985uh,Chan:1986jq,Cheyette:1987qz,Henning:2014wua}. The heart of CDE is to recover all non-gauge-invariant terms to total derivatives in the expansion and to transform all open derivatives into `closed forms', i.e. derivatives acting on only one term. A pedagogical description is presented in Appendix B of \cite{Cohen:2019btp}, which is briefly reviewed here. To be general, a supertrace term can be written as
\eqal{{\rm STr} Q(P_\mu, U(\phi, (D_\mu\phi)) = \pm \int {\rm d}^4x \langle x | {\rm tr} Q(P_\mu, U) | x \rangle ,}
where ``${\rm tr}$'' here takes trace only on the internal space, and $P_\mu = iD_\mu$ is an hermitian version of the covariant derivative. Inserting the identity operator $\int {\rm d}^4 p /(2\pi)^4 |p \rangle \langle p |$, one obtains
\eqal{{\rm STr} Q(P_\mu, U(\phi, (D_\mu\phi))  & = \pm \int {\rm d}^4x \int \frac{{\rm d}^4 p}{(2\pi)^4} \langle x | p \rangle \langle p |{\rm tr} Q(P_\mu, U) | x \rangle \nonumber \\
& = \pm \int {\rm d}^4x \int \frac{{\rm d}^4 p}{(2\pi)^4} e^{ip\cdot x} {\rm tr} Q(P_\mu, U(x)) e^{-ip\cdot x} \nonumber \\
& = \pm \int {\rm d}^4x \int \frac{{\rm d}^4 p}{(2\pi)^4} {\rm tr} Q(P_\mu + p_\mu, U(x)).}
Now the trick here is to put the trace between a pair of operators:
\eqal{{\rm STr} Q(P_\mu, U(\phi, (D_\mu\phi)) = \pm \int {\rm d}^4x \int \frac{{\rm d}^4 p}{(2\pi)^4} e^{-P\cdot \partial_p} {\rm tr} Q(P_\mu + p_\mu, U(x)) e^{P\cdot \partial_p},}
where $\partial_p^\mu$ is the abbreviation of $\partial/\partial p_\mu$. Note that after Taylor expansion of the exponential terms, anything but the unity vanishes because of integration by parts. Therefore, by moving the operator $e^{-P\cdot \partial_p} $ from one side to the other, and applying the Baker-Campbell-Hausdorff formula, one can write each term in $Q$ as
\eqal{e^{-P\cdot \partial_p} (p_\mu + P_\mu) e^{P\cdot \partial_p} & =: p_\mu + i {\cal X}_{\mu\nu} \partial^\nu_p,\\
e^{-P\cdot \partial_p} U e^{P\cdot \partial_p} & = \sum_{n = 0} \frac{(-i)^n}{n!} \left(D_{\alpha_1} \cdots D_{\alpha_n} U\right) \partial_p^{\alpha_1}\cdots \partial_p^{\alpha_n} =: \tilde{U},}
with
\eqal{{\cal X}_{\mu\nu} := \sum_{n = 0} \frac{(-i)^n}{(n+2)n!} \left(D_{\alpha_1} \cdots D_{\alpha_n} X_{\mu\nu}\right) \partial_p^{\alpha_1}\cdots \partial_p^{\alpha_n},\label{eq:defofcalX}}
and $X_{\mu\nu}$ is the gauge field strength tensor. Note that now all terms in the supertraces are free of any open derivatives:
\eqal{{\rm STr} Q(P_\mu, U(\phi, (D_\mu\phi)) = \pm \int {\rm d}^4x \int \frac{{\rm d}^4 p}{(2\pi)^4} {\rm tr} Q(p_\mu + i {\cal X}_{\mu\nu} \partial^\nu_p, \tilde{U}) , }
such that one can further Taylor expand $\Delta$ for evaluating these supertraces, with all open derivatives translated into ${\cal X}_{\mu\nu}$.

\subsection{From Green's basis to the Warsaw basis}\label{subsec:GreenToWarsaw}

The result of functional matching can be practically complicated. With identities from the Lorentz and the gauge groups, as well as integration by part, it can be turned into the Green's basis \cite{Jiang:2018pbd,Gherardi:2020det,Chala:2021pll, Chala:2021cgt}
, which is the simplest complete off-shell basis. However, for process where all particles are on-shell, the Green's basis would be over-complete and would inherit redundancy from EOM, meaning that two or more operators in the Green's basis can be related through EOM. Removing this redundancy is nontrivial. For this reason, we explain here how to eliminate this over-completeness with one example. To put it in a general way, we use $H$, $\psi$ and $X_\mu$ for scalar, fermion and gauge bosons, respectively, in the following. 
On one hand, the EOM can be applied quite straightforwardly when $D_\mu D^\mu H$, ${D\!\!\!\!/} \, \psi$ and $D^\mu X_{\mu\nu}$ show up in the operators. On the other hand, in other cases, $D_\mu \gamma_\nu \psi$ and $D_\mu X_{\nu\rho}$ for example, one would need the symmetry between the Lorentz indices inside the derivatives and other conditions to remove the redundancy. Let us take the $\psi^2 H D^2$ for illustration. The building blocks of Lorentz indices are $g^{\mu\nu}$ and $g_{\mu\nu}$, the antisymmetric tensor $\epsilon_{\mu\nu\rho\sigma}$, and the gamma matrices $\gamma_\mu, \sigma_{\mu\nu} = i \gamma_{[\mu} \gamma_{\nu]}$. To produce an even number of Lorentz indices, the gamma matrices between fermions can only be $\mathbf{1}$ or $\sigma_{\mu\nu}$. Complication arises when two derivatives act on a fermion and the Higgs doublet, respectively. For the following example, it can be rewritten as
\eqal{\bar{\psi} D_\mu \psi D^\mu H &= \bar{\psi} D^\nu \frac{1}{2}(\gamma_\mu \gamma_\nu + \gamma_\nu \gamma_\mu) \psi D^\mu H \nonumber\\
& = \frac{1}{2} \bar{\psi} (\gamma_\mu {D\!\!\!\!/} \, - \overset{\leftarrow}{D\!\!\!\!/} \, \gamma_\mu) \psi D^\mu H - \frac{1}{2} \bar{\psi} \gamma_\nu \gamma_\mu \psi D^\nu D^\mu H.}
Recall that $\gamma_\mu \gamma_\nu = g_{\mu\nu} - i\sigma_{\mu\nu}$, one can further reduce the last term to an EOM of the Higgs and a commutator of covariant derivatives $[D_\mu, D_\nu] = -iX_{\mu\nu}$. We comment that for all dimension-6 SMEFT operators, derivatives acting on a fermion can always be reduced by its EOM, which, however, may not be the case for higher dimensional operators.

%% file: type123seesaw.tex
\section{Type-I, II, III seesaw models and matching}\label{sec:seesaw123}
As is mentioned above, throughout this work, we assume neutrino masses that are responsible for neutrino oscillations are generated through the dimension-5 Weinberg operator\,\cite{Weinberg:1979sa}. At tree-level, there are only three UV completions for this operator that are well-known as the type-I, II and III seesaw models. In this section, we will first briefly review the details of these models, and then present the induced SMEFT operators, up to dimension-6 at the one-loop level, from the functional method discussed in section\,\ref{sec:functionalmethod}. We should mention that the one-loop matching for the type-I seesaw model has been investigated by several groups recently~\cite{Coy:2018bxr,Coy:2021hyr,Zhang:2021jdf,Ohlsson:2022hfl}. After utilizing the CDE technique to obtain the Green's and Warsaw basis operators, we perform a very careful cross-check against their results and find agreement. 




\subsection{Model setup}

To verify that there are no other models which can generate neutrino masses through the dimension-5 Weinberg operator, one can draw all tree-level processes of $\ell\ell \rightarrow HH$. The Lorentz and gauge representations of two leptons and two Higgs place restrictions on the mediating propagator in the $s$ channel to be an $\rm SU(2)_L$ triplet scalar with $-1$ hypercharge, which is just the type-II seesaw model. Similarly, the type-I and -III seesaw models can be identified from the gauge quantum numbers of the mediating fermion in $t$-channel processes. In this subsection, we will firstly define these models in our notations, and then present our matching results up to one-loop order.

Throughout this paper, the SM Lagrangian is parameterized as
\eqal{{\cal L}_{\rm SM} = & - \frac{1}{4}G_{\mu\nu}^A G^{A\mu\nu} - \frac{1}{4} W_{\mu\nu}^I W^{I\mu\nu} - \frac{1}{4} B_{\mu\nu} B^{\mu\nu} + (D_\mu H)^\dagger (D^\mu H) + \mu_H^2 H^\dagger H - \lambda_H (H^\dagger H)^2 \nonumber \\
& + \sum_{f = q,l,u,d,e} \bar{f} i D\!\!\!\!/ \, f - \left( \bar{\ell}_L Y_e e_R H + \bar{q}_L Y_u u_R \tilde{H} + \bar{q}_L Y_d d_R H + {\rm h.c.} \right) ,}
where $D_\mu = \partial_\mu - ig_1YB_\mu - ig_2 {\sigma^I}W_\mu^I/{2} - i g_3 T^A G_\mu^A$ for fields in fundamental representation of gauge groups ${\rm SU(3)}_C\times {\rm SU(2)}_L\times {\rm U(1)}_Y$, with $\sigma^I$ the Pauli matrices and $T^A$ related to the Gell-Mann matrices $\lambda^A$ by $T^A = \lambda^A/2$.

\subsubsection{Type-I seesaw}

In the type-I seesaw model\,\cite{Minkowski:1977sc,Yanagida:1979as,Mohapatra:1979ia,GellMann:1980vs,Mohapatra:1986aw,levy2013quarks}, heavy right-handed gauge singlet fermions $N_R$ are introduced into the SM, which will result in the following extra Lagrangian terms
\eqal{\Delta\mathcal{L} & = \overline{N}_R i\partial\!\!\!/ N_R - \left(\frac{1}{2}\overline{N_R^c} M_N N_R + \overline{\ell}_L Y_\nu \tilde{H} N_R + {\rm h.c.} \right) \nonumber\\
& = \frac{1}{2} \overline{N} (i\partial\!\!\!/ - M_N) N - \frac{1}{2}\overline{N} \left( Y_\nu^\dagger \tilde{H}^\dagger  \ell_L + Y_\nu^T \tilde{H}^T \ell_L^c \right) - \frac{1}{2} \left(\bar{\ell}_L \tilde{H} Y_\nu + \bar{\ell}_L^c \tilde{H}^* Y_\nu^* \right)N.}
Here $N = N_R + N_R^c$, and flavor indices are all omitted. We consider the case that all heavy neutrinos $N_p$ have the same mass $M_N = {\rm diag}(M)$.

\subsubsection{Type-II seesaw}

There is a new scalar triplet $\Delta^I$ in the type-II seesaw model\,\cite{Mohapatra:1980yp,Konetschny:1977bn,Cheng:1980qt,Lazarides:1980nt,Schechter:1980gr}, which transforms as the $(1,3)_{1}$ representation of the SM gauge group. The scalar potential is modified to
\eqal{V(H,\Delta) & =  -\mu_H^2 H^\dagger H + \lambda_H (H^\dagger H)^2  + M^2 {\rm tr}(\Delta^\dagger \Delta) + \left[\frac{\mu}{\sqrt{2}} H^Ti\sigma^2\Delta^\dagger H + \rm h.c. \right] \nonumber \\
& + \lambda_2 [{\rm tr}(\Delta^\dagger \Delta)]^2 + \lambda_3 {\rm tr}(\Delta^\dagger \Delta \Delta^\dagger \Delta) + \lambda_4 H^\dagger H {\rm tr}(\Delta^\dagger \Delta) + 2\lambda_5 H^\dagger \Delta \Delta^\dagger H,}
and there is an extra Yukawa term for neutrino mass generation:
\eqal{\Delta {\cal L} =  - \left(\frac{Y_\nu^{ij}}{\sqrt{2}} \overline{\ell_{Li}^c} i\sigma^2\Delta\ell_{Lj}+\rm h.c.\right),}
where $\Delta = \sigma^I \Delta^I / \sqrt{2}$. 
We also point out that the Yukawa matrix ${Y_\nu}$ is symmetric in this case.

\subsubsection{Type-III seesaw}

Similar to the type-I seesaw model, new heavy right-handed fermions $\Sigma^I$ are introduced into the SM to obtain the type-III seesaw model\,\cite{Foot:1988aq,Barr:2003nn}. The difference here is that $\Sigma^I$ is an ${\rm SU(2)}_L$ triplet, and the Lagrangian can be written as
\eqal{\Delta\mathcal{L} & = \overline{\Sigma}_R^I iD\!\!\!\!/ \, \Sigma_R^I - \left(\frac{1}{2}\overline{\Sigma}_R^I M_\Sigma \Sigma_R^{I,c} + \overline{\Sigma}_R^I Y_\Sigma \tilde{H}^\dagger \sigma^I \ell_L + {\rm h.c.} \right) \nonumber \\
& = \frac{1}{2} \overline{\Sigma}^I (iD\!\!\!\!/ \, - M_\Sigma) \Sigma^I - \frac{1}{2} \left[ \overline{\Sigma}^I \left( Y_\Sigma \tilde{H}^\dagger \sigma^I \ell_L + Y_\Sigma^* H^\dagger \sigma^I i\sigma^2 \ell_L^c \right) + {\rm h.c.} \right].}
Here $\Sigma^I = \Sigma^I_R + \Sigma^{I,c}_R$. Same as the case in the type-I model, we consider that all new fermions $\Sigma_p$ have the same mass $M_\Sigma = {\rm diag}(M)$.

\subsection{Tree-level matching}\label{subsec:treematching}

\begin{table}[h!]
  \centering
  \resizebox{\textwidth}{!}{  
\begin{tabular}{|c||l|l|l|}
\hline\multicolumn{4}{|c|}{{\color{blue} Tree-level matching in the Green's basis}} \\
\hline Operator & Type-I & Type-II & Type-III \\
\hline$\left(H^{\dagger} H\right)^{2}$ & & $\frac{\mu^{2}}{2 M^{2}}$ & \\
\hline$\left(\bar{\ell}_{L p}^{c} \tilde{H}^{*}\right)\left(\tilde{H}^{\dagger} \ell_{L r}\right)$ & $\frac{1}{2M}\left(Y_{\nu}^{*} Y_{\nu}^{\dagger}\right)^{pr}$ & $-\frac{\mu}{2 M^{2}} Y_{\nu}^{pr}$ & $\frac{1}{2M}\left(Y_{\Sigma}^{T} Y_{\Sigma}\right)^{pr}$ \\
\hline $\mathcal{O}_{H D}$ & & $\frac{\mu^{2}}{M^{4}}$ & \\
\hline $\mathcal{R}_{H D}'$ & & $\frac{\mu^{2}}{M^{4}}$ & \\
\hline $\mathcal{O}_{H}$ &  & $-\frac{\mu^{2}}{2 M^{4}}\left(\lambda_{4}+2 \lambda_{5}\right)$ & \\
\hline $\mathcal{O}_{\ell \ell, p r s r}$ &  & $\frac{1}{4M^2}Y_\nu^{*ps}Y_\nu^{rt}$ & \\
\hline $\mathcal{O}_{H \ell}^{(1)}$ & $\frac{1}{4M^2}\left(Y_{\nu} Y_{\nu}^{\dagger}\right)^{pr}$ & & $\frac{3}{4M^2}\left(Y_{\Sigma}^{\dagger} Y_{\Sigma}\right)^{pr}$ \\
\hline $\mathcal{R}_{H \ell}^{'(1)}$ & & & $\frac{3}{4M^2}\left(Y_{\Sigma}^{\dagger} Y_{\Sigma}\right)^{pr}$ \\
\hline $\mathcal{O}_{H \ell}^{(3)}$ & $-\frac{1}{4M^2}\left(Y_{\nu} Y_{\nu}^{\dagger}\right)^{pr}$ & & $\frac{1}{4}\left(Y_{\Sigma}^{\dagger} M^{-2} Y_{\Sigma}\right)^{pr}$ \\
\hline $\mathcal{R}_{H \ell}^{'(3)}$ & & & $\frac{1}{4M^2}\left(Y_{\Sigma}^{\dagger} Y_{\Sigma}\right)^{pr}$ \\
\hline\multicolumn{4}{|c|}{{\color{blue} Tree-level matching in the Warsaw basis}} \\
\hline Operator & Type-I & Type-II & Type-III \\
\hline $\left(H^{\dagger} H\right)^{2}$ & & $\frac{\mu^{2}}{2 M^{2}}\left(1-\frac{2 \mu_{H}^{2}}{M^{2}}\right)$ & \\
\hline $\left(\bar{\ell}_{L p}^{c} \tilde{H}^{*}\right)\left(\tilde{H}^{\dagger} {\ell}_{L r}\right)$ & $\frac{1}{2M}\left(Y_{\nu}^{*} Y_{\nu}^{\dagger}\right)^{pr}$ & $-\frac{\mu}{2 M^{2}} Y_{\nu}^{pr}$ & $\frac{1}{2M}\left(Y_{\Sigma}^{T} Y_{\Sigma}\right)^{pr}$ \\
\hline $\mathcal{O}_{H \square}$ & & $\frac{\mu^2}{2M^4}$ & \\
\hline $\mathcal{O}_{H D}$ & & $\frac{ \mu^{2}}{M^{4}}$ & \\
\hline $\mathcal{O}_{H}$ & & $-\frac{\mu^{2}}{2 M^{4}}\left(\lambda_{4}+2 \lambda_{5}-4 \lambda_{H}\right)-\frac{\mu^4}{M^6}$ & \\
\hline $\mathcal{O}_{e H, pr}$ & & $\frac{\mu^{2}}{2 M^{4}} Y_{e}^{pr}$ & $\frac{1}{M^2}(Y_\Sigma^\dagger Y_\Sigma Y_e)^{pr}$\\
\hline $\mathcal{O}_{u H, pr}$ & & $\frac{\mu^{2}}{2 M^{4}} Y_{u}^{pr}$ & \\
\hline $\mathcal{O}_{d H, pr}$ & & $\frac{\mu^{2}}{2 M^{4}} Y_{d}^{pr}$ & \\
\hline $\mathcal{O}_{H \ell}^{(1)}$ & $\frac{1}{4M^2}\left(Y_{\nu} Y_{\nu}^{\dagger}\right)^{pr}$ & & $\frac{3}{4M^2}\left(Y_{\Sigma}^{\dagger} Y_{\Sigma}\right)^{pr}$ \\
\hline $\mathcal{O}_{H \ell}^{(3)}$ & -$\frac{1}{4M^2}\left(Y_{\nu} Y_{\nu}^{\dagger}\right)^{pr}$ & & $\frac{1}{4M^2}\left(Y_{\Sigma}^{\dagger} Y_{\Sigma}\right)^{pr}$ \\
\hline $\mathcal{O}_{\ell \ell, p r s t}$ &  & $\frac{1}{4M^2}Y_\nu^{*ps}Y_\nu^{rt}$ & \\
\hline
\end{tabular}
}\caption{Tree-level matching for the three seesaw models using the functional method. Results are presented in both the Green's and the Warsaw bases. Empty cells imply the specific model does not generate the corresponding operator. The specific forms of the dimension-6 operators are listed in table\,\ref{tab:operatorbasis}.}\label{tab:treematching}
\end{table}

Following the matching method in section\,\ref{subsec:ProcOfFuncMatch}, we can substitute all heavy fields in the Lagrangian by their equations of motion to obtain tree-level matching results. In different seesaw models, their EOMs are given respectively by
\eqal{N_c = & -\frac{1}{M} \left(\tilde{H}^\dagger Y_\nu^\dagger \ell_L + \tilde{H}^T Y_\nu^T \ell_L^c \right) - \frac{1}{M^2} i\partial\!\!\!/ \left(\tilde{H}^\dagger Y_\nu^\dagger \ell_L + \tilde{H}^T Y_\nu^T \ell_L^c \right) + O(M^{-3}),}
\eqal{\Delta^I_c = & -\frac{\mu}{2M^2}H^T i\sigma^2 \sigma^I H + \frac{(Y_\nu^*)^{ij}}{2M^2}\bar{\ell}_{L,j} \sigma^I  i\sigma^2  \ell_{L,i}^c \nonumber \\
 & + \frac{\mu}{2M^4} \left[\delta^{IJ}(D_\mu D^\mu - \lambda_4 H^\dagger H) - \lambda_5 H^\dagger \sigma^J\sigma^I H\right] \left(H^T  i\sigma^2  \sigma^J H\right) + O({\rm dim.} \ge 5),\\
\Sigma_c^I = & - \frac{1}{M}\left(Y_\Sigma \tilde{H}^\dagger \sigma^I \ell_L + Y_\Sigma^* H^\dagger \sigma^I  i\sigma^2  \ell_L^c \right) - \frac{1}{M^2} iD\!\!\!\!/ \, \left(Y_\Sigma \tilde{H}^\dagger \sigma^I \ell_L + Y_\Sigma^* H^\dagger \sigma^I  i\sigma^2  \ell_L^c \right) + O(M^{-3}),}
where the subscript $c$ means the classical solutions. Terms that have dimensions higher than 4 are omitted since they only contribute to operators with dimension higher than 6 in SMEFT, which will not be covered in this work. The tree-level EFT Lagrangian is listed in table \ref{tab:treematching} in both the Green's and the Warsaw bases. Throughout this paper, all operators in Warsaw basis are denoted by ${\cal O}_i$, while the redundant operators from EOM in the Green's basis are denoted by ${\cal R}_i$.

As is shown in table\,\ref{tab:treematching}, the effective operators generated by new fields are quite different due to the spins of new fields. 
To be concrete, except the dimension-5 Weinberg operator that is required by neutrino oscillations, none of the operators in the Green's basis can be simultaneously generated by the ferminoic type-I/-III model and the bosonic type-II seesaw model. This difference comes from the different topologies of the Feynman diagrams in generating these effective operators and their restrictions on the spins of the mediating propagators. 
Although after eliminating the redundancy an operator ${\cal O}_{eH}$ exists for both type-II and -III models, bosonic operators could still be used as a possible type-II seesaw model identifier.

Another issue needs to be discussed here is that, although the sign ahead of the dimension-5 Weinberg operator is positive in type-I and -III seesaw models, we can always redefine the lepton fields to produce a correct neutrino mass term. After electroweak symmetry breaking, the neutrino sector can be written as
\eqal{\Delta {\cal L}_\nu = \bar{\nu}_{Li} i\partial\!\!\!/ \nu_{Li} + \frac{1}{2} v^2 C_5^{ij} \bar{\nu}_{Li}^c\nu_{Lj} + \frac{1}{2} v^2 C_5^{*ij} \bar{\nu}_{Lj}\nu_{Li}^c,}
where $C_5^{ij}$ is the coefficient of the Weinberg operator ${\cal O}_5 = \left( \bar{\ell}_{Li}^c \tilde{H}^* \right)\left( \tilde{H}^\dagger \ell_{Lj}\right)$. 
One can always redefine the neutrino field $\nu_{Li} \rightarrow i \nu_{L,i}$ to reverse the sign of the mass term without touching the kinetic term, and thereafter diagonalize the mass matrix. 
This sign-reversing procedure can also be done after the diagonalization of the mass matrix, which is a complex-orthogonal diagonalization of a complex-symmetric matrix:
\eqal{\nu_{L,i} \rightarrow U_{i}^{\phantom{i} j} \nu_{L,j},~ \bar{\nu}_{L,i}^c \rightarrow \bar{\nu}_{L,j}^c (U^T)^{j}_{\phantom{j} i}.}
Redefining the phase of each neutrino will absorb the imaginary part of each mass term in the Lagrangian.

\subsection{One-loop matching}

The one-loop functional matching relies on the calculation of supertrace in the power-type terms, which can be systematically done by the CDE method discussed in section\, \ref{subsec:EvalOfSupertrace}. It has two inputs: the functional derivative of the kinematic part and the interaction part of action, noted as $\Delta^{-1}$ and $X$. The first part, $\Delta^{-1}$, only depends on the masses and the gauge quantum numbers of the fields and is thus free of any couplings in the UV model. The second part, $X$, has mass dimension of at least one and contains the interacting information. In practice, one only needs to calculate the $X$ terms, which is illustrated by the following example in the type-III model. Taking the second functional variation on the action and extracting the remaining term, one obtains
\eqal{-\delta^2{\cal L} \supset & -\delta^2 \left(  -\frac{1}{2}\bar{\Sigma}_p^I ig_2\epsilon^{IJK} W_\mu^J \gamma^\mu \Sigma_p^K \right) \nonumber \\
\supset & (\delta \bar{\Sigma}_p^I) \left( ig_2\epsilon^{IJK} \gamma^\mu \Sigma_p^K \right) (\delta W_\mu^J) + (\delta W_\mu^J) \left( \bar{\Sigma}_p^I ig_2\epsilon^{IJK} \gamma^\mu \right) (\delta \Sigma_p^K) \\
\Rightarrow X_{\Sigma W} {:=} & - \frac{\delta^2 {\cal L}}{\delta{\bar{\Sigma}_p^I}\delta W_\nu^J} = ig_2 \epsilon^{IJK} \gamma^\nu \Sigma_p^K, ~ X_{W\Sigma} {:=} - \frac{\delta^2 {\cal L}}{\delta W_\mu^I \delta \Sigma_P^J} = ig_2 \epsilon^{KIJ} \bar{\Sigma}_p^K \gamma^\mu,}
where $(T^I)_{JK} = -i \epsilon^{IJK}$ is the $\rm SU(2)$ generator of the adjoint representation. The first line comes from the kinematic part of the type-III model, and in the second line, the second functional variation to $\Sigma$ is omitted because it is contained in the kinematic part of $\Sigma$. In this work, after obtaining all $X$ terms by hand, we use the {\tt Mathematica} package {\tt SuperTracer} \cite{Fuentes-Martin:2020udw} for CDE evaluation.

\begin{figure}
  \centering{
    \begin{adjustbox}{max width = \textwidth}
  \begin{tabular}{cc}
  \includegraphics[width=0.3\textwidth]{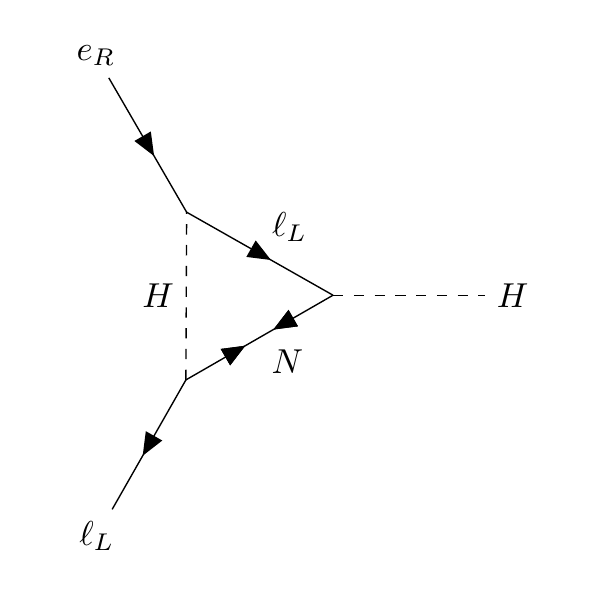} & 
    \includegraphics[width=0.3\textwidth]{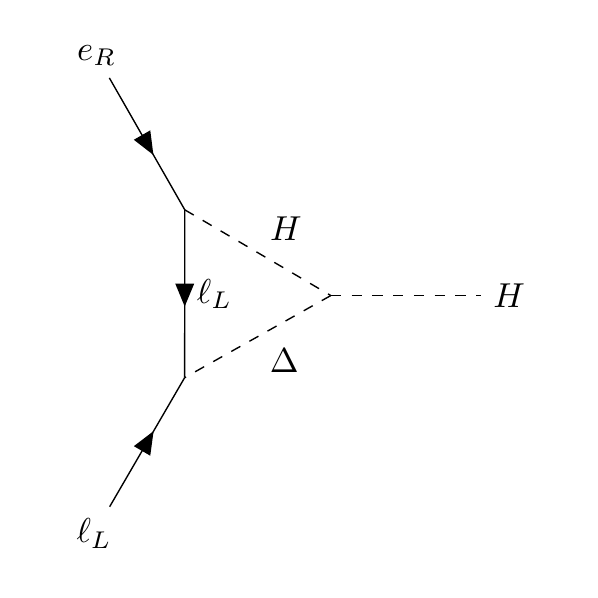}\\
  \end{tabular}
    \end{adjustbox}
    }\caption{Left: An illustrating diagram for generating operators $\bar{\ell}_L e_R H, {\cal O}_{eX}, {\cal R}_{eHDn}$ in the type-I seesaw model. Gauge bosons in operators ${\cal O}_{eX}$ and covariant derivatives in ${\cal R}_{eHDn}$ are originated from heavy-mass expansion of the propagators in the loop. Right: A seemingly contributing but vanishing -- due to the 
   inconsistency in the fermion flow -- diagram that could possible translate ${\ell}_L$ into $e_R$ in the type-II model.}\label{fig:diagramyukawa}
\end{figure}

\begin{figure}
  \centering{
    \begin{adjustbox}{max width = \textwidth}
  \begin{tabular}{cc}
  \includegraphics[width=0.3\textwidth]{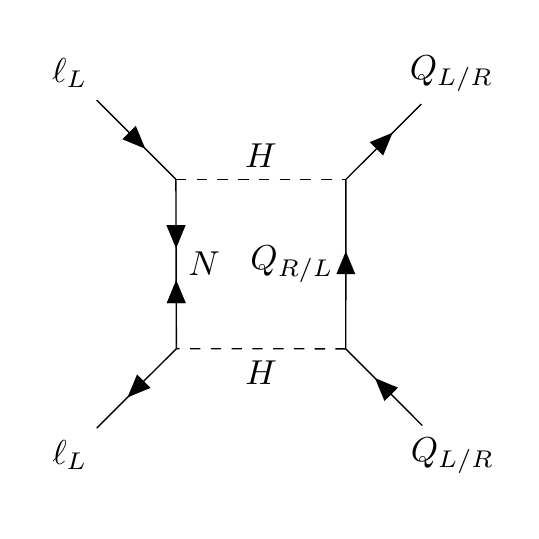} & 
    \includegraphics[width=0.3\textwidth]{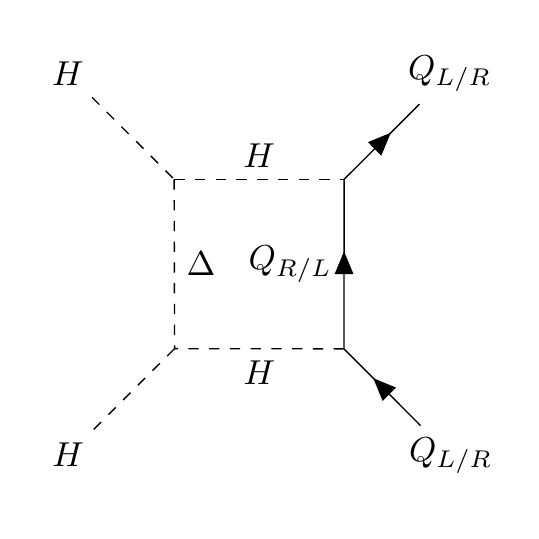}\\
  \end{tabular}
    \end{adjustbox}
    }\caption{Left: A diagram that contributes to the operator ${\cal O}_{lQ}$ in type-I seesaw model. Right: A diagram that contributes to the operator ${\cal O}_{HQ} ({\cal R}_{HQ})$ in type-II seesaw model. ${\cal O}_{QH}$ can be obtained by substituting one of the Higgs propagators for a gauge boson and adding one more Higgs propagator.}\label{fig:diagramquark}
\end{figure}

\begin{figure}
  \centering{
    \begin{adjustbox}{max width = \textwidth}
  \begin{tabular}{cc}
  \includegraphics[width=0.3\textwidth]{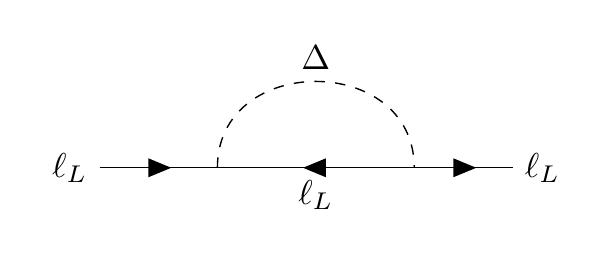} & 
    \includegraphics[width=0.3\textwidth]{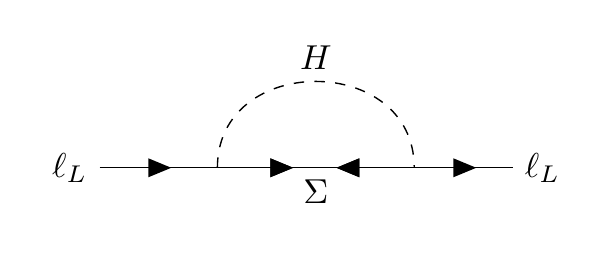}\\
  \end{tabular}
    \end{adjustbox}
    }\caption{The diagrams that contributes to effective operators ${\cal R}_{\tilde{X}\ell}'$ in the type-II and -III models respectively. The dual tensors in the effective operators can originate from the gamma matrices and covariant derivatives in the fermionic propagators in the CDE method. See the text for details.}\label{fig:diagramxl}
\end{figure}

The whole one-loop matching results are presented in the following pages.
One can find more distinctions than that at tree level through these tables. As mentioned in section\,\ref{subsec:EvalOfSupertrace}, the results can be divided into log-type and power-type contributions. Since the log-type terms only depend on the gauge quantum numbers of the heavy fields, the resulting EFT from the type-I seesaw model receives vanishing contributions and that of the type-III seesaw model only has operators containing $W$ bosons. For power-type contributions, in Green's basis, there are seven classes of operators having distinctions, which are $\bar{\ell}_L e_R H, {\cal O}_{eX}, {\cal R}_{eHDn}, {\cal O}_{HQ} ({\cal R}_{HQ}), {\cal O}_{QH}, {\cal O}_{lQ}$ and ${\cal R}_{\tilde{X}\ell}'$. Here $X$ and $Q$ stand for gauge bosons, and both left- and right-handed quarks, respectively, with $n = 1,2,3,4$. Lack of the first three classes in the EFT from the type-II model can be understood diagrammatically: 
For the type-I seesaw model, $\ell_L$ can be converted to $e_R$ as depicted in figure\,\ref{fig:diagramyukawa}, while it does not exist in the type-II model due to the inconsistency in the fermion flow as shown in the right panel of figure\,\ref{fig:diagramyukawa}.
For the generation of operators that contain quarks, ${\cal O}_{HQ} ({\cal R}_{HQ}), {\cal O}_{QH}$ and ${\cal O}_{lQ}$ for example, one can consider the scattering process $XX \rightarrow \bar{q}q$. Since only the Higgs and the gauge bosons can interact with quarks, the initial states $XX$ can only be connected by heavy particles. As a consequence, the type-I and -III models have the ${\cal O}_{lQ}$ contributions but not ${\cal O}_{HQ} ({\cal R}_{HQ})$ and ${\cal O}_{QH}$, while the type-II model predicts the opposite. Figure\,\ref{fig:diagramquark} depicts two examples in the type-I and -II model respectively.
All these six classes of operators can be adopted for distinguishing the type-II model from the fermionic type-I and -III models. A special case where the type-III model is different from the type-I model is in the ${\cal R}_{\tilde{X}\ell}'$, $\bar{\psi}\gamma^\mu \overset{\leftrightarrow}{D^\nu} \psi \tilde{X}_{\mu\nu} $ operators, which contain dual tensors $\tilde{X}_{\mu\nu} = \frac{1}{2}\epsilon_{\mu\nu\rho\sigma}X^{\rho\sigma}$. 
The antisymmetric tensor $\epsilon_{\mu\nu\rho\sigma}$ $(\epsilon_{0123}=+1)$ required by the dual tensors can be deduced by simplification of three gamma matrices:
\eqal{\gamma^\mu \gamma^\nu \gamma^\rho = g^{\mu\nu}\gamma^\rho + g^{\nu\rho}\gamma^\mu - g^{\mu\rho}\gamma^\nu - i\epsilon^{\mu\nu\rho\sigma}\gamma_\sigma\gamma_5.}
In the CDE method of the functional matching, the expansion of a fermionic propagator will give rise to three gamma matrices with a gauge field strength,
\eqal{\frac{1}{iD\!\!\!\!/\, - M} \overset{\rm CDE}{\longrightarrow} \frac{p\!\!\!/ + M}{p^2 - M^2}\left(-i\gamma^\mu {\cal X}_{\mu\nu} \frac{\partial}{\partial p_\nu}\right)\frac{p\!\!\!/ + M}{p^2 - M^2} + \dots,}
where ${\cal X}_{\mu\nu}$ is defined in eq.\,\eqref{eq:defofcalX}. As a consequence, a diagram like figure\,\ref{fig:diagramxl} would contribute to ${\cal R}_{\tilde{X}\ell}'$ with $X_{\mu\nu}$ the field strength tensor of the fermion in the loop. EFT of the type-I model receives null contribution since $N$ is a gauge singlet. The $\rm SU(2)_L$ triplet $\Delta$ of the type-II model results in both ${\cal R}_{\tilde{W}\ell}'$ and ${\cal R}_{\tilde{B}\ell}'$, while EFT of the type-III model only contains ${\cal R}_{\tilde{W}\ell}'$ due to the vanishing hypercharge of $\Sigma$.

This diagrammatic way of depicting the origin of effective operators are spoiled by the EOM reduction when we consider that all particles are in their physical states, which leaves no distinct operator between different seesaw models in the power-type contribution.
However, the sign and the magnitude of the Wilson coefficients can still release some information of the UV physics. A more detailed description on the relation between the difference in effective operators and experiments is presented in section\,\ref{sec:pheno}.

\begin{table}[h!]
  \centering
  {  
\begin{tabular}{|c||l|l|l|}
\hline\multicolumn{4}{|c|}{{\color{blue}One-loop log-type operators}} \\
\hline Operator & Type-I & Type-II & Type-III \\
\hline$W_{\mu \nu}^{I} W^{I \mu \nu}$ & & $-\frac{g_{2}^{2}}{6} L$ & $-\frac{g_{2}^{2}}{3}  N_\Sigma L$ \\
\hline$B_{\mu \nu} B^{\mu \nu}$ & & $-\frac{g_{1}^{2}}{4} L $ & \\
\hline $\mathcal{O}_{W}$ & & $\frac{g_{2}^{3}}{90 M^{2}}$ & $-\frac{g_{2}^{3}}{90 M^{2}} N_\Sigma$ \\
\hline $\mathcal{R}_{2 W}$ & & $\frac{g_{2}^{2}}{15 M^{2}}$ & $\frac{4 g_{2}^{2}}{15 M^{2}} N_\Sigma$ \\
\hline $\mathcal{R}_{2 B}$ & & $\frac{g_{1}^{2}}{10 M^{2}}$ & \\
\hline
\end{tabular}
}\caption{Log-type terms from one-loop matching in the Green's basis, where $N_\Sigma$ counts the number of flavors for $\Sigma$, which we take to be 3 for our numerical study. We define $L\equiv\log\left({\mu_{\text{'t Hooft}}^2}/{M^2}\right)$ for convenience. For matching at one loop, we always leave the factor $1/(16\pi^2)$ implicit in the Wilson coefficients. The specific forms of the dimension-6 operators are listed in table\,\ref{tab:operatorbasis}.}\label{tab:1looplogtypeGreen}
\end{table}

\begin{table}[h!]
  \centering
  \resizebox{\textwidth}{!}{  
\begin{tabular}{|c||l|l|l|}
\hline\multicolumn{4}{|c|}{{\color{blue}One-loop power-type operators in $\mathcal{L}_{\rm SM}$}} \\
\hline Operator & Type-I & Type-II & Type-III \\
\hline$H^{\dagger} H$ & $-2{\rm tr}(Y_\nu^\dagger Y_\nu)M^2(1+L)$ & $3(\mu^2+\lambda_4M^2+\lambda_5M^2)(1+L)$ & $-6{\rm tr}(Y_\Sigma^\dagger Y_\Sigma)M^2(1+L)$ \\
\hline$\left(D_{\mu} H\right)^{\dagger}\left(D^{\mu} H\right)$ & $\frac{1}{2}{\rm tr}(Y_\nu^\dagger Y_\nu)(1+2L)$ & $\frac{3\mu^2}{2M^2}$ & $\frac{3}{2}{\rm tr}(Y_\Sigma^\dagger Y_\Sigma)(1+2L)$ \\
\hline$\left(H^{\dagger} H\right)^{2}$ & $\begin{array}{ll} {\rm tr}(Y_\nu Y_\nu^T Y_\nu^* Y_\nu^\dagger ) \\ - {\rm tr}(Y_\nu Y_\nu^\dagger Y_\nu Y_\nu^\dagger ) L \\\end{array}$ & $\begin{array}{ll}  \frac{1}{2}\left(3\lambda_4^2+6\lambda_4\lambda_5+5\lambda_5^2\right)L - \frac{5\mu^4}{2M^4} (2+L) \\
+ \frac{\mu^2}{M^2} [\left(10 \lambda_H +4 \lambda_2 +4 \lambda_3 \right) (1+L)] \\ -\frac{\mu^2}{M^2} [(\lambda_4(5+2L) +\lambda_5(11+6L)] \end{array}$ & $\begin{array}{ll} -4{\rm tr}(Y_\Sigma^\dagger Y_\Sigma Y_e Y_e^\dagger)(1+L) \\+ {\rm tr}(Y_\Sigma^\dagger Y_\Sigma^* Y_\Sigma^T Y_\Sigma) \\ -5 {\rm tr}(Y_\Sigma^\dagger Y_\Sigma Y_\Sigma^\dagger Y_\Sigma) L \\\end{array}$  \\
\hline $\bar{\ell}_{L p} i {D\!\!\!\!/\,} \ell_{L r}$ & $\frac{1}{4}(Y_\nu Y_\nu^\dagger)^{p r}(3+2L)$ & $\frac{3}{4}(1+2 L)\left(Y_{\nu}^{\dagger} Y_{\nu}\right)^{p r}$ & $\frac{3}{4}\left(Y_{\Sigma}^{\dagger} Y_{\Sigma}\right)^{p r}(3+2 L)$ \\
\hline $\bar{\ell}_{L p} e_{R r} H$ & $-\left(Y_{\nu} Y_{\nu}^{\dagger} Y_{e}\right)^{p r}(1+L)$ & & $3\left(Y_{\Sigma}^{\dagger} Y_{\Sigma} Y_{e}\right)^{p r}(1+L)$ \\

\hline\multicolumn{4}{|c|}{{\color{blue}{\bf Type}: $\psi^2H^2$}} \\
\hline
\end{tabular}
}\caption{Power-type terms from one-loop matching in the Green's basis. We only show the matching results for $\mathcal{L}_{\rm SM}$ and the Weinberg operator in this table. Higher dimensional power-type operators from one-loop matching can be found in following tables.}\label{tab:1looppowtypeGreen}
\end{table}

\begin{table}[h!]
  \centering
  \resizebox{\textwidth}{!}{  
\begin{tabular}{|c||l|l|l|}
\hline\multicolumn{4}{|c|}{{\color{blue}{\bf Type}: $X^2H^2$}} \\
\hline Operator & Type-I & Type-II & Type-III \\
\hline $\mathcal{O}_{H W}$ & $\frac{g_{2}^{2}}{24M^2} {\rm tr}\left(Y_{\nu} Y_{\nu}^{\dagger}\right) $ & $\frac{g_{2}^{2}}{12 M^{4}}\left(2 \lambda_{4} M^{2}+2 \lambda_{5} M^{2}-\mu^{2}\right)$ & $\frac{7 g_{2}^{2}}{24 M^{2}}{\rm tr}\left(Y_{\Sigma}^{\dagger} Y_{\Sigma}\right) $ \\
\hline $\mathcal{O}_{H B}$ & $\frac{g_{1}^{2}}{24M^2} {\rm tr}\left(Y_{\nu}  Y_{\nu}^{\dagger}\right) $ & $\frac{g_{1}^{2}}{4 M^{4}}\left(\lambda_{4} M^{2}+\lambda_{5} M^{2}-\mu^{2}\right)$ & $\frac{g_{1}^{2}}{8 M^{2}}{\rm tr}\left(Y_{\Sigma}^{\dagger} Y_{\Sigma}\right) $ \\
\hline $\mathcal{O}_{H W B}$ & $\frac{g_{1} g_{2}}{12M^2} {\rm tr}\left(Y_{\nu} Y_{\nu}^{\dagger}\right) $ & $\frac{g_{1} g_{2}}{12 M^{4}}\left(4 \lambda_{5} M^{2}-5 \mu^{2}\right)$ & $-\frac{g_{1} g_{2}}{4 M^{2}}{\rm tr}\left(Y_{\Sigma}^{\dagger} Y_{\Sigma}\right) $ \\
\hline\multicolumn{4}{|c|}{{\color{blue}{\bf Type}: $H^2XD^2$}} \\
\hline Operator & Type-I & Type-II & Type-III \\
\end{tabular}

\hline\multicolumn{4}{|c|}{{\color{blue}{\bf Type}: $H^6$}} \\
\hline Operator & Type-I & Type-II & Type-III \\
\hline $\mathcal{O}_{H}$ & 
$\begin{array}{ll}  \frac{1}{3M^2}{\rm tr}(Y_\nu Y_\nu^\dagger Y_\nu Y_\nu^\dagger Y_\nu Y_\nu^\dagger) \\ -\frac{1}{M^2}{\rm tr}( Y_\nu Y_\nu^\dagger Y_\nu Y_\nu^T Y_\nu^* Y_\nu^\dagger) \\\end{array}$ 
&
$\begin{array}{ll} -\frac{1}{2M^2}(\lambda_4^3+3\lambda_4^2\lambda_5+5\lambda_4\lambda_5^2+3\lambda_5^3) \\ +\frac{\mu^2}{2M^4}[4g_2^2\lambda_H(1+L) + 58\lambda_4\lambda_5]\\
+\frac{\mu^2}{2M^4}[88\lambda_H^2 +13 \lambda_4^2 + 63\lambda_5^2]\\
+\frac{\mu^2}{2M^4}[- (64\lambda_H + 16\lambda_2 + 12\lambda_3)(\lambda_4 + 2\lambda_5)]\\ 
+\frac{\mu^2}{2M^4}[88\lambda_H^2 + 6\lambda_4^2 + 17\lambda_5^2]L\\
+\frac{\mu^2}{2M^4}[- 44\lambda_H\lambda_4 -8\lambda_2\lambda_4 -6\lambda_3\lambda_4]L\\
+\frac{\mu^2}{2M^4}[- 46\lambda_H\lambda_5 -11\lambda_2\lambda_5 -9 \lambda_3\lambda_5 + 14\lambda_4\lambda_5]L\\
-\frac{\mu^4}{2M^6}[2g_2^2(1+L) +\left(88 \lambda_H-25 \lambda_4-50 \lambda_5\right)L]\\
-\frac{\mu^4}{2M^6}[124 \lambda_H+10 \lambda_2+6 \lambda_3-48 \lambda_4-97 \lambda_5 ]\\
+\frac{\mu^6}{2M^8}(37+22L) \\\end{array}$ 
&
$\begin{array}{ll} \frac{1}{M^2}{\rm tr}(8Y_\Sigma^\dagger Y_\Sigma Y_\Sigma^\dagger Y_\Sigma Y_e Y_e^\dagger) \\ -\frac{4(1+L)}{M^2}{\rm tr}(Y_\Sigma^\dagger Y_\Sigma Y_e Y_e^\dagger Y_e Y_e^\dagger) \\ +\frac{3}{M^2}{\rm tr}(Y_\Sigma^\dagger Y_\Sigma Y_\Sigma^\dagger Y_\Sigma Y_\Sigma^\dagger Y_\Sigma) \\ - \frac{1}{M^2}{\rm tr}(Y_\Sigma^\dagger Y_\Sigma Y_\Sigma^\dagger Y_\Sigma^* Y_\Sigma^T Y_\Sigma )  \\\end{array}$ \\
\hline\multicolumn{4}{|c|}{{\color{blue}{\bf Type}: $\psi^2D^3$}} \\
\hline Operator & Type-I & Type-II & Type-III \\
\hline $\mathcal{R}_{\ell D}$ & $-\frac{1}{3M^2}(Y_\nu Y_\nu^\dagger)^{p r}$ & $-\frac{1}{2M^2}(Y_\nu^\dagger Y_\nu)^{p r}$ & $-\frac{1}{M^2}(Y_\Sigma^\dagger Y_\Sigma)^{p r}$ \\
\hline
\end{tabular}
}\caption{Higher dimensional operators in the Green's basis from one-loop matching. We classify the operators by their types as specified in blue. The specific forms of these operators are listed in table\,\ref{tab:operatorbasis}.}\label{tab:1loopGreen1}
\end{table}

\begin{table}[h!]
  \centering
  \resizebox{0.95\textwidth}{!}{  
\begin{tabular}{|c||l|l|l|}
\hline \multicolumn{4}{|c|}{{\color{blue}{\bf Type}: $\psi^2XD$}} \\
\hline Operator & Type-I & Type-II & Type-III \\
\hline $\mathcal{R}_{W \ell}$ & $-\frac{g_{2}}{72M^2} \left(Y_{\nu}Y_{\nu}^{\dagger}\right)^{p r}\left(11+6 L\right)$ & $\frac{5+4 L}{24 M^{2}} g_{2}\left(Y_{\nu}^{\dagger} Y_{\nu}\right)^{p r}$ & $\frac{25+2 L}{24 M^{2}} g_{2}\left(Y_{\Sigma}^{\dagger} Y_{\Sigma}\right)^{p r}$ \\
\hline $\mathcal{R}_{\tilde{W} \ell}^{'}$ & & -$\frac{g_{2}}{4 M^{2}}\left(Y_{\nu}^{\dagger} Y_{\nu}\right)^{p r}$ & $\frac{g_{2}}{M^{2}}\left(Y_{\Sigma}^{\dagger} Y_{\Sigma}\right)^{p r}$ \\
\hline $\mathcal{R}_{B \ell}$ & $\frac{g_{1}}{72M^2}\left(Y_{\nu}Y_{\nu}^{\dagger}\right)^{p r}\left(11+6 L\right)$ & $-\frac{11+12 L}{24 M^{2}} g_{1}\left(Y_{\nu}^{\dagger} Y_{\nu}\right)^{p r}$ & $\frac{11+6 L}{24 M^{2}} g_{1}\left(Y_{\Sigma}^{\dagger} Y_{\Sigma}\right)^{p r}$ \\
\hline $\mathcal{R}_{\tilde{B} \ell}^{\prime}$ & & $\frac{3 g_{1}}{4 M^{2}}\left(Y_{\nu}^{\dagger} Y_{\nu}\right)^{p r}$ & \\
\hline\multicolumn{4}{|c|}{{\color{blue}{\bf Type}: $\psi^2XH$}} \\
\hline Operator & Type-I & Type-II & Type-III \\
\hline
\end{tabular}
}\caption{Table\,\ref{tab:1loopGreen1} continued.}\label{tab:1loopGreen2}
\end{table}

\begin{table}[h!]
  \centering
  \resizebox{\textwidth}{!}{  
\begin{tabular}{|c||l|l|l|}
\hline\multicolumn{4}{|c|}{{\color{blue}{\bf Type}: $\psi^4$}} \\
\hline Operator & Type-I & Type-II & Type-III \\
\hline $(\mathcal{O}_{\ell\ell})_{prst}$ & 
$\begin{array}{ll} -\frac{1}{4M^2}(2+L)(Y_\nu Y_\nu^T)^{ps}(Y_\nu^* Y_\nu^\dagger)^{rt} \\ - \frac{1}{16M^2}(3+2L)(Y_\nu Y_\nu^\dagger)^{pr} (Y_e Y_e^\dagger)^{st} \\ - \frac{1}{16M^2}(3+2L) (Y_e Y_e^\dagger)^{pr} (Y_\nu Y_\nu^\dagger)^{st} \\ + \frac{1}{16M^2}(3+2L)(Y_\nu Y_\nu^\dagger)^{pt} (Y_e Y_e^\dagger)^{sr} \\ + \frac{1}{16M^2}(3+2L)(Y_e Y_e^\dagger)^{pt} (Y_\nu Y_\nu^\dagger)^{sr} \\ - \frac{1}{8M^2}(Y_\nu Y_\nu^\dagger)^{pt} (Y_\nu Y_\nu^\dagger)^{sr} \\\end{array}$ 
&
$\begin{array}{ll} -\frac{1}{8M^2}(Y_\nu^\dagger Y_\nu)^{pr}(Y_\nu^\dagger Y_\nu)^{st} \\ - \frac{1}{2M^2}(Y_\nu^\dagger Y_\nu)^{pt}(Y_\nu^\dagger Y_\nu)^{sr} \\ + \frac{1+L}{2M^2}(4\lambda_2+3\lambda_3)Y_\nu^{*sp}Y_\nu^{rt} \\\end{array}$ 
&
$\begin{array}{ll} -\frac{1}{4M^2}(2+L)(Y_\Sigma^\dagger Y_\Sigma^*)^{ps}(Y_\Sigma^T Y_\Sigma)^{rt} \\ - \frac{1}{16M^2}[(3+2L)(Y_\Sigma^\dagger Y_\Sigma)^{pr} (Y_e Y_e^\dagger)^{st}] \\ - \frac{1}{16M^2}[(3+2L)(Y_e Y_e^\dagger)^{pr} (Y_\Sigma^\dagger Y_\Sigma)^{st}] \\ - \frac{1}{16M^2}[8 (Y_\Sigma^\dagger Y_\Sigma)^{pr} (Y_\Sigma^\dagger Y_\Sigma)^{st}] \\ - \frac{1}{16M^2}[(3+2L)(Y_\Sigma^\dagger Y_\Sigma)^{pt} (Y_e Y_e^\dagger)^{sr} ] \\  - \frac{1}{16M^2}[(3+2L)(Y_e Y_e^\dagger)^{pt} (Y_\Sigma^\dagger Y_\Sigma)^{sr}] \\ - \frac{1}{16M^2}[2 (Y_\Sigma^\dagger Y_\Sigma)^{pt} (Y_\Sigma^\dagger Y_\Sigma)^{sr}] \\\end{array}$ \\
\hline $(\mathcal{O}_{\ell e})_{prst}$   & $\frac{1}{8M^2}(3+2L)(Y_\nu Y_\nu^\dagger)^{pr}(Y_e^\dagger Y_e)^{st}$ & $-\frac{3}{8M^2}(3+2L)(Y_\nu^\dagger Y_e^*)^{ps}(Y_e^TY_\nu)^{tr}$ & $\frac{3}{8M^2}(3+2L)(Y_\Sigma^\dagger Y_\Sigma)^{pr}(Y_e^\dagger Y_e)^{st}$ \\
\hline
\end{tabular}
}\caption{Table\,\ref{tab:1loopGreen2} continued.}\label{tab:1loopGreen3}
\end{table}

\begin{table}[h!]
  \centering
  \resizebox{\textwidth}{!}{  
\begin{tabular}{|c||l|l|l|}
\hline\multicolumn{4}{|c|}{{\color{blue}{\bf Type}: $\mathcal{L}_{\rm SM}$}} \\
\hline Operator & Type-I & Type-II & Type-III \\
\hline$W_{\mu \nu}^{I} W^{I \mu \nu}$ & & $-\frac{g_{2}^{2}}{6} L$ & $-\frac{g_{2}^{2}}{3}  N_\Sigma L$ \\
\hline$B_{\mu \nu} B^{\mu \nu}$ & & $-\frac{g_{1}^{2}}{4} L $ & \\

\hline$\left(D_{\mu} H\right)^{\dagger}\left(D^{\mu} H\right)$ & $\frac{1}{2}{\rm tr}(Y_\nu^\dagger Y_\nu)(1+2L)$ & $\frac{3\mu^2}{2M^2}$ & $\frac{3}{2}{\rm tr}(Y_\Sigma^\dagger Y_\Sigma)(1+2L)$ \\

\hline $\bar{\ell}_{L p} i {D\!\!\!\!/\,}\ell_{L r}$ & $\frac{1}{4}\left(Y_{\nu} Y_{\nu}^{\dagger}\right)^{pr}(3+2 L)$ & $\frac{3}{4}(1+2 L)\left(Y_{\nu}^{\dagger} Y_{\nu}\right)^{p r}$ & $\frac{3}{4}\left(Y_{\Sigma}^{\dagger} Y_{\Sigma}\right)^{p r}(3+2 L)$ \\
\hline $H^\dagger H$ & $\left[-2M^2(1+L)+\frac{\mu_H^4}{3M^2}\right]\operatorname{tr}(Y_\nu^\dagger Y_\nu)$  & $\frac{\mu ^2 \mu _H^4}{2 M^4}+3 (L+1) \left(\mu ^2+\lambda _4 M^2+\lambda _5 M^2\right)$ & $\left[-6M^2(1+L)+\frac{\mu_H^4}{M^2}\right]\operatorname{tr}(Y_{\Sigma } Y^\dagger_{\Sigma 
   })$\\
\hline
\end{tabular}
}\caption{Modifications to $\mathcal{L}_{\rm SM}$ and the Weinberg operator from one-loop matching in the Warsaw basis. Fields are not renormalized for simplicity and corrections to the kinetic terms are retained.}\label{tab:1loopWarsaw1}
\end{table}


\begin{table}[h!]
  \centering
  \resizebox{\textwidth}{!}{  
\begin{tabular}{|c||l|l|l|}
\hline\multicolumn{4}{|c|}{{\color{blue}{\bf Type}: $X^3$}} \\
\hline Operator & Type-I & Type-II & Type-III \\
\hline $\mathcal{O}_W$ & $ $ & $\frac{g_2^3}{90 M^2}$ & $-\frac{g_2^3}{90 M^2}N_\Sigma$\\
\hline\multicolumn{4}{|c|}{{\color{blue}{\bf Type}: $H^6$ and $H^4D^2$}} \\
\hline Operator & Type-I & Type-II & Type-III \\
\hline
\end{tabular}
}\caption{Higher dimensional operators in the Warsaw basis from one-loop matching. Once again, we classify the operators by their types as specified in blue, and the specific forms of these operators are listed in table\,\ref{tab:operatorbasis}.}\label{tab:1loopWarsaw2}
\end{table}


\begin{table}[h!]
  \centering
  \resizebox{\textwidth}{!}{  
\begin{tabular}{|c||l|l|l|}
\hline\multicolumn{4}{|c|}{{\color{blue}{\bf Type}: $\psi^2HD$}} \\
\hline Operator & Type-I & Type-II & Type-III \\
\hline $\mathcal{O}_{Hl,pr}^{(1)}$ & 
$\begin{array}{ll} -\frac{1}{2M^2}(Y_\nu Y_\nu^\dagger Y_\nu Y_\nu^\dagger)^{pr} - \frac{1+L}{2M^2}(Y_\nu Y_\nu^T Y_\nu^* Y_\nu^\dagger)^{pr}\\+\frac{11+6L}{288M^2}(11g_1^2 + 27g_2^2)(Y_\nu Y_\nu^\dagger)^{pr} \\+\frac{5+6L}{72M^2}g_1^2{\rm tr}(Y_\nu^\dagger Y_\nu)\delta^{pr}\\ -\frac{3+2L}{16M^2}(Y_\nu Y_\nu^\dagger Y_e Y_e^\dagger + Y_e Y_e^\dagger Y_\nu Y_\nu^\dagger)^{pr}  \\\end{array}$  
   &
   $\begin{array}{ll} -\frac{3}{4M^2}(1+L)(Y_\nu^\dagger Y_e^* Y_e^T Y_\nu)^{pr}\\-\frac{3\mu^2}{16M^4}(5+2L)(2Y_\nu^\dagger Y_\nu + Y_e Y_e^\dagger)^{pr}\\+\frac{g_1^4}{40M^2}\delta^{pr}-\frac{g_1^2\mu^2}{48M^4}(19+6L)\delta^{pr}\\ -\frac{5+3L}{12M^2}g_1^2(Y_\nu^\dagger Y_\nu)^{pr} \\\end{array}$ 
   &
   $\begin{array}{ll} -\frac{3}{2M^2}(Y_\Sigma^\dagger Y_\Sigma Y_\Sigma^\dagger Y_\Sigma)^{pr} - \frac{15+6L}{16M^2}(Y_\Sigma^\dagger Y_\Sigma^* Y_\Sigma^T Y_\Sigma)^{pr}\\+\frac{1}{96M^2}[5(11+6L)g_1^2 + 27(1+2L)g_2^2](Y_\Sigma^\dagger Y_\Sigma)^{pr} \\ +\frac{g_1^2}{24M^2}(5+6L)(Y_\Sigma^\dagger Y_\Sigma)\delta^{{pr}}\\ +\frac{9+6L}{16M^2}(Y_\Sigma^\dagger Y_\Sigma Y_e Y_e^\dagger +  Y_e Y_e^\dagger Y_\Sigma^\dagger Y_\Sigma)^{pr} \\\end{array}$ \\
\hline
\end{tabular}
}\caption{Table\,\ref{tab:1loopWarsaw2} continued.}\label{tab:1loopWarsaw3}
\end{table}

\clearpage

%% file: EWSB.tex
\section{Radiative symmetry breaking in seesaw models}\label{sec:ewsb}
The one-loop matching discussed in previous section could possibly give rise to large threshold effects on the Higgs potential, and thus significantly modifies the pattern of EWSB. One such interesting pattern is EWSB originated from radiative corrections in these seesaw models, which would be the topic of this section. To that end, we obtain the RGEs for each model with the help of ${\tt PyR@ TE 3}$\,\cite{Sartore:2020gou}, a python package based on methods developed in\,\cite{Jack:1982hf,Jack:1982sr,Machacek:1983tz,Machacek:1983fi,Machacek:1984zw,Jack:1984vj,Luo:2002ti,Poole:2019kcm} to systematically calculate the RGEs up to 3-loop order. Specifically, we solve the RGEs of the SM below the matching scale, and those of the UV models above the matching scale. While at the matching scale, we include the threshold effects from the both tree-level and one-loop matching. These loop effects from the matching may na\"ively seem negligible, which is not true in general as we will see below especially for the type-II seesaw model.


\subsection{Type-I and -III}
In the type-I seesaw model, neutrino masses are generated via introducing 3 generations of heavy right-handed neutrinos $N_R$, which transform as $(1,1,0)$ under the SM gauge group. At tree level, the resulting Weinberg operator can be written as
\eqal{{\cal L}^{(5)}=-\frac{1}{4M}\left(Y_\nu^*Y_\nu^\dagger\right)^{pr}\left(\overline{\tilde{\ell}}_{Lp}\vec{\tau}\ell_{Lr}\right)\left(\tilde{\phi}^\dagger\vec{\tau}\phi\right) + h.c.,}
with $p$, $r$ the lepton flavors, $Y_\nu$ the neutrino Yukawa couplings, $M$ the mass scale of $N_R$ and $\vec{\tau}$ the three Pauli matrices. Non-vanishing neutrino masses can be naturally generated after electroweak spontaneous symmetry breaking when the neutral component of the Higgs doublet $\phi$ gets a non-vanishing vev $v$. Ignoring the flavor indices, the neutrino mass matrix can then be written as\footnote{The overall negative sign has been absorbed by field redefinition as discussed in section\,\ref{subsec:treematching}.}
\eqal{m_\nu=\frac{v^2}{2M}\left(Y_\nu^*Y_\nu^\dagger\right).\label{eq:typeivmass}}

Due to the smallness of the neutrino masses, it is well known that $M$ needs to be above the GUT scale to avoid the naturalness problem. In this work, we will relax the naturalness assumption and allow $Y_\nu$'s to be small. Specifically, we consider three representative benchmark points in this subsection to illustrate the role of $N_R$ in the evolution of the Higgs potential. These benchmark values are summarized in table\,\ref{tab:seesaw13BM}, which are obtained by assuming a diagonal $(m_{\nu})_{pr}$ and fixing $m_\nu=0.01\rm\,eV$ for each generation from the cosmological consideration on the sum of neutrino masses\,\cite{ParticleDataGroup:2020ssz}.\footnote{It will be fine to assume equal-mass neutrinos since the phenomenology on neutrino oscillations are never touched in this work. Even with the inclusion of neutrino oscillations, since the mass square differences are very tiny, the equal-mass approximation here still remains valid for the discussion below.} For simplicity, we adopt this assumption throughout this work and comment on that the off-diagonal elements of $(m_{\nu})_{pr}$ will not have a large impact on our conclusions below due to their smallness and also the suppression from the lepton-flavor mixing matrix. 

\begin{table}[h!]
  \centering
  \resizebox{0.5\textwidth}{!}{  
\begin{tabular}{|c|c|c|}
\hline
\diagbox{Benchmark}{Parameters} & $M$ [GeV] & $Y_\nu$ \\
\hline\hline
BM1 & $10^3$ & $4.13\times10^{-7}$ \\
\hline
BM2 & $10^6$ & $1.35\times10^{-5}$ \\
\hline
BM3 & $10^{12}$ & $1.37\times10^{-2}$ \\
\hline
\end{tabular}
}\caption{Three representative benchmark points for the type-I and -III seesaw models for illustrating radiative generated electroweak spontaneous symmetry breaking.}\label{tab:seesaw13BM}
\end{table}

\begin{figure}[t]
\centering{
  \begin{adjustbox}{max width = \textwidth}
\begin{tabular}{ccc}
\includegraphics[width=0.8\textwidth]{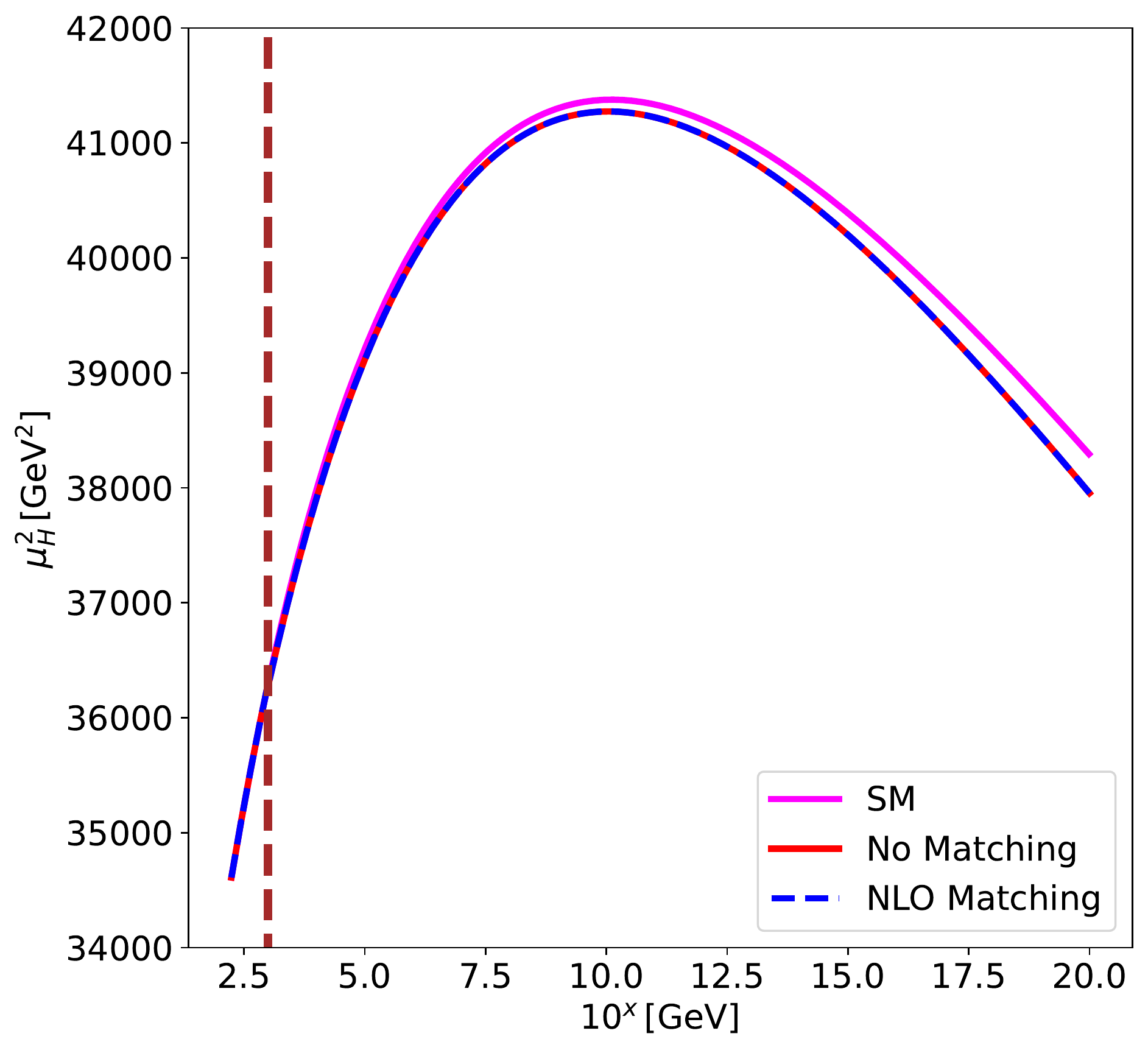} & 
	\includegraphics[width=0.8\textwidth]{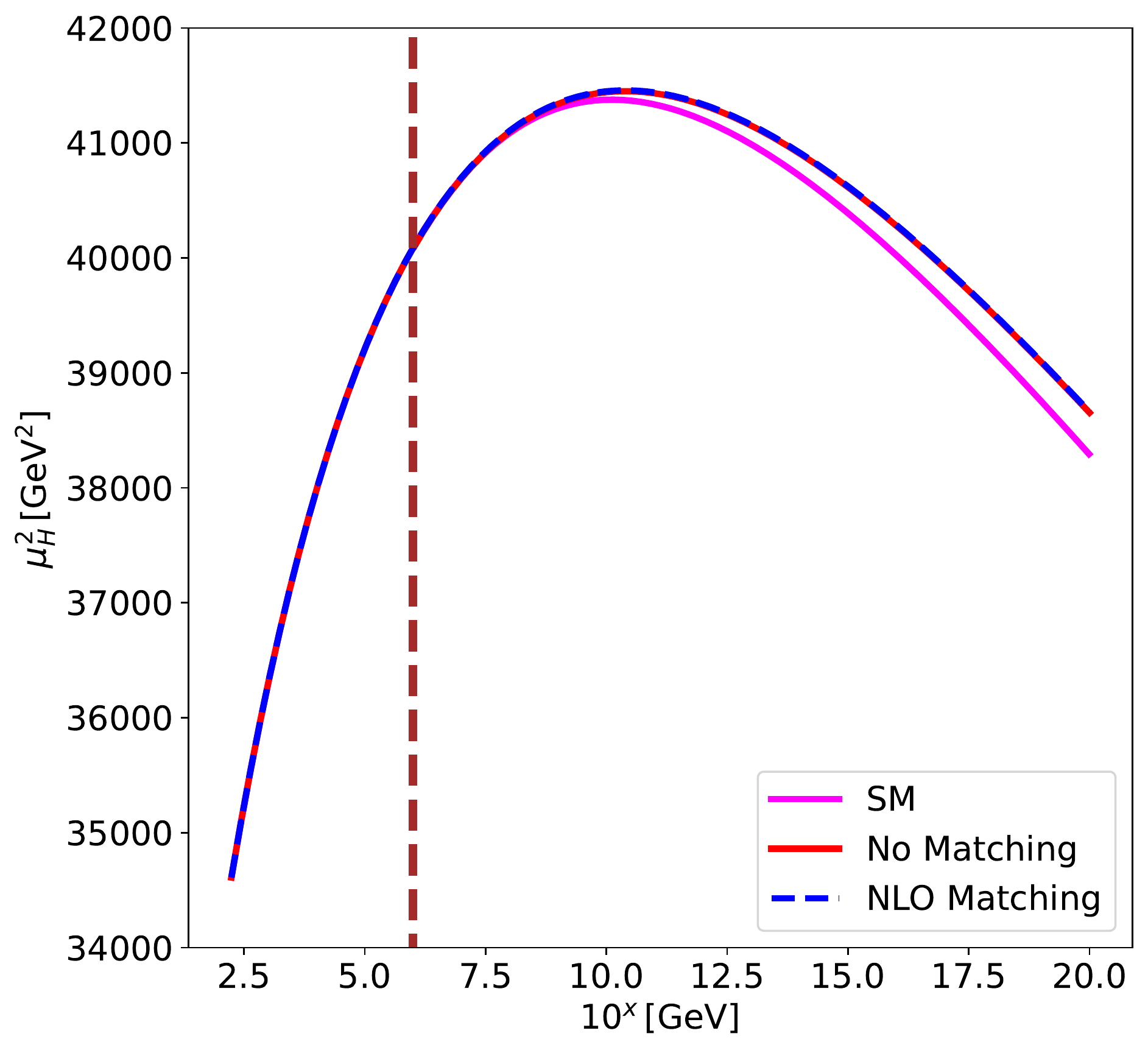} & 
	\includegraphics[width=0.76\textwidth]{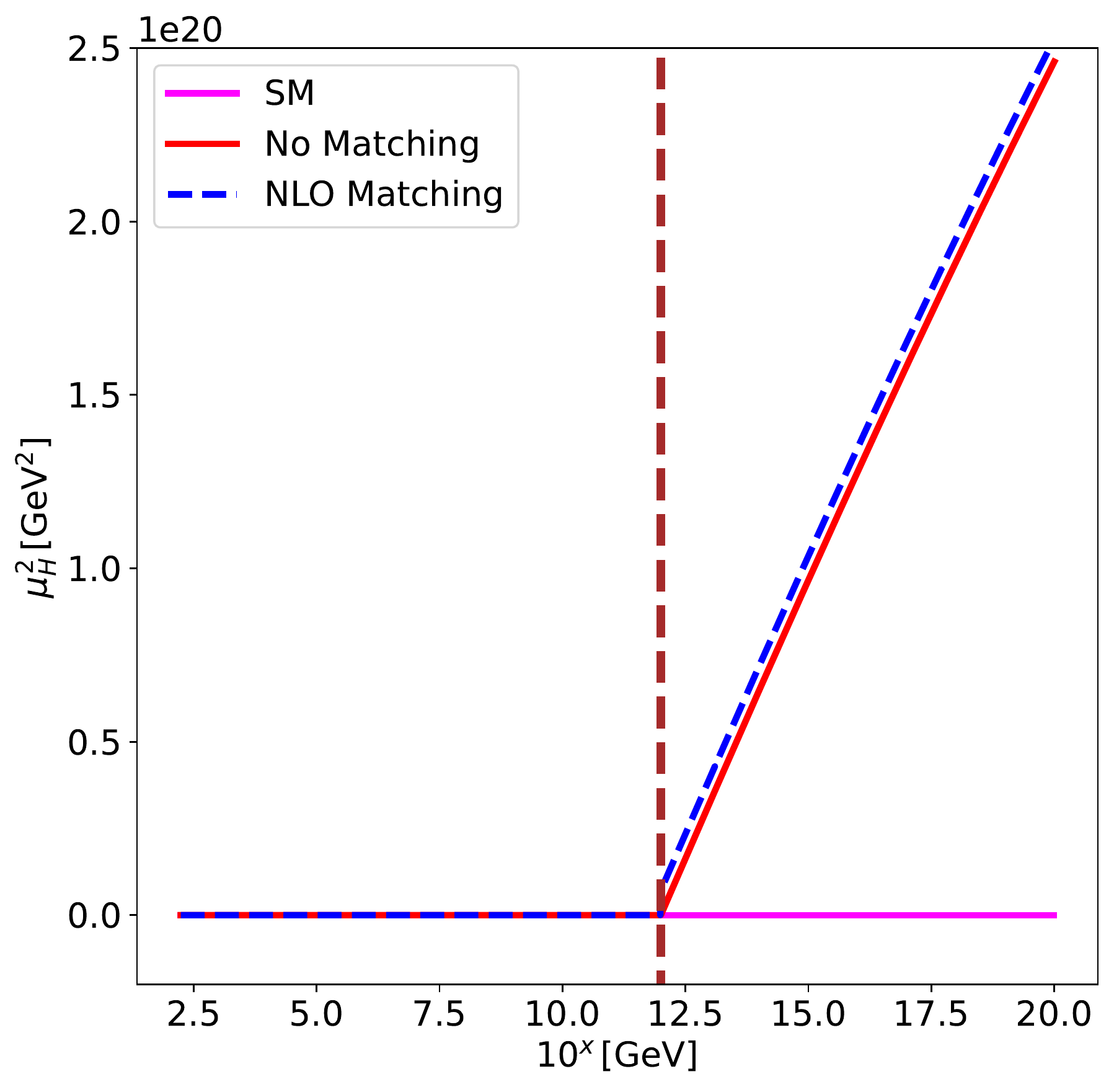}\\
	& \Huge{(Type-I seesaw model)} & \\ 
\includegraphics[width=0.8\textwidth]{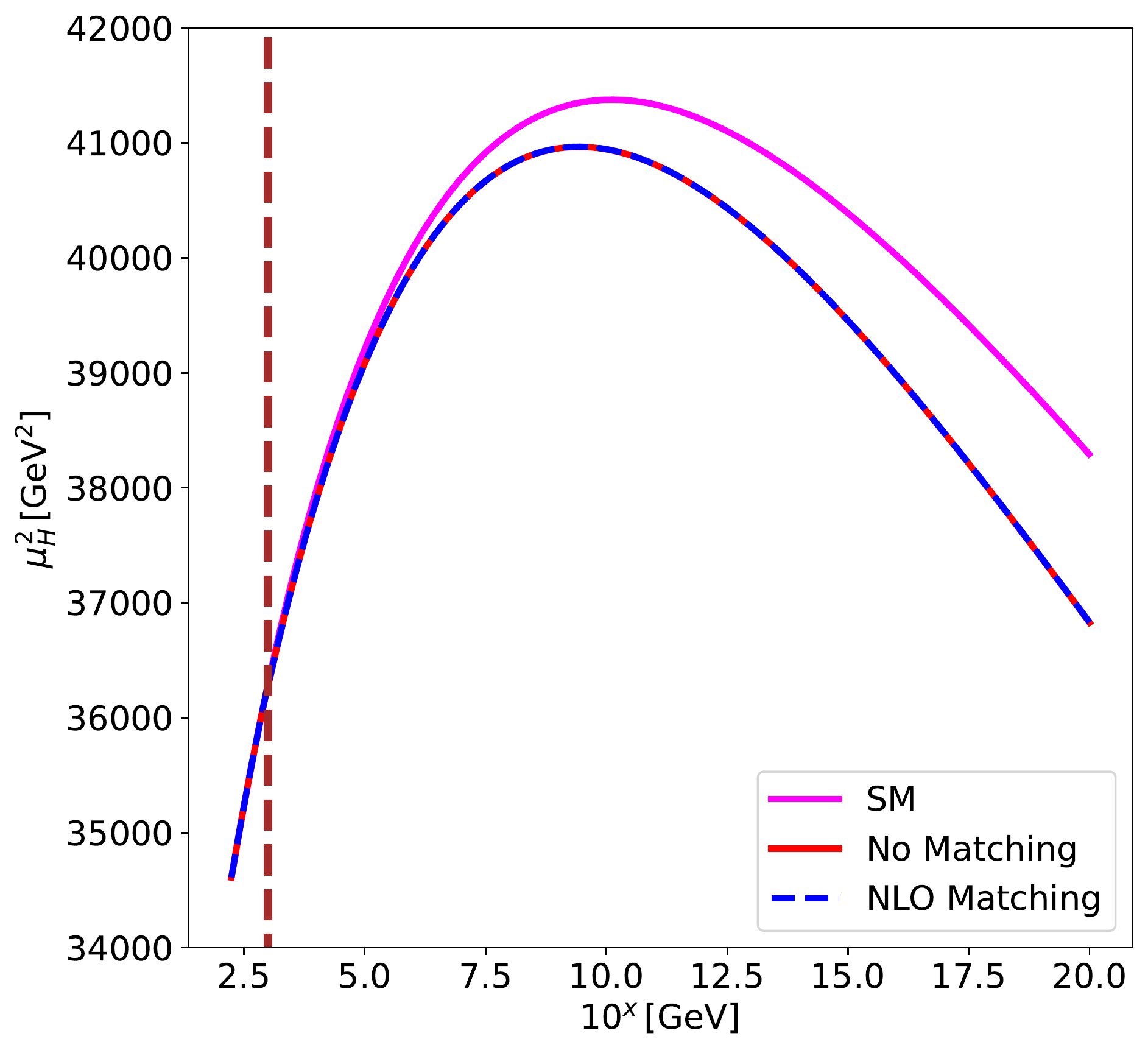} & 
	\includegraphics[width=0.8\textwidth]{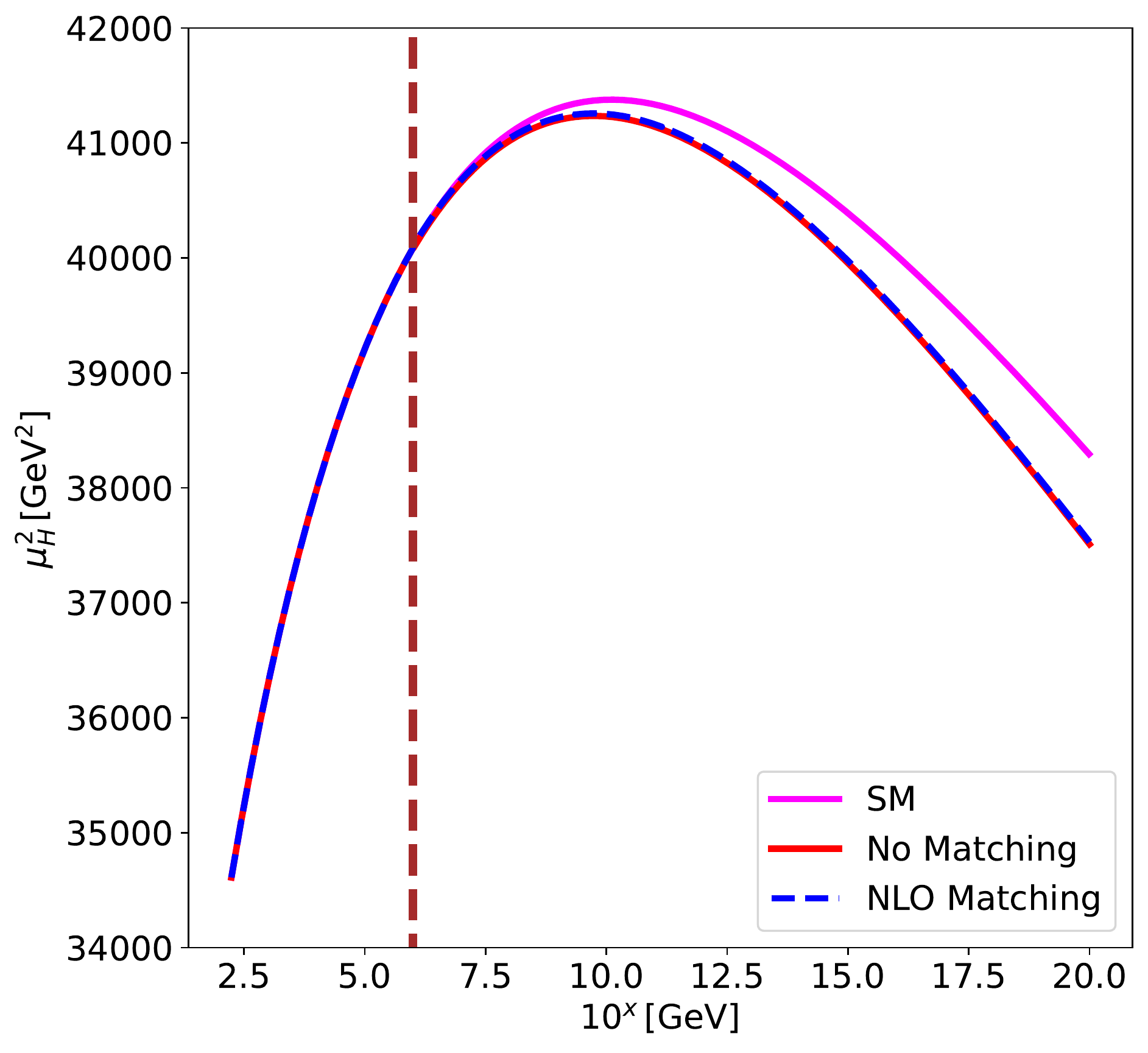} & 
	\includegraphics[width=0.75\textwidth]{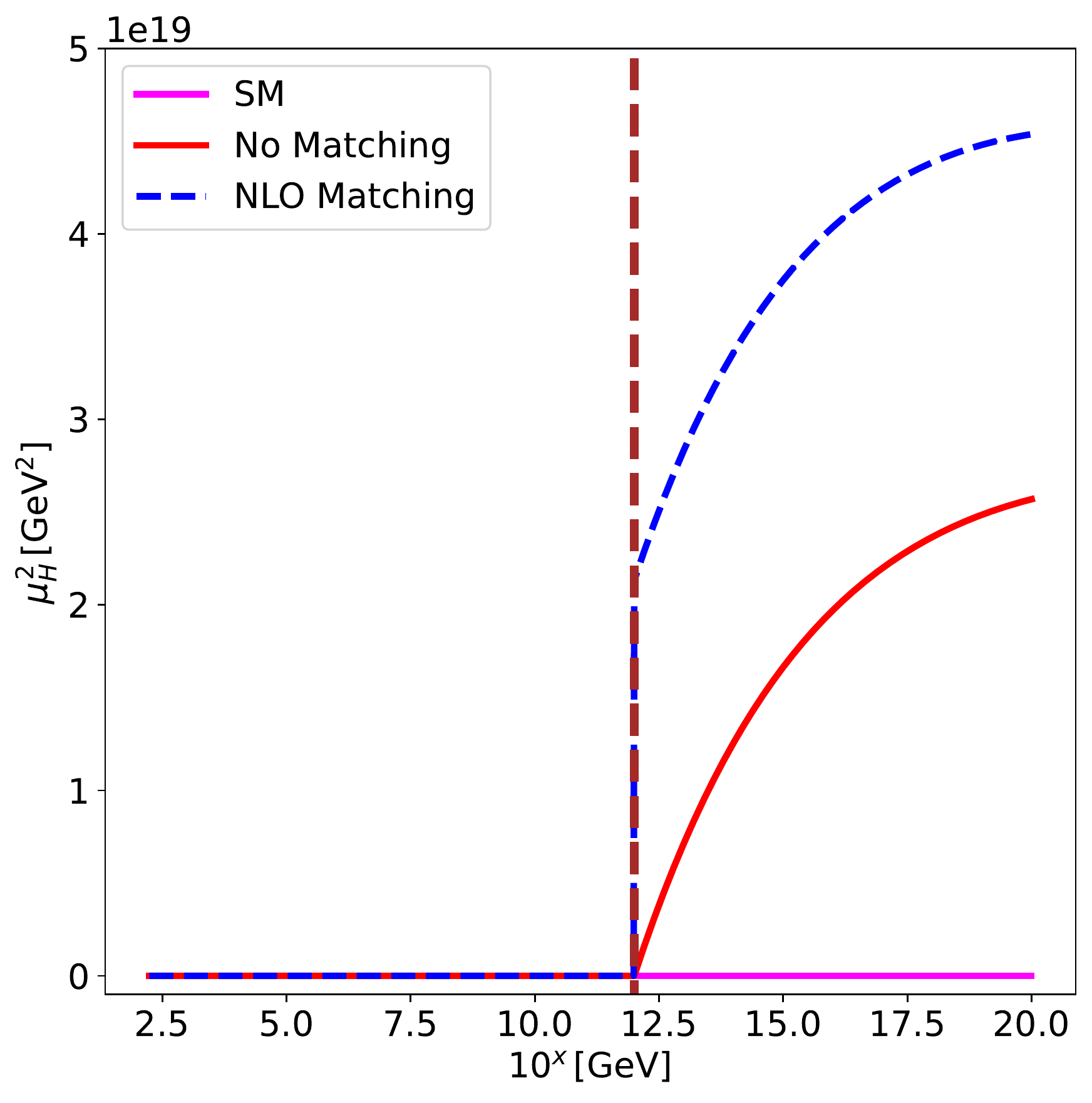}\\
	& \Huge{(Type-III seesaw model)} & \\ 
\end{tabular}
  \end{adjustbox}
  }\caption{Running of $\mu_H^2$ in the type-I (first row) and -III (second row) seesaw models. The first, second and thirs plots in each row are for BM1, BM2 and BM3 in table\,\ref{tab:seesaw13BM}, respectively. In each plot, we use the vertical dashed line in brown to indicate the matching scale. The magenta curve is obtained from solving the SM RGEs up to the Planck scale. The red (dashed blue) curve is for the running of $\mu_H^2$ without (with) the inclusion of one-loop matching.}\label{fig:seesaw1mu}
\end{figure}

\begin{figure}[t]
\centering{
  \begin{adjustbox}{max width = \textwidth}
\begin{tabular}{ccc}
  \Huge{~~~~~~~$M = 10^3\rm\,GeV$}	& \Huge{~~~~~~~$M = 10^6\rm\,GeV$} & \Huge{~~~~~~~$M = 10^{12}\rm\,GeV$} \\ 
\includegraphics[width=0.8\textwidth]{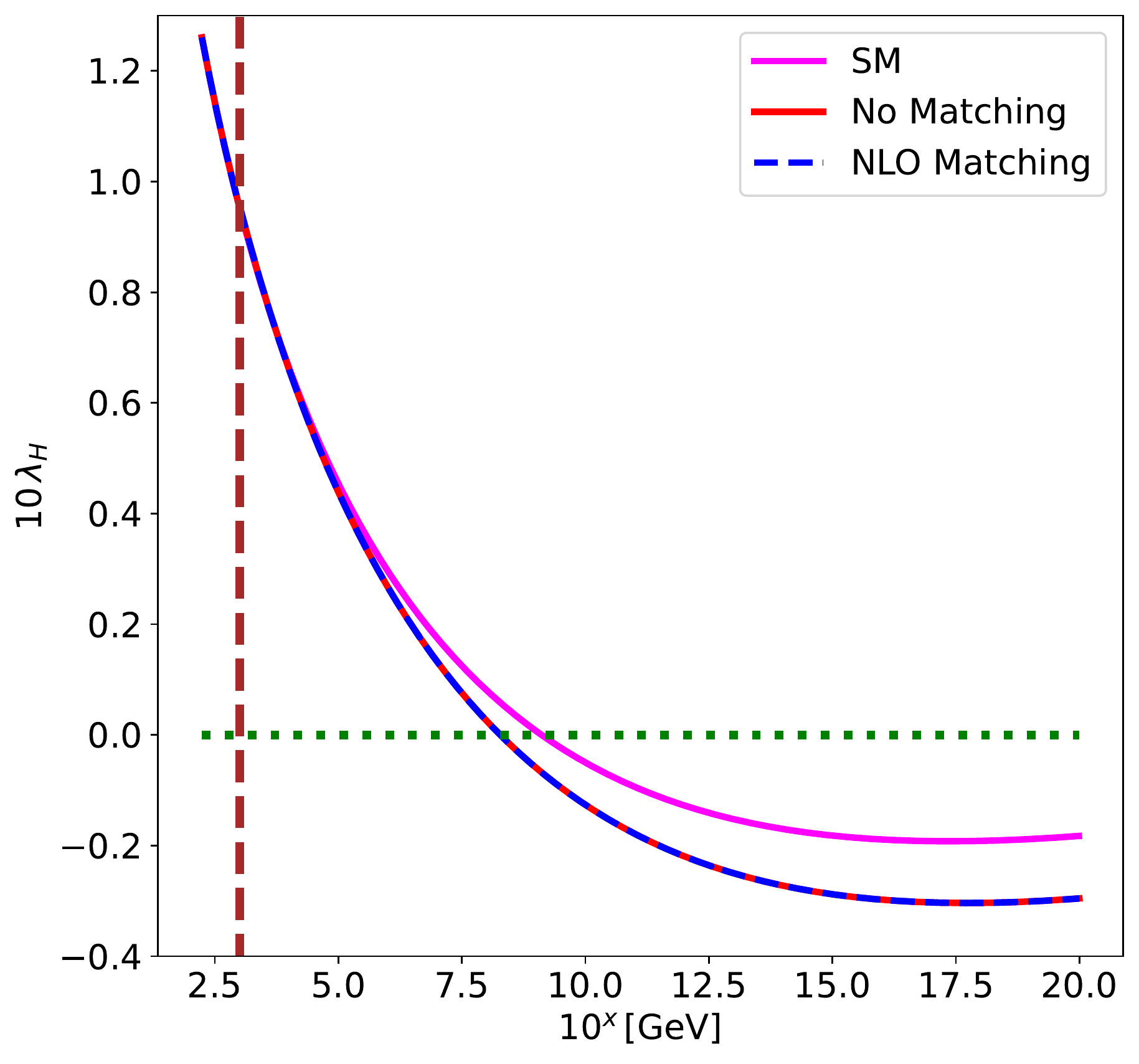} & 
	\includegraphics[width=0.8\textwidth]{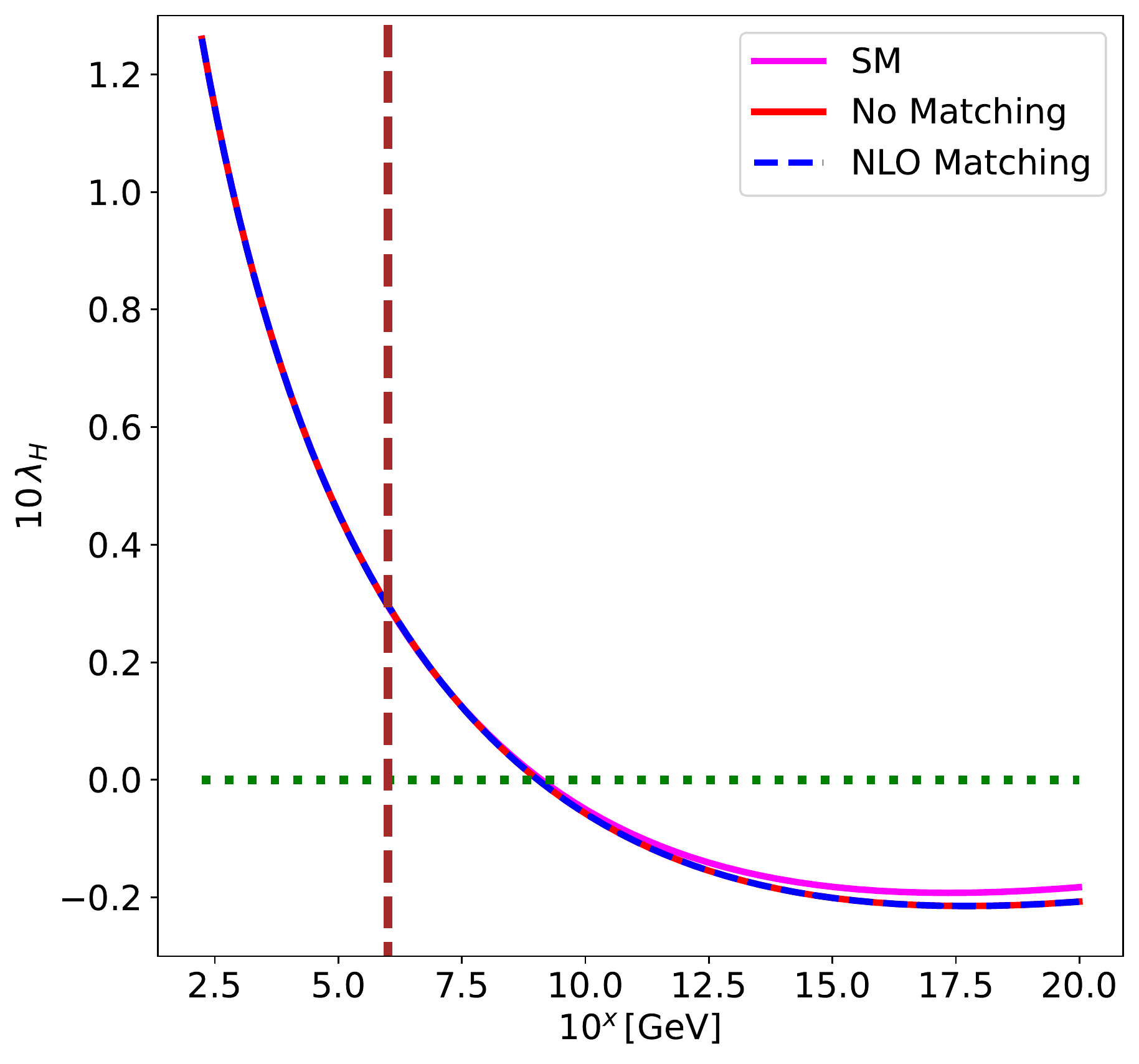} & 
	\includegraphics[width=0.8\textwidth]{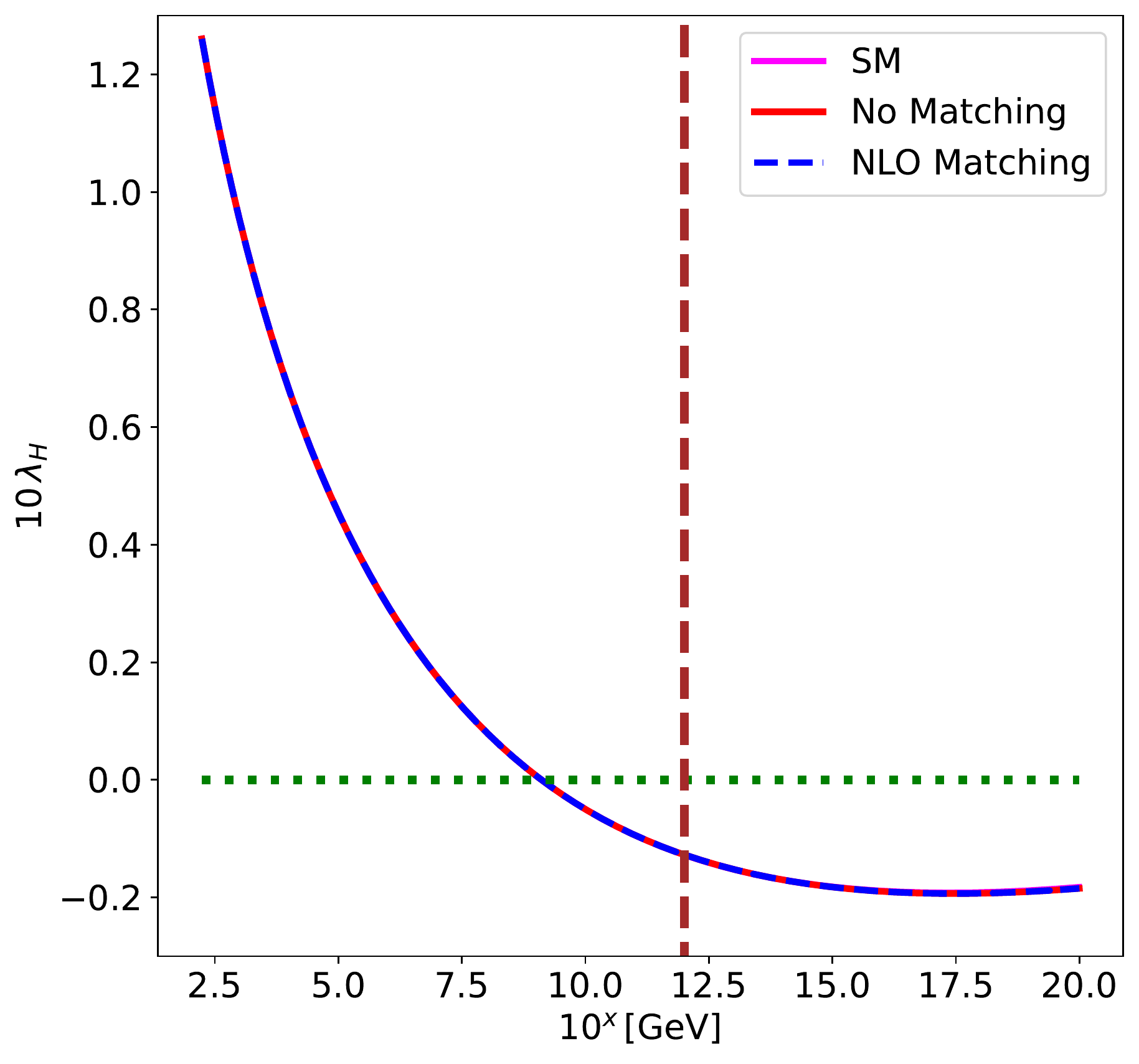}\\
	& \Huge{(Type-I seesaw model)} & \\ 
\includegraphics[width=0.8\textwidth]{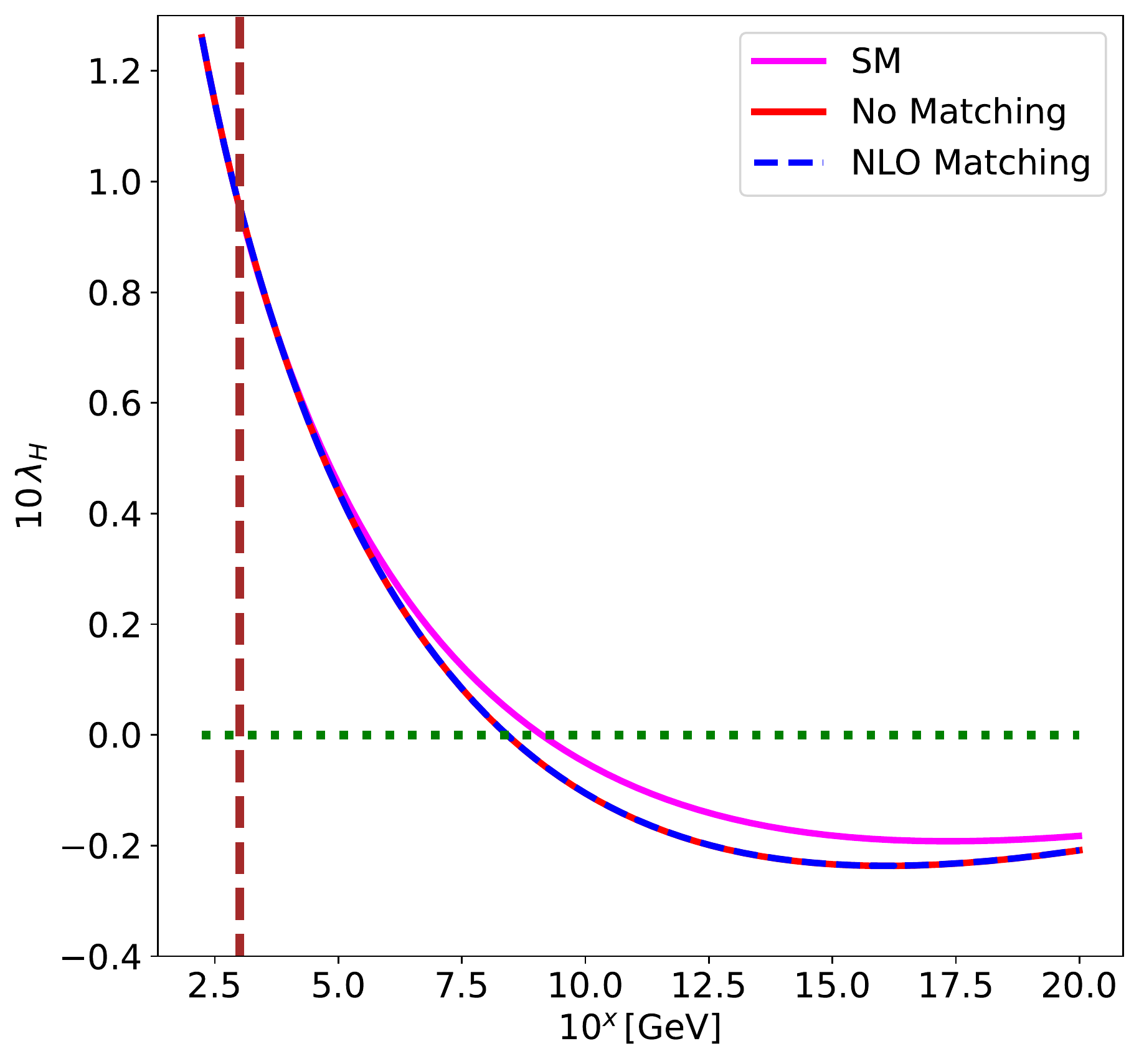} & 
	\includegraphics[width=0.8\textwidth]{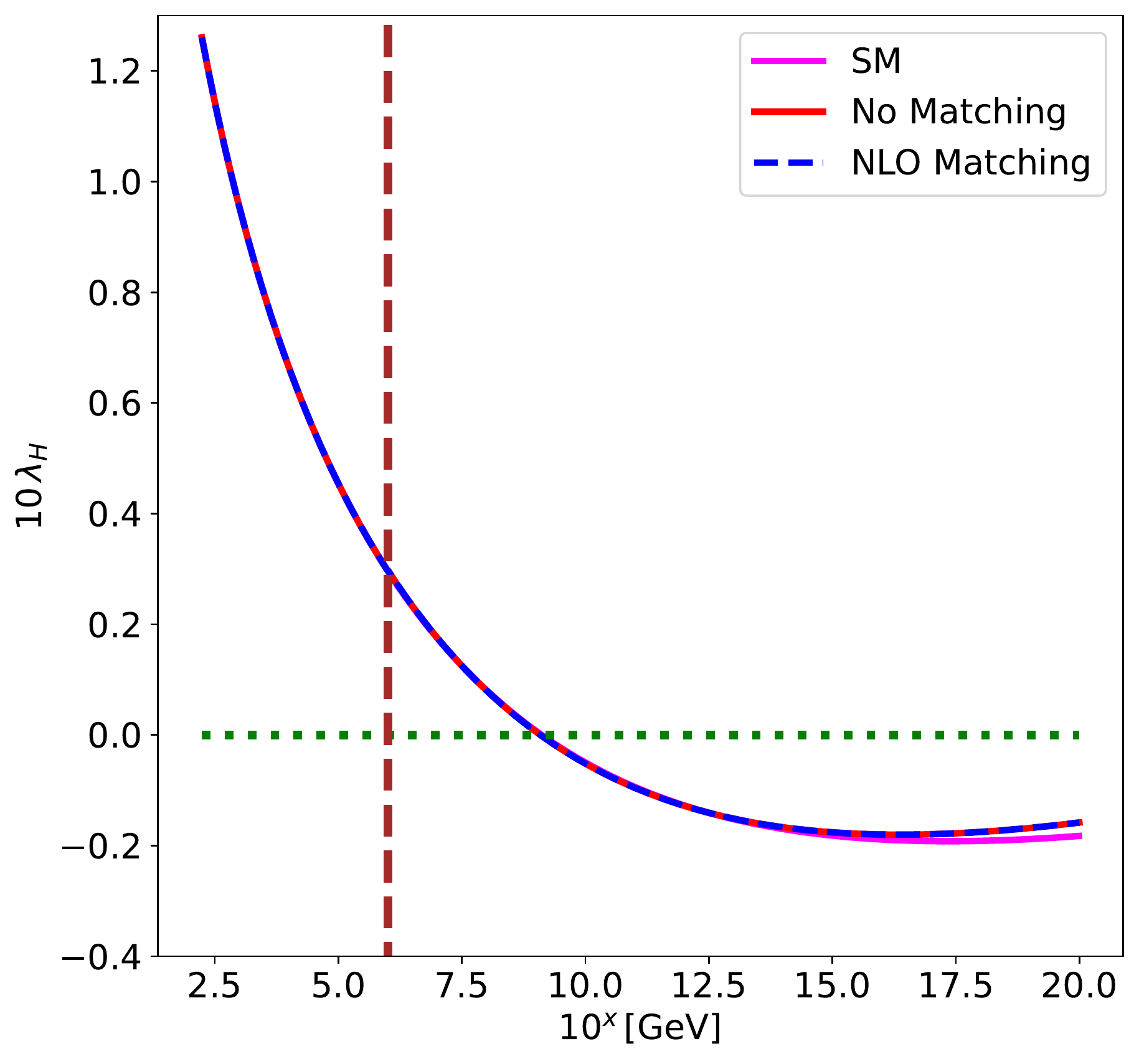} & 
	\includegraphics[width=0.8\textwidth]{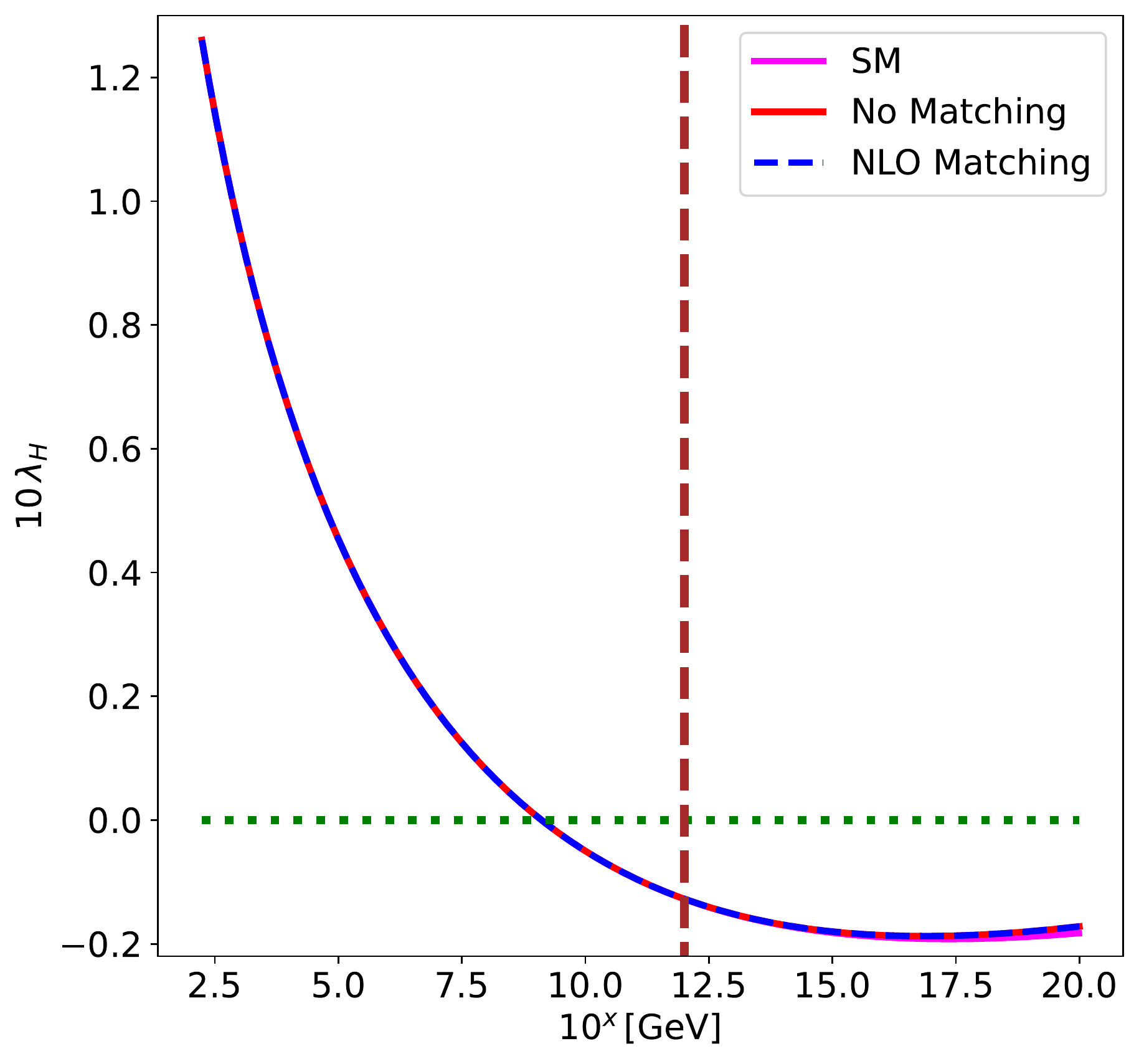}\\
	& \Huge{(Type-III seesaw model)} & \\ 
\end{tabular}
  \end{adjustbox}
  }\caption{Same as figure\,\ref{fig:seesaw1mu} but for the running of $\lambda_H$ in the type-I (first row) and -III (second row) seesaw models. We also introduce a dotted horizontal line in green for $\lambda_H=0$ to help the eyes locate when it runs to negative numbers.}\label{fig:seesaw1lambda}
\end{figure}

To solve the RGEs for the type-I seesaw model, we assume diagonal $Y_\nu$ and $M$ for simplicity and perform the matching at scale $M$. Below $M$, we simply obtain the RG evolution from pure SM, and above $M$, we solve the full RGEs in the presence of $N_R$. The numerical results are then shown in the upper row of figure\,\ref{fig:seesaw1mu} for the running of $\mu_H^2$ and the upper row of figure\,\ref{fig:seesaw1lambda} for that of $\lambda_H$. In each plot, we use the vertical dashed line in brown to indicate the matching scale. The magenta curve is obtained from solving the SM RGEs up to the Planck scale. The red (dashed blue) curve is for the type-I seesaw model without (with) the inclusion of one-loop matching. Note that below the matching scale, we use the SM RGEs such that the curves overlap with that from the SM. While at the matching scale, the inclusion of the one-loop matching terms in the RGEs generically leads to a either a positive or a negative shift in $\mu_H^2$ and/or $\lambda_H$ due to the threshold effects. As we will see below and also in section\,\ref{subsec:type2}, this shift from tree- and one-loop matching would play an essential role in radiatively triggering EWSB. Finally, above the matching scale, we use the full RGEs from the specific UV model.

From our numerical solution as presented in the upper row of figure\,\ref{fig:seesaw1mu}, we find $N_R$ generically introduces a positive shift in $\mu_H^2$ such that the Higgs mass stays positive throughout the range we consider, and radiatively generated EWSB is absent. This can be easily understood from the one-loop matching result presented in section\,\ref{sec:seesaw123}, where one finds, in the Warsaw basis,
\eqal{(\delta\mu_H^2)_{\rm Warsaw} &= \frac{1}{16\pi^2}\frac{1}{3M^2}\left[ \mu_{H,0}^4 - 6\left(1+\log\frac{\mu_{\text{'t Hooft}}^2}{M^2}\right)M^4 \right]{\rm Tr}(Y_\nu^{pr}\bar{Y}_\nu^{pr})\nonumber\\
&\simeq\frac{m_\nu}{24\pi^2M v^2}\left[ \mu_{H,0}^4 - 6\left(1+\log\frac{\mu_{\text{'t Hooft}}^2}{M^2}\right)M^4 \right],\label{eq:type1deltam}}
with $\mu_{\text{'t Hooft}}$ the 't Hooft scale, $\mu_{H,0}$ the value of $\mu_H^2$ at the matching scale, and ``$\rm Tr$'' the trace operation. Note that eq.\,\eqref{eq:typeivmass} has been applied to obtain the last equality, and $\mu_H^2$ in the SM is of $\mathcal{O}(10^4)\rm\,GeV^2$ as seen from the magenta curve in the first plot in the first row of figure\,\ref{fig:seesaw1mu} for example. Since $\mu_{H,\rm UV}^2 = \mu_{H,\rm SM}^2 - (\delta\mu_H^2)_{\rm Warsaw}$, from above equation, it is clear that one gains a negative shift to $\mu_H^2$ from the second term in eq.\,\eqref{eq:type1deltam} only when $M<\mu_{H,0}/\sqrt[4]{6}$. However, this can not be true for the validity of EFTs, which explains the absence of radiative EWSB in the type-I seesaw model up to one-loop order. 
We comment on that the difference between our conclusion and that in Ref.~\cite{Brivio:2017dfq, Brivio:2018rzm}
seeds in the sign difference in corrections to the Higgs mass term from one-loop matching. The opposite result 
can be understood as a consequence of choosing different mass-independent subtraction schemes with the same matching scale $M$. In Ref.~\cite{Brivio:2017dfq, Brivio:2018rzm}, the matching between theories at different scales is performed with the effective potential method $V(\phi)$, in which a different subtraction scheme is chosen that a term  $-\frac{3}{2}\frac{m^4(\phi)}{64\pi^2}$ is absorbed into the counterterms, thus resulting in a different threshold effect.~\footnote{A similar argument exists in Ref.~\cite{Masina:2015ixa}, where the Veltmann condition also depends on the scheme choice, although the $\overline{\textrm{MS}}$ scheme was chosen in their work.} We refer the readers to these articles for further discussion on this point, and in this work we consistently choose the $\overline{\textrm{MS}}$ scheme with the 't Hooft matching scale $\mu=M$ in the following discussion.

On the other hand, for the Higgs potential, we find that it becomes more unstable in the presence of $N_R$. This can be observed from the first row of figure\,\ref{fig:seesaw1lambda}. As indicated by the red and the blue curves, due to the extra fermionic loop contributions from $N_R$, the threshold effects generically introduces a negative shift to $\lambda_H$ at the matching scale such that $\lambda_H$ becomes more negative above the matching scale. 

A similar conclusion can be drawn for the type-III seesaw model, which introduces a fermionic triplet to the SM. For our numerical results, see the bottom row of figure\,\ref{fig:seesaw1mu} for the running of $\mu_H^2$ and that of figure\,\ref{fig:seesaw1lambda} for the running of $\lambda_H$. 

\subsection{Type-II}\label{subsec:type2}
As seen in last subsection, the Higgs potential becomes more unstable due to the presence of extra fermionic degrees of freedom in the type-I and -III seesaw models. Now to remedy this Higgs potential stability issue, one expects extra contributions from bosonic loops as for the type-II seesaw model. The type-II seesaw model is obtained by introducing an extra complex scalar triplet to the SM. While this has been studied in\,\cite{Kobakhidze:2013pya,Babu:2016gpg}, the novelty of our work is that we also include one-loop matching in our study, whose effects on the breaking of the electroweak symmetry turn out to be very important as we will see below.

Different from the type-I and -III seesaw models, the type-II seesaw model contains more free parameters that would be constrained from both the theoretical and the experimental sides which we will briefly discuss here. From section\,\ref{sec:seesaw123}, we find the tree-level matching introduces the following shift to $\lambda_H$:
\eqal{
\lambda_H^{\rm eff\,, LO} &\equiv\lambda_{H} - \delta\lambda_H^G= \lambda_{H,0} - \frac{\mu^2}{2M^2},\quad\quad\quad\quad\quad\quad\quad\,\,\text{(Green's basis)},\\
\lambda_H^{\rm eff\,, LO} &\equiv\lambda_{H} - \delta\lambda_H^W= \lambda_{H,0} - \frac{\mu^2}{2M^2}\left(1-\frac{2\mu_{H,0}^2}{M^2}\right),\quad\text{(Warsaw basis)},
}
where again we use $\mu_{H,0}$ and $\lambda_{H,0}$ for their values at the matching scale. Note that the $\mu$ parameter is not independent. It can be related to other model parameters through the minimizing condition of the triplet potential, leading to\,\cite{Du:2018eaw}
\eqal{
\mu = \sqrt{2}v_\Delta\left(\frac{M^2}{v_\Phi^2} + \lambda_{23} \frac{v_\Delta^2}{v_\Phi^2} + \frac{\lambda_{45}}{2}\right),
}
where $v_{\Delta\,(\Phi)}$ is the vacuum expectation value of the triplet (doublet) and $\lambda_{ij}\equiv\lambda_i+\lambda_j$.

\begin{table}[h!]
  \centering
  \resizebox{\textwidth}{!}{  
\begin{tabular}{|c||c|cccccc|}
\hline
\diagbox{Benchmark}{Parameters} & $v_\Delta$ [GeV] & $M$ [GeV] & $\lambda_1$ & $\lambda_2$ & $\lambda_3$ & $\lambda_4$ & $\lambda_5$ \\
\hline\hline
\multirow{3}{*}{\rm Tiny Yukawa} & \multirow{3}{*}{\rm 1} & \multirow{3}{*}{\rm $10^3$} & \multirow{3}{*}{\rm 0.0957} & \multirow{3}{*}{\rm 0.3} &  \multirow{3}{*}{\rm 0.3} & -0.1 & -0.1 \\
 &  &  &   &  &  & 0.1 & 0.1 \\
 &  &  &   &  &  & 0.5 & 0.5 \\
\hline\hline
\multirow{6}{*}{\rm Modest Yukawa} & \multirow{6}{*}{\rm $10^{-6}$} &  \multirow{3}{*}{\rm $10^3$} & \multirow{3}{*}{\rm 0.0957} & \multirow{3}{*}{\rm 0.3} & \multirow{3}{*}{\rm 0.3} & -0.1 & -0.1 \\
 &  &  &   &  &  & 0.1 & 0.1 \\
 &  &  &   &  &  & 0.5 & 0.5 \\
  \cline{3-8}
 &  & \multirow{3}{*}{\rm $10^6$} & \multirow{3}{*}{\rm 0.0298} & \multirow{3}{*}{\rm 0.3} & \multirow{3}{*}{\rm 0.3} & -0.1 & -0.1 \\
 &  &  &   &  &  & 0.1 & 0.1 \\
 &  &  &   &  &  & 0.5 & 0.5 \\
\hline\hline
\multirow{9}{*}{\rm Large Yukawa} & \multirow{9}{*}{\rm $10^{-11}$} & \multirow{3}{*}{\rm $10^3$}  & \multirow{3}{*}{\rm 0.0957} & \multirow{3}{*}{\rm 0.3} & \multirow{3}{*}{\rm 0.3} & -0.1 & -0.1 \\
 &  &  &   &  &  & 0.1 & 0.1 \\
 &  &  &   &  &  & 0.5 & 0.5 \\
  \cline{3-8}
 &  & \multirow{3}{*}{\rm $10^6$}  & \multirow{3}{*}{\rm 0.0298} & \multirow{3}{*}{\rm 0.3} & \multirow{3}{*}{\rm 0.3} & -0.1 & -0.1 \\
 &  &  &   &  &  & 0.1 & 0.1 \\
 &  &  &   &  &  & 0.5 & 0.5 \\
  \cline{3-8}
 &  & \multirow{3}{*}{\rm $10^{12}$}  & \multirow{3}{*}{\rm -0.0127} & \multirow{3}{*}{\rm 0.3} & \multirow{3}{*}{\rm 0.3} & -0.1 & -0.1 \\
 &  &  &   &  &  & 0.1 & 0.1 \\
 &  &  &   &  &  & 0.5 & 0.5 \\
\hline\hline
\end{tabular}
}\caption{Benchmark points for the type-II seesaw model for illustrating radiative generated electroweak spontaneous symmetry breaking.}\label{tab:seesaw2BM}
\end{table}

The most recent precision measurements of the $\rho$ parameter gives $\rho=1.00038\pm0.00020$\,\cite{ParticleDataGroup:2020ssz}, resulting in $0\le v_\Delta \lesssim 2.56 {\rm\,GeV}\ll v_\Phi$. Therefore, we obtain the following approximate expression based on this hierarchy
\eqal{\delta\lambda_H^{G,W}\approx\frac{\mu^2}{2M^2} \approx\left(\frac{v_\Delta M}{v_\Phi^2}\right)^2\lesssim\mathcal{O}(1),\label{eq:vMbound}}
where the last inequality is from the consideration of perturbative unitarity and perturbativity, which then provides the upper bound for $v_\Delta M$. On the other hand, neutrino masses from the type-II seesaw model is directly proportional to $v_\Delta$, i.e., $(m_\nu)_{\alpha\beta}=\sqrt{2}(Y_\nu)_{\alpha\beta}v_\Delta$ with $Y_{\nu}$ the neutrino Yukawa in the type-II seesaw model. Therefore, using the same approximation for $m_\nu$ as for the type-I seesaw model and taking the upper bound on $v_\Delta M$ into account, we choose our benchmark values as enumerated in table\,\ref{tab:seesaw2BM}. A few comments on the numbers we choose are in order:
\begin{itemize}
\item Depending on $v_\Delta$, the neutrino Yukawa can be either tiny, or modest, or even as large as $\mathcal{O}(1)$. Since collider signatures of the triplet model are very distinct for different neutrino Yukawa couplings, see for example figure 7 of Ref.\,\cite{Du:2018eaw}, we consider three scenarios where the Yukawa couplings $Y_\nu$ can be either tiny with $v_\Delta=1$\,GeV, or modest with $v_\Delta=10^{-6}$\,GeV, or large with $v_\Delta=10^{-11}$\,GeV.
\item Given current results from collider searches\,\cite{ATLAS:2017xqs} and the upper bound on $v_\Delta M$ from eq.\,\eqref{eq:vMbound}, we consider three cases for the scale of the triplet: $M=10^3$\,GeV, $10^6$\,GeV and $10^{12}$\,GeV, respectively. While the last two scenarios are beyond the reach of current and/or future colliders, the lightest scenario is testable at the HL-LHC and/or future colliders.
\item The value of $\lambda_1$ in the fourth column is the corresponding value of $\lambda_H$ at the matching scale but without including the tree- and one-loop matching shifts discussed in section\,\ref{sec:seesaw123}.\footnote{The shift from tree- and one-loop matching can be easily calculated by using our analytical results presented in section\,\ref{sec:seesaw123}, which are included in solving the RGEs in practice.} Note that $\lambda_1$ already becomes negative in the last row due to the large matching scale.
\item $\lambda_{2,3}$ barely have any phenomenological effects due to suppression from $v_\Delta$. For this reason, we fix $\lambda_2=\lambda_3=0.3$ and comment on that, in practice, we find different values for $\lambda_{2,3}$ only slightly change the concavity or convexity of the running of $\mu_H^2$ and $\lambda_H$ at very large energy scales.
\item The value of $\lambda_{4,5}$ are chosen based on tree-level vacuum stability and perturbative unitarity. See Ref.\,\cite{Du:2018eaw} and references therein for a detailed discussion. Phenomenologically, these two couplings play a key role in explaining the observed baryon number asymmetry of the universe through electroweak baryogenesis. For a recent study on phase transition as well as the gravitational waves from the type-II seesaw model, see Ref.\,\cite{Zhou:2022mlz}.
\end{itemize}

\begin{figure}[t]
\centering{
  \begin{adjustbox}{max width = \textwidth}
\begin{tabular}{ccc}
  \Huge{~~~~~$\lambda_4=-0.1$, $\lambda_5=-0.1$}	& \Huge{~~~~~$\lambda_4=0.1$, $\lambda_5=0.1$} & \Huge{~~~~~$\lambda_4=0.5$, $\lambda_5=0.5$} \\ 
\includegraphics[width=0.78\textwidth]{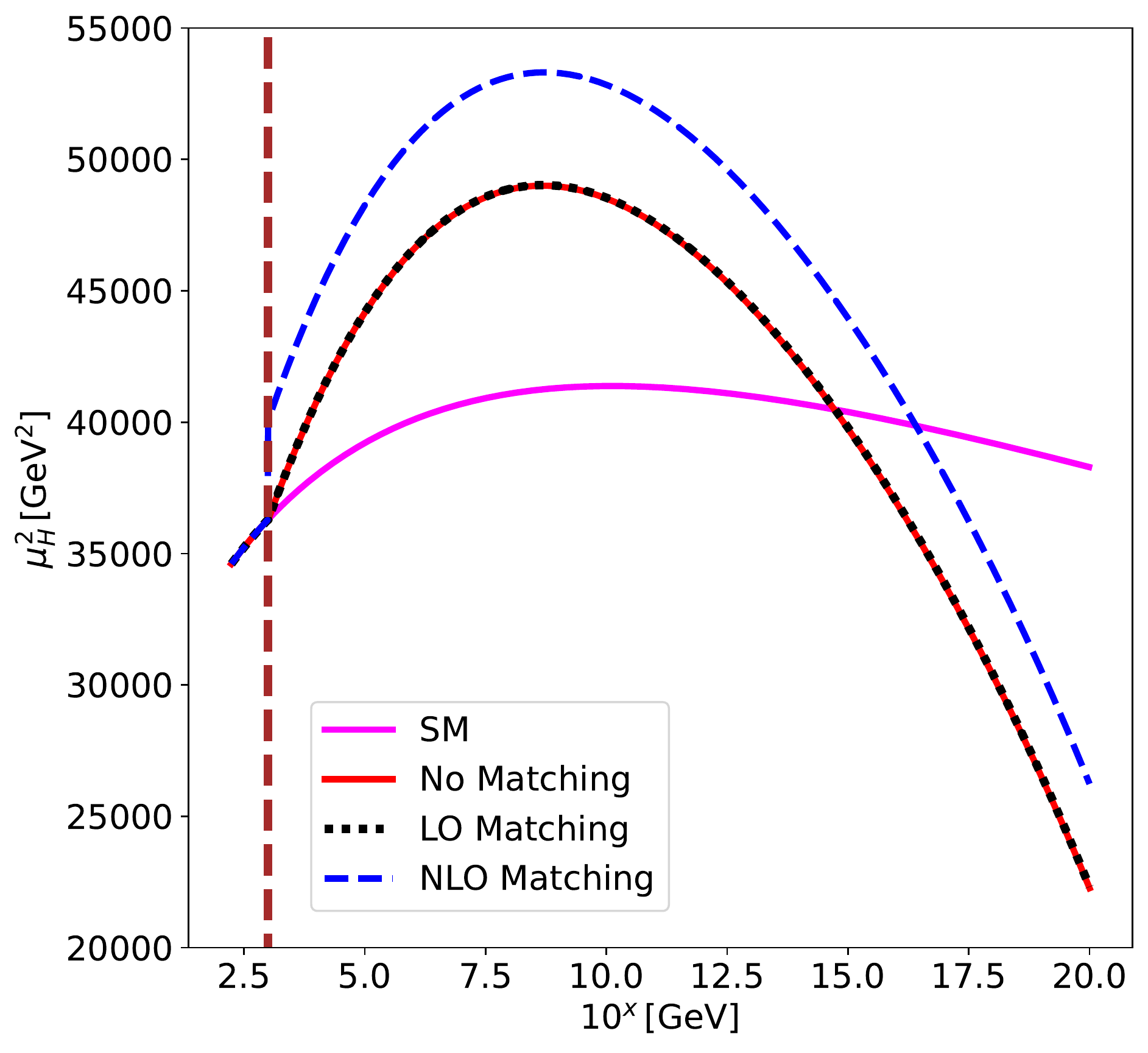} & 
	\includegraphics[width=0.8\textwidth]{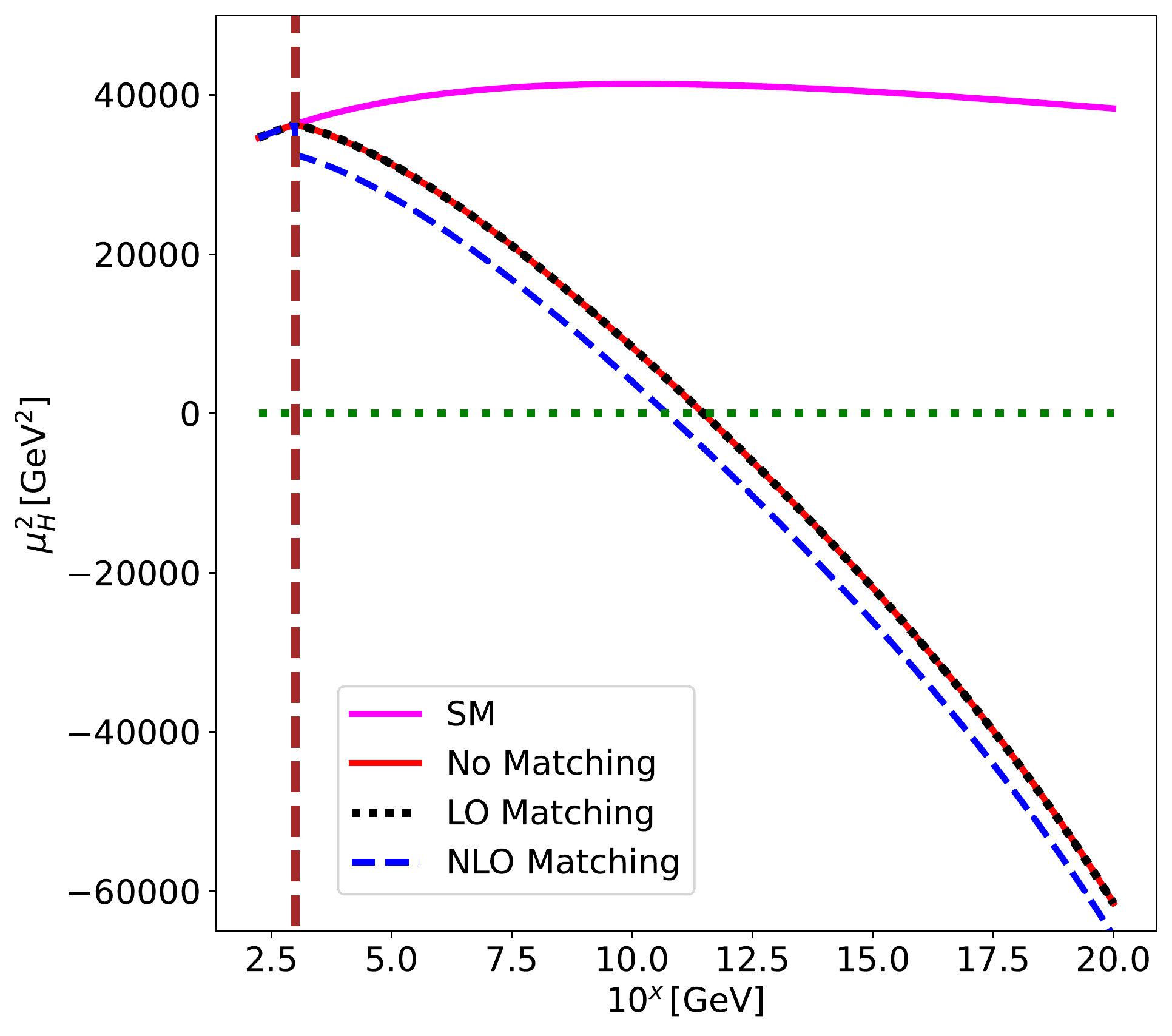} & 
	\includegraphics[width=0.8\textwidth]{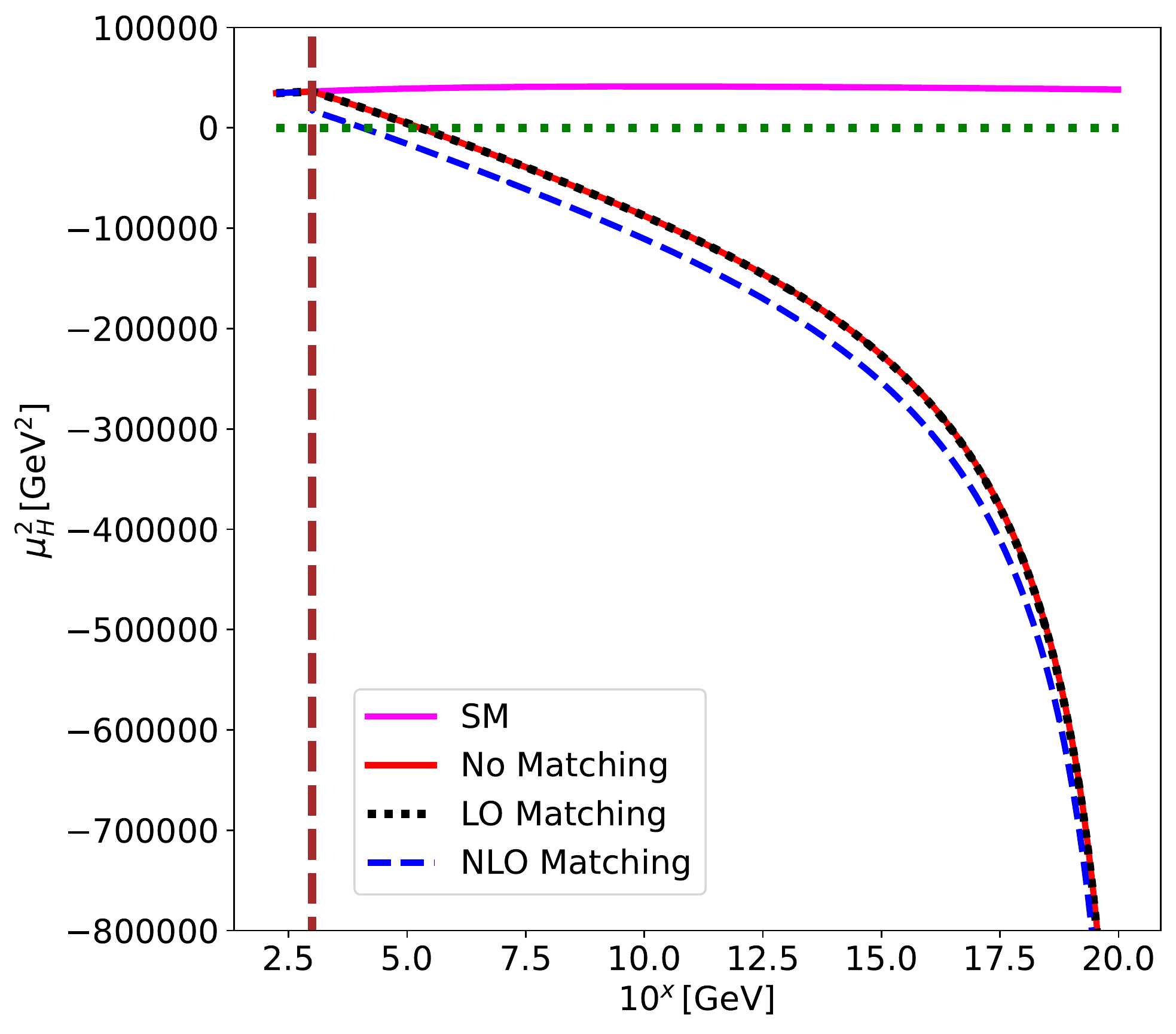}\\
	& \Huge{(Tiny Yukawa with $M=10^3$\,GeV)} & \\ 
\includegraphics[width=0.8\textwidth]{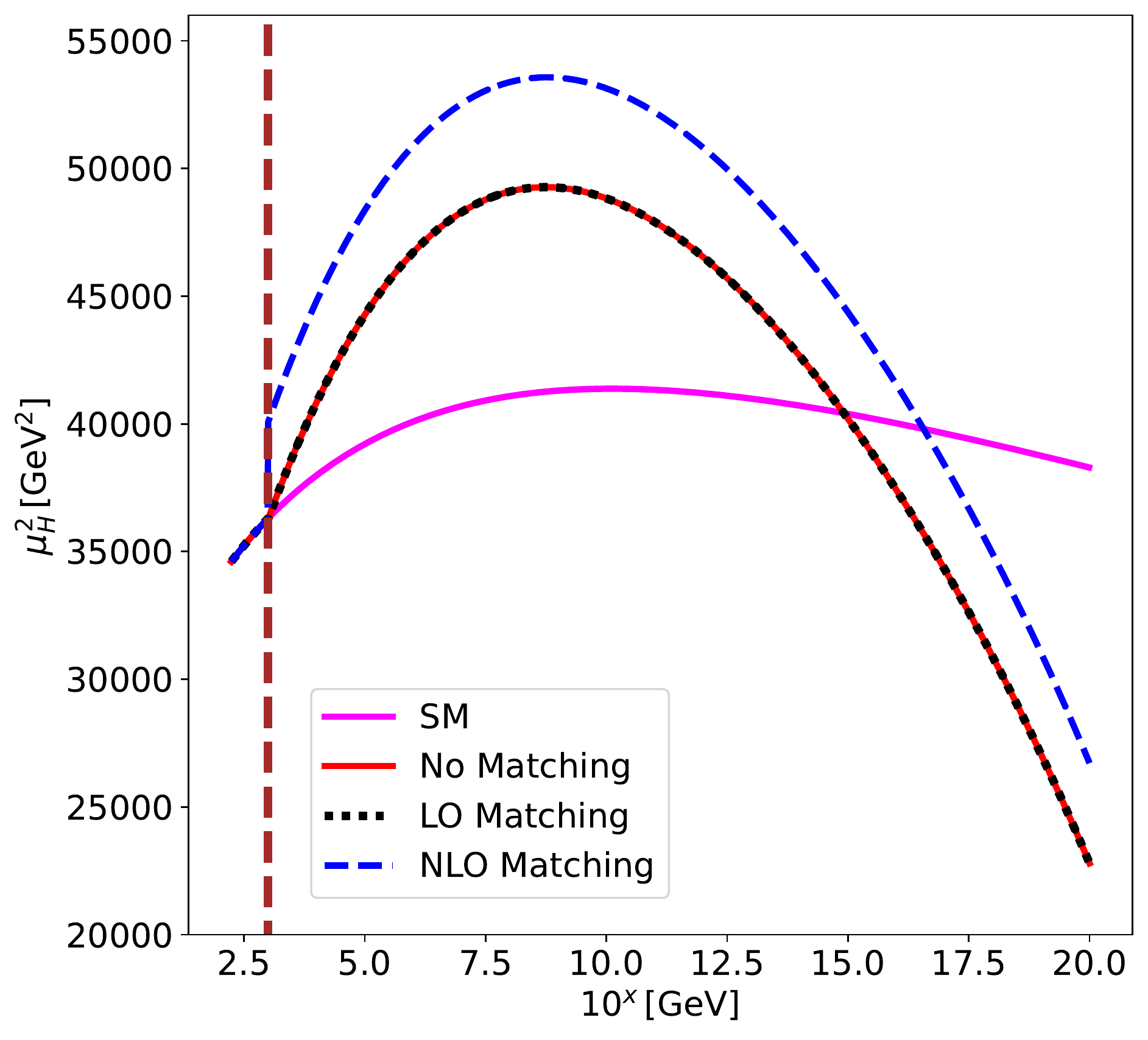} & 
	\includegraphics[width=0.8\textwidth]{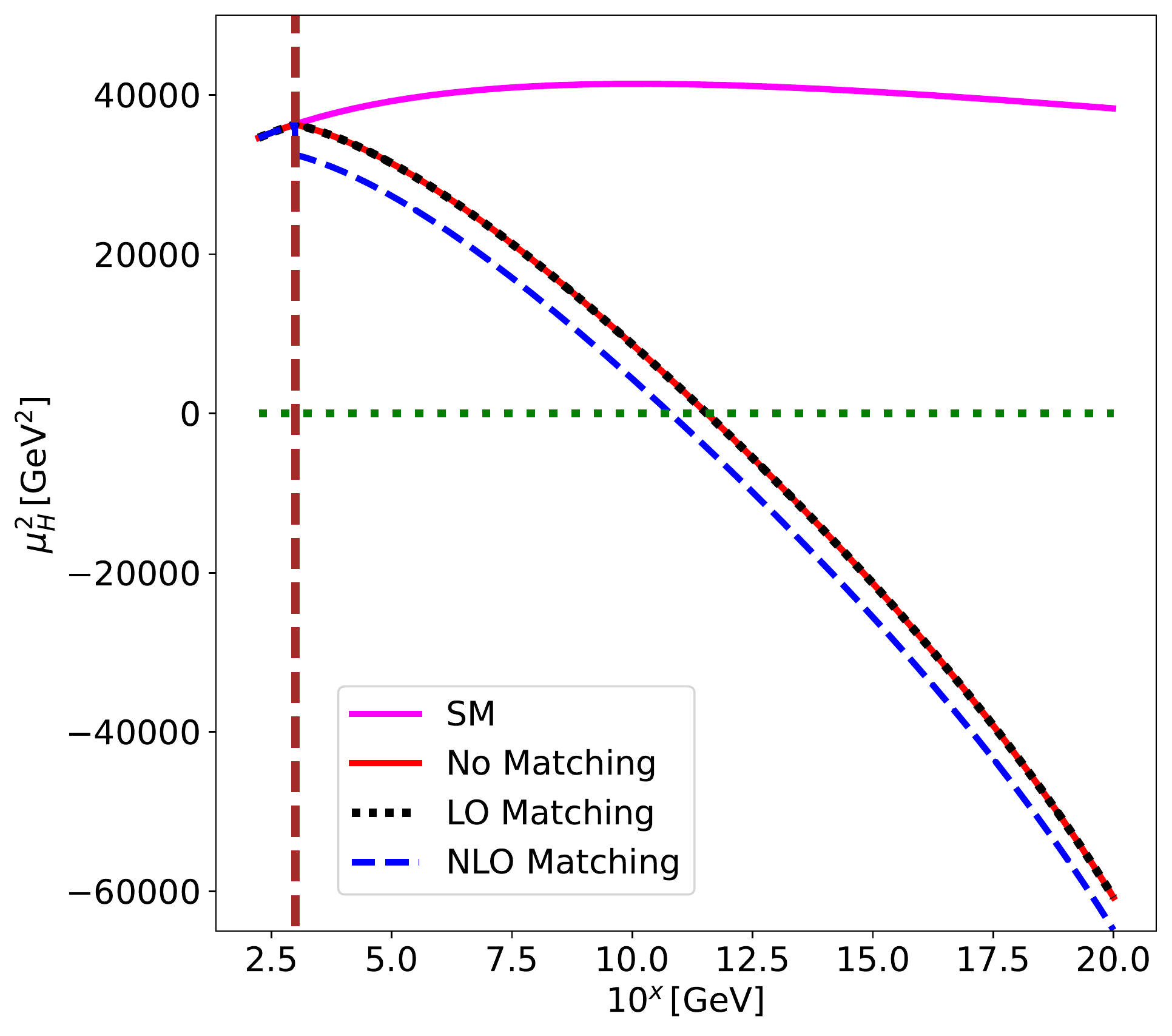} & 
	\includegraphics[width=0.8\textwidth]{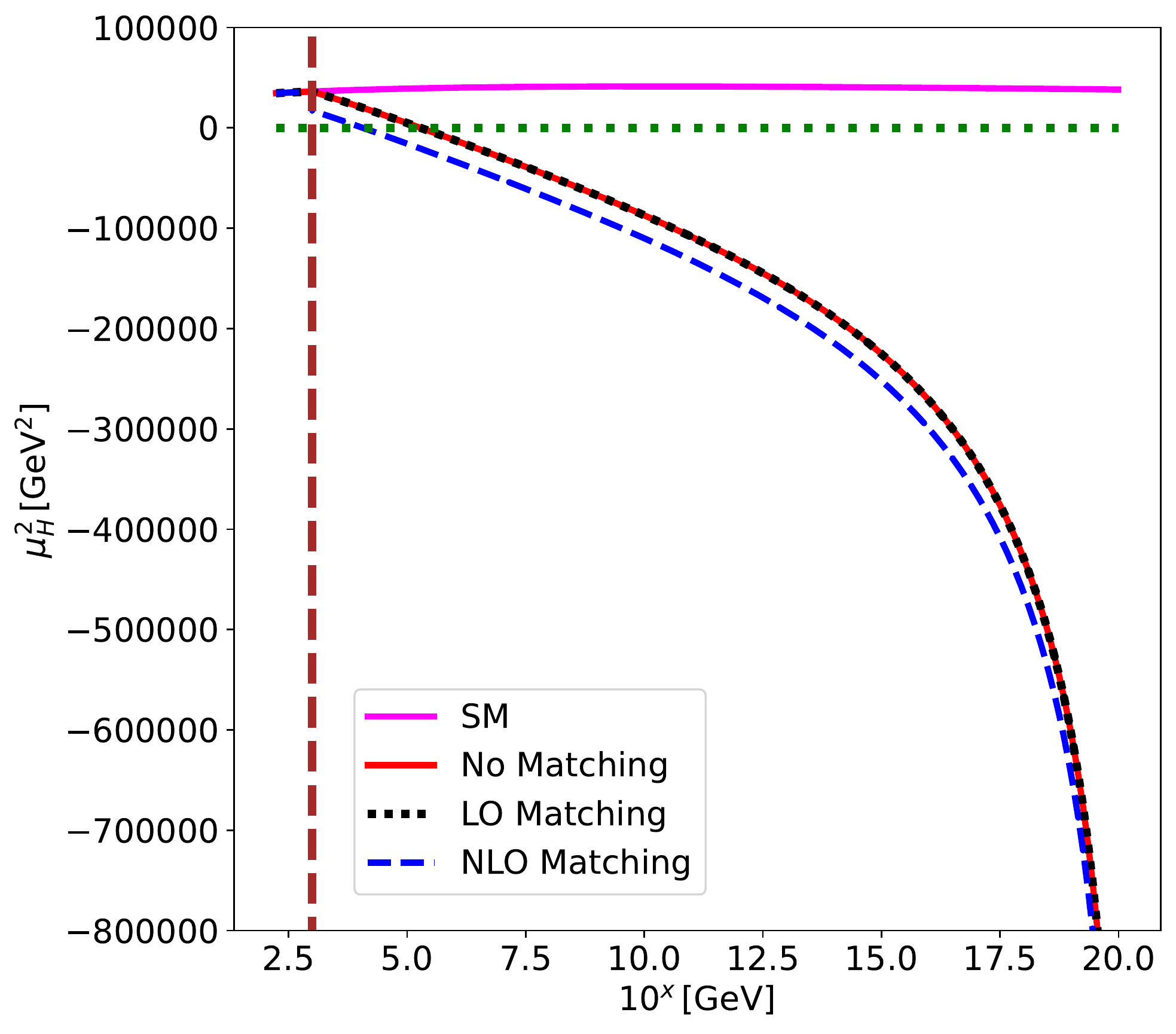}\\
	& \Huge{(Modest Yukawa with $M=10^3$\,GeV)} & \\ 
\includegraphics[width=0.76\textwidth]{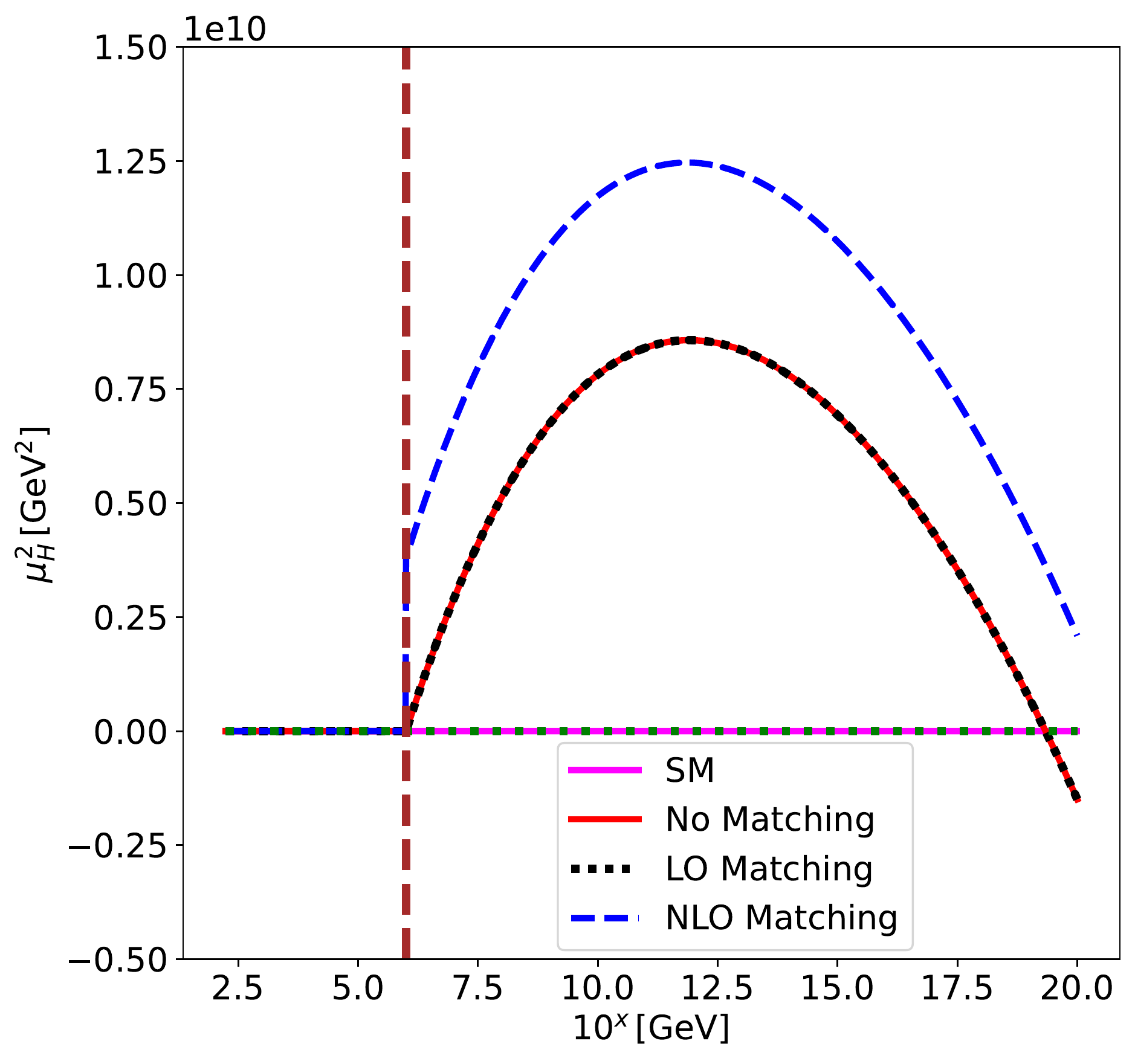} & 
	\includegraphics[width=0.76\textwidth]{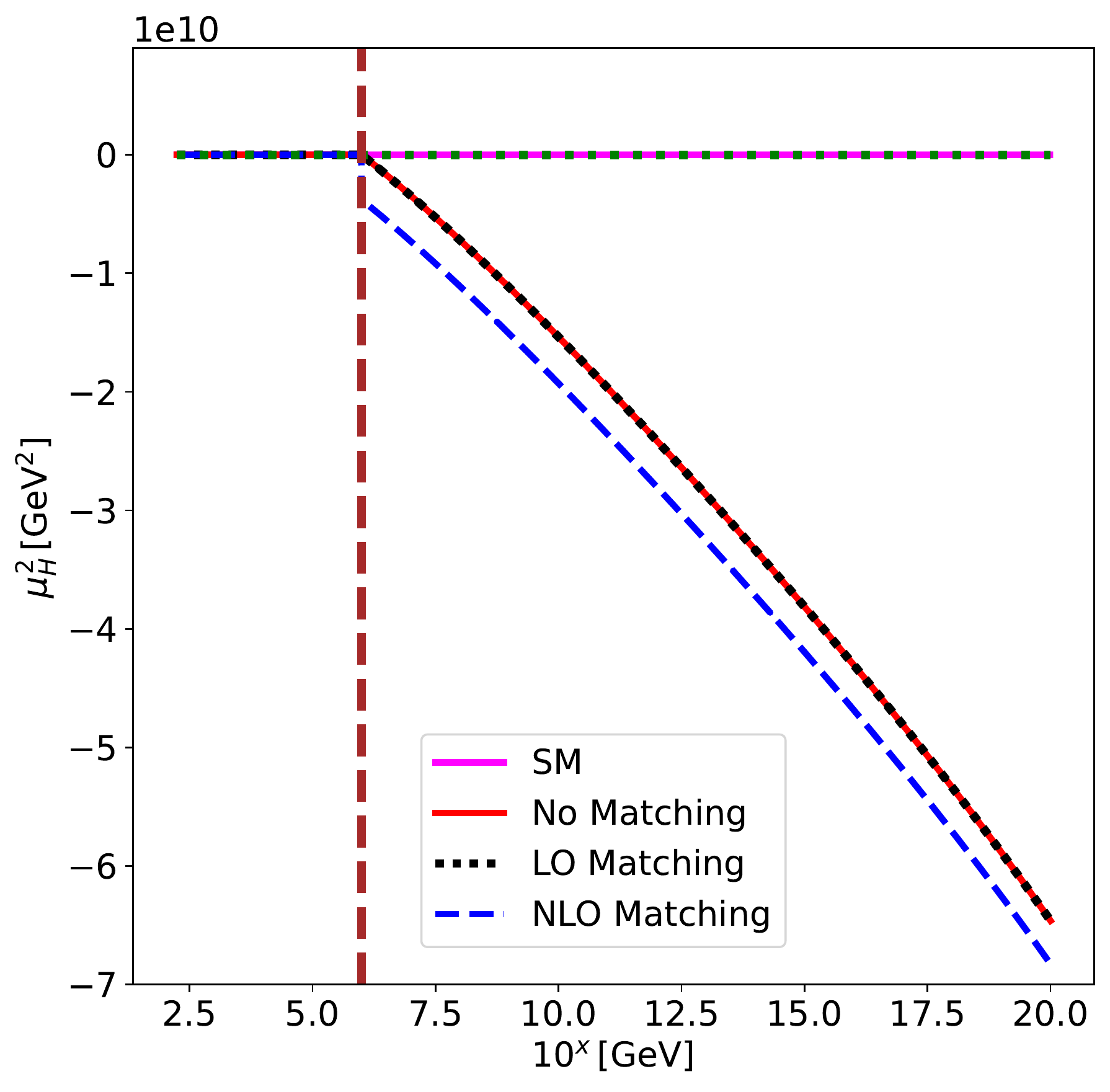} & 
	\includegraphics[width=0.78\textwidth]{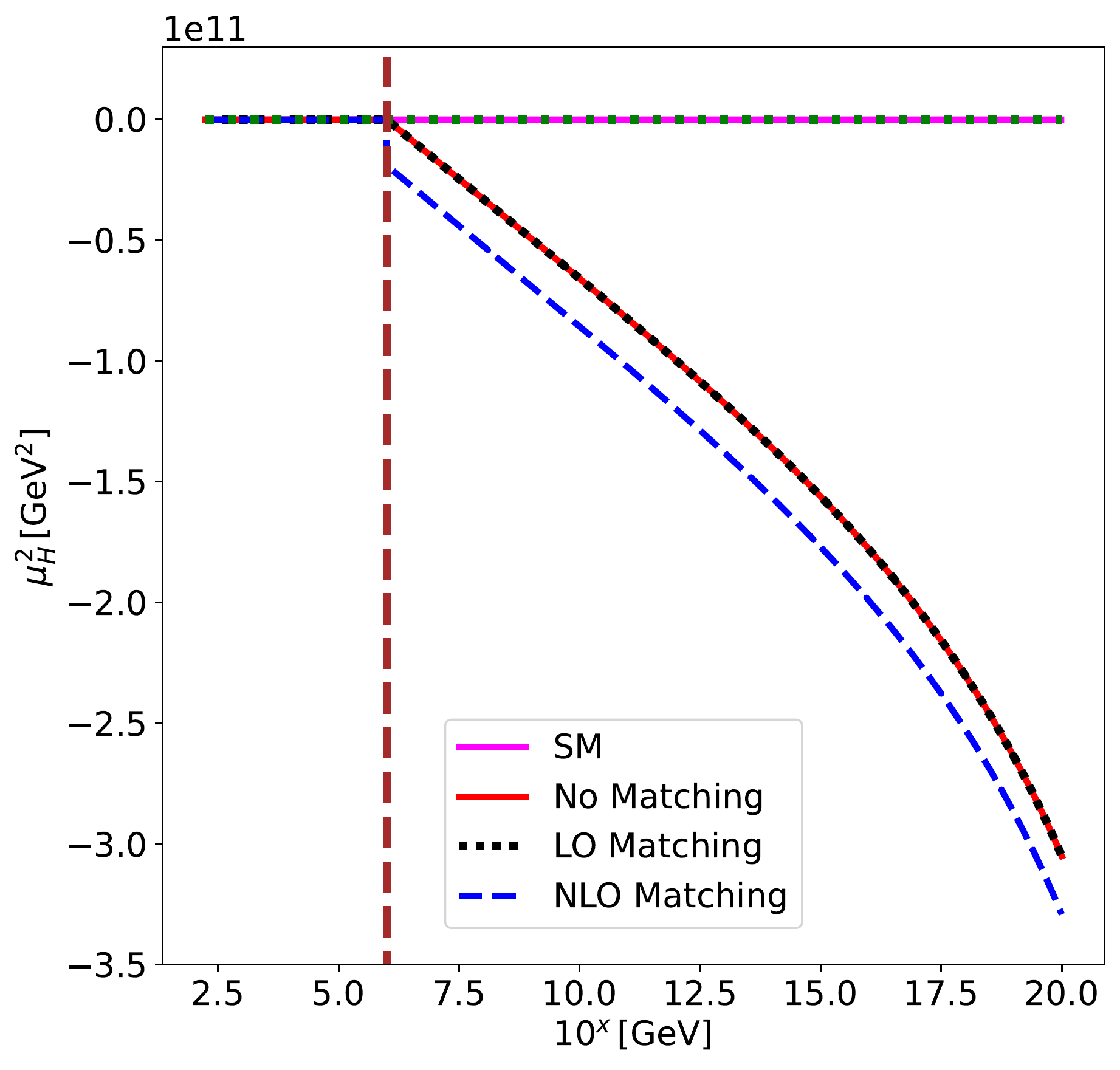}\\
	& \Huge{(Modest Yukawa with $M=10^6$\,GeV)} & \\ 	
\end{tabular}
  \end{adjustbox}
  }\caption{Running of $\mu_H^2$ in the type-II seesaw model, where we use the same color codes as those in figure\,\ref{fig:seesaw1mu} with the addition of the dotted line in black for the tree-level matching. The first row is for the tiny Yukawa case, and the last two rows are for the modest Yukawa case in table\,\ref{tab:seesaw2BM}.}\label{fig:seesaw2mu}
\end{figure}

\begin{figure}[t]
\centering{
  \begin{adjustbox}{max width = \textwidth}
\begin{tabular}{ccc}
  \Huge{~~~~~$\lambda_4=-0.1$, $\lambda_5=-0.1$}	& \Huge{~~~~~$\lambda_4=0.1$, $\lambda_5=0.1$} & \Huge{~~~~~$\lambda_4=0.5$, $\lambda_5=0.5$} \\ 
\includegraphics[width=0.8\textwidth]{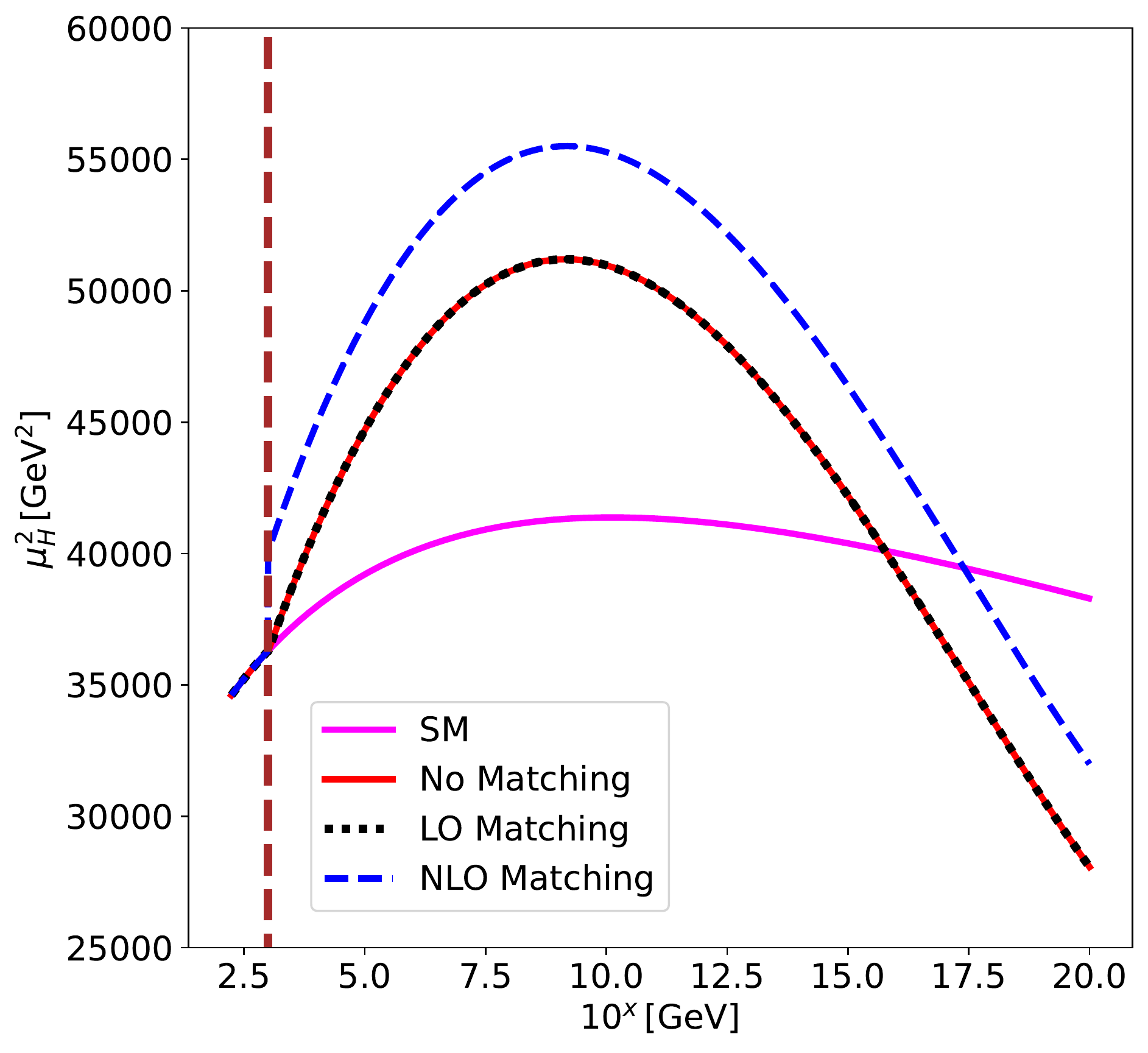} & 
	\includegraphics[width=0.82\textwidth]{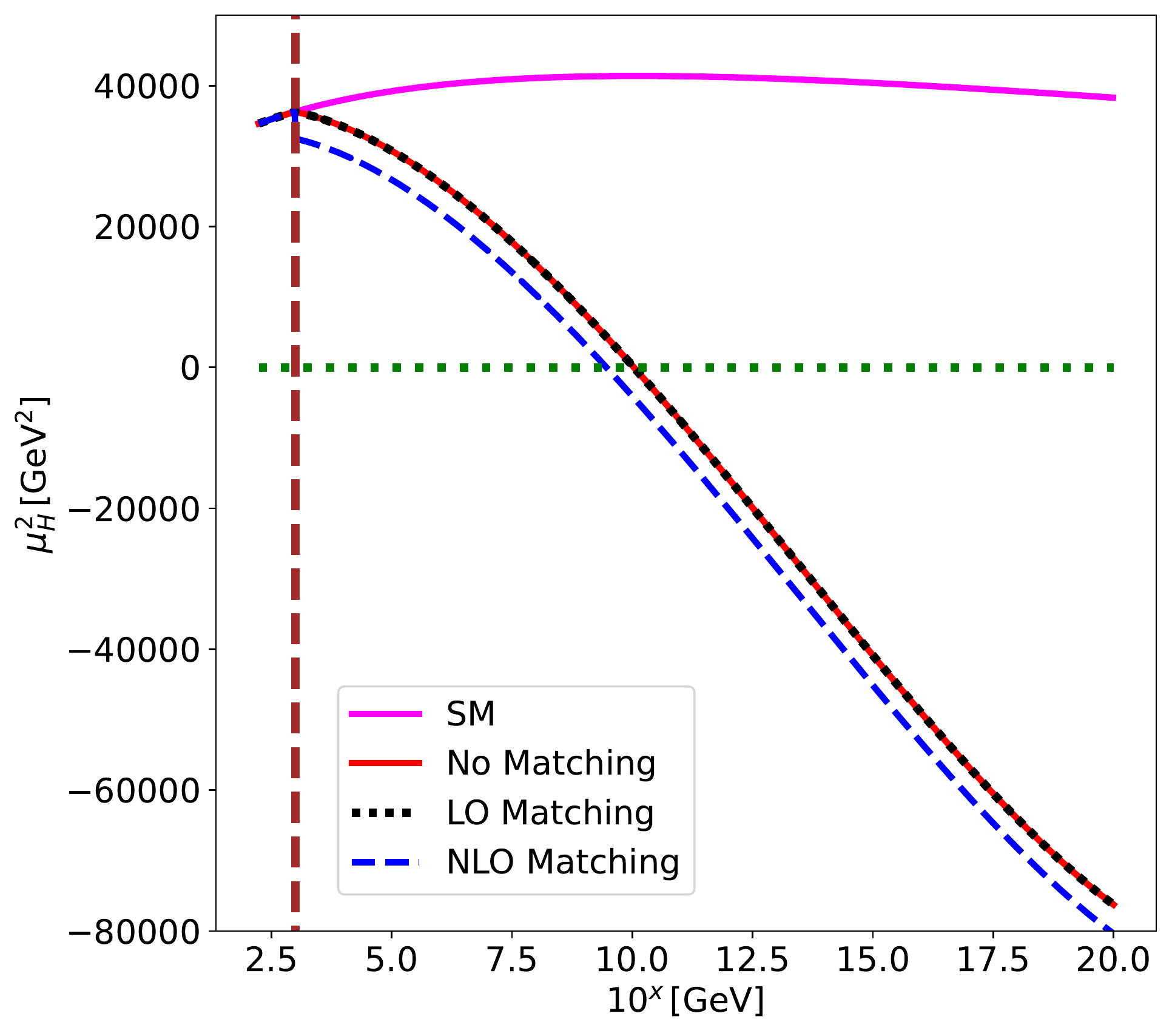} & 
	\includegraphics[width=0.84\textwidth]{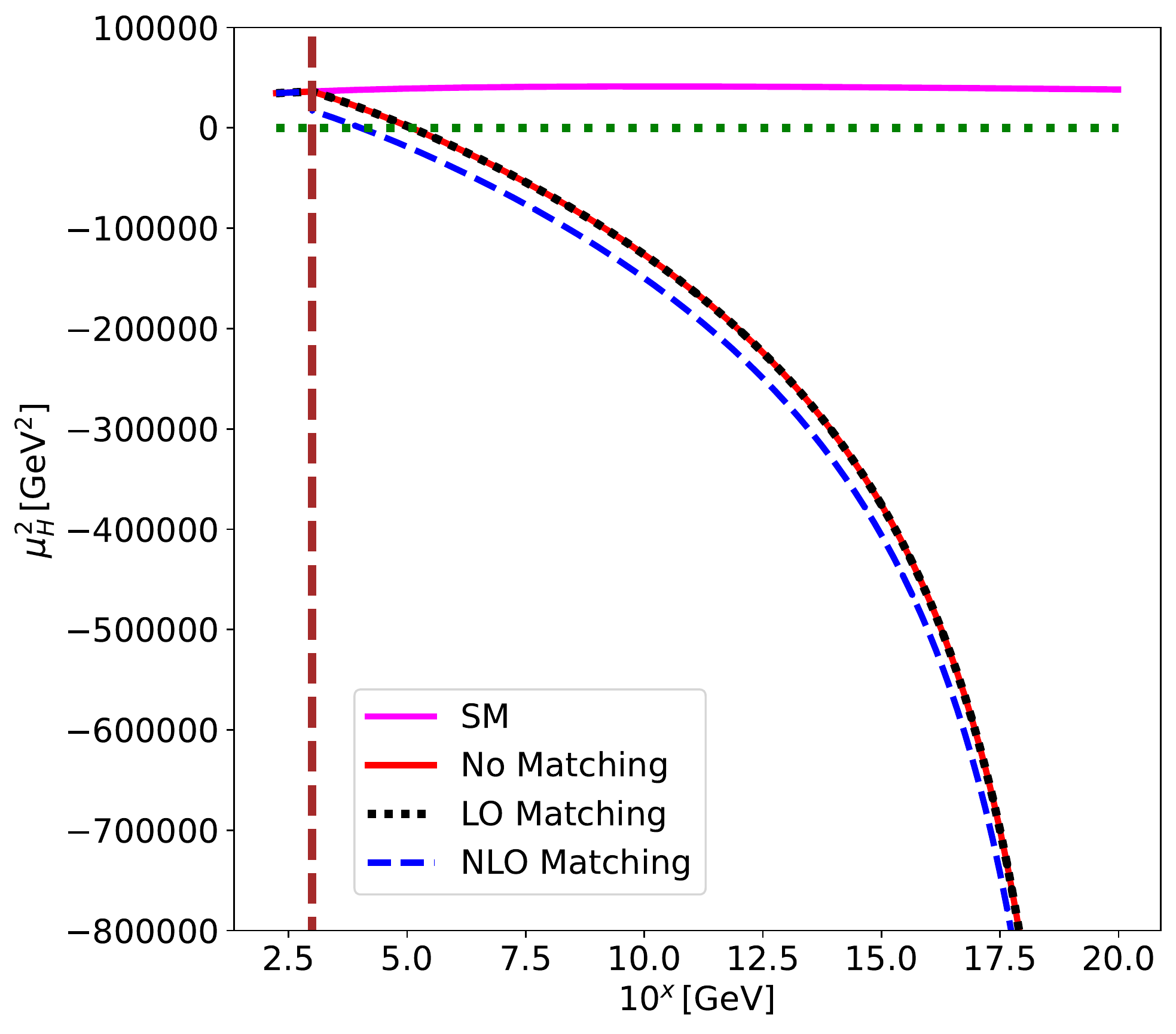}\\
	& \Huge{(Large Yukawa with $M=10^3$\,GeV)} & \\ 
\includegraphics[width=0.8\textwidth]{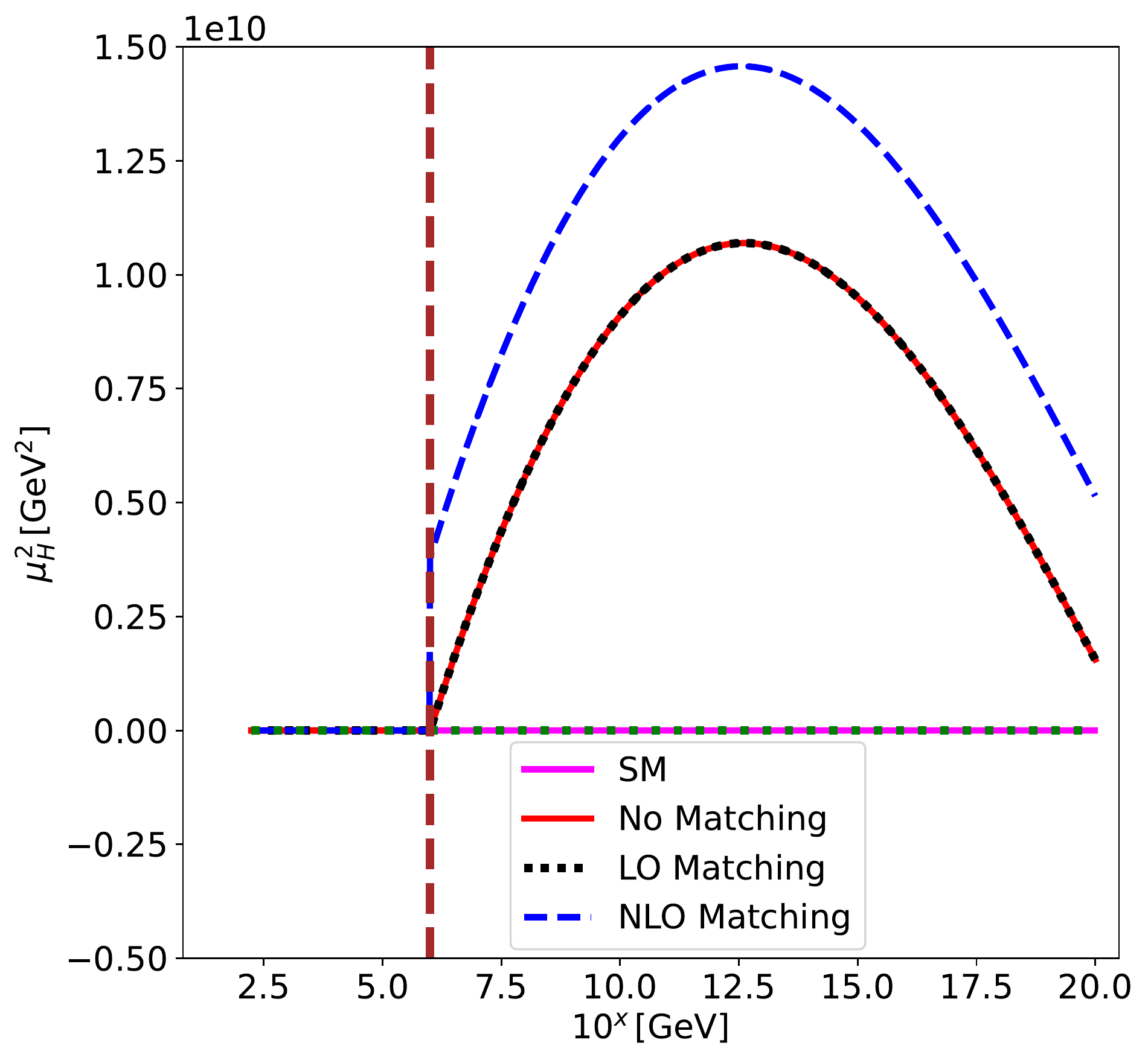} & 
	\includegraphics[width=0.78\textwidth]{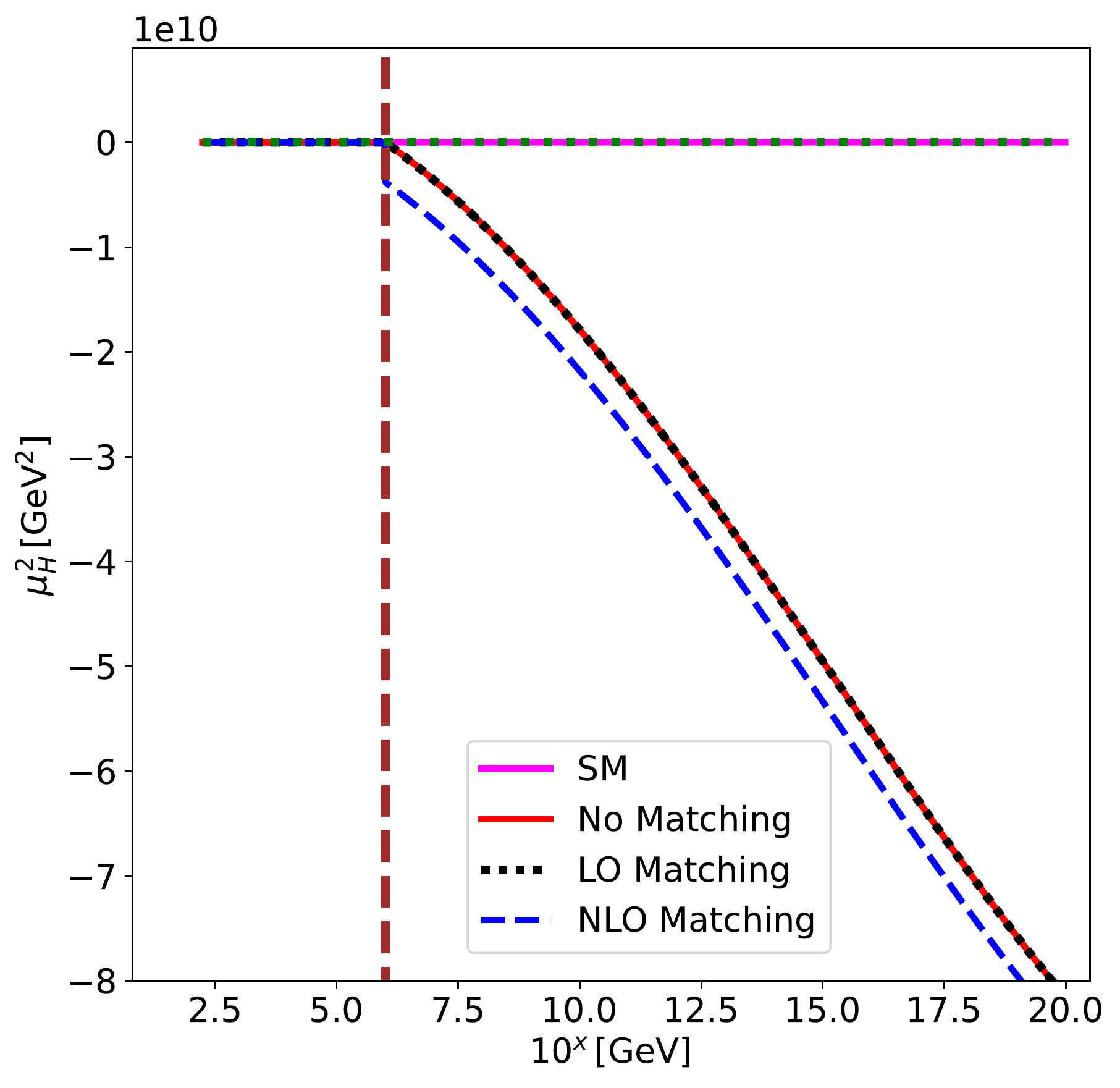} & 
	\includegraphics[width=0.78\textwidth]{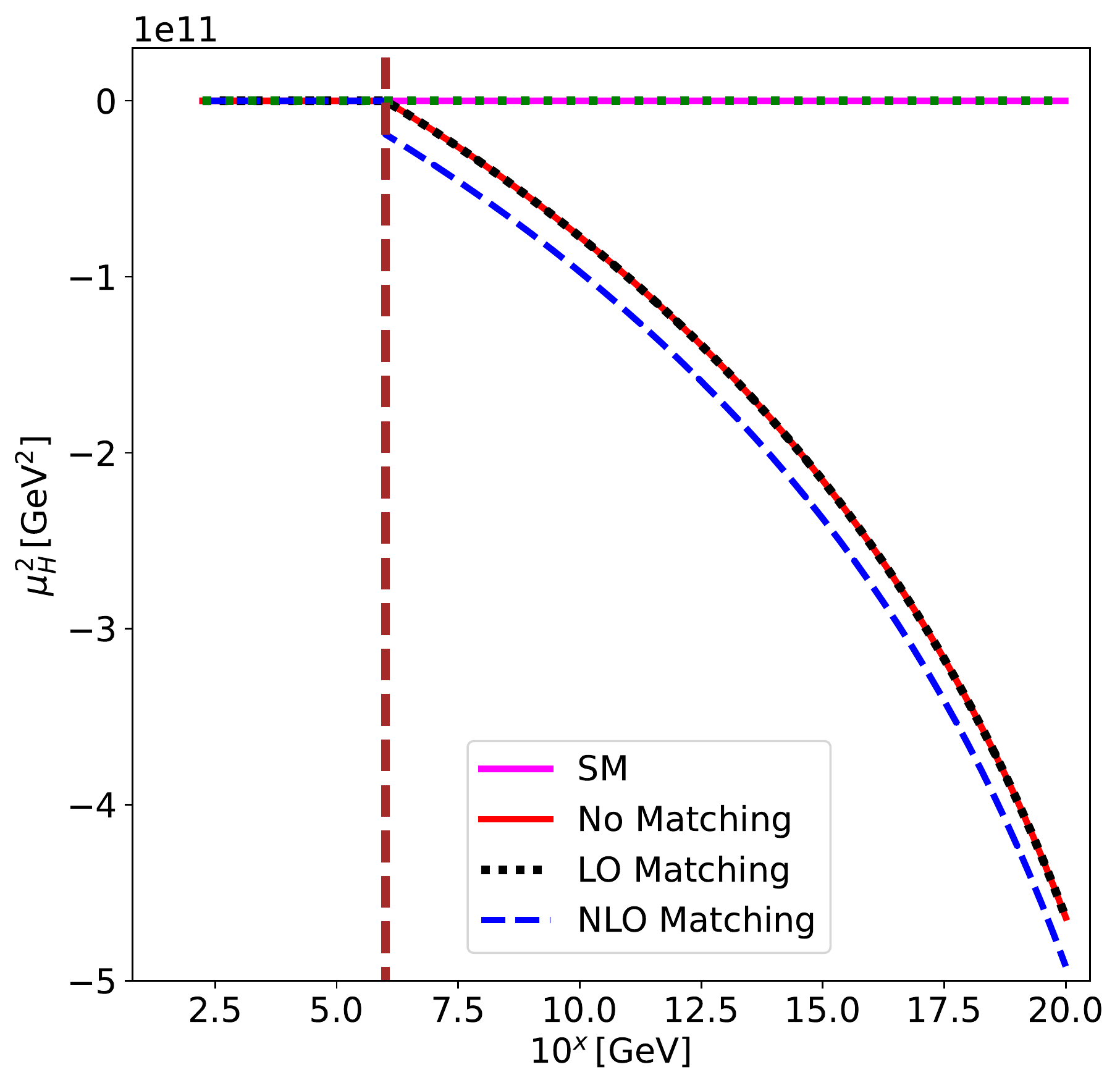}\\
	& \Huge{(Large Yukawa with $M=10^6$\,GeV)} & \\ 
\includegraphics[width=0.8\textwidth]{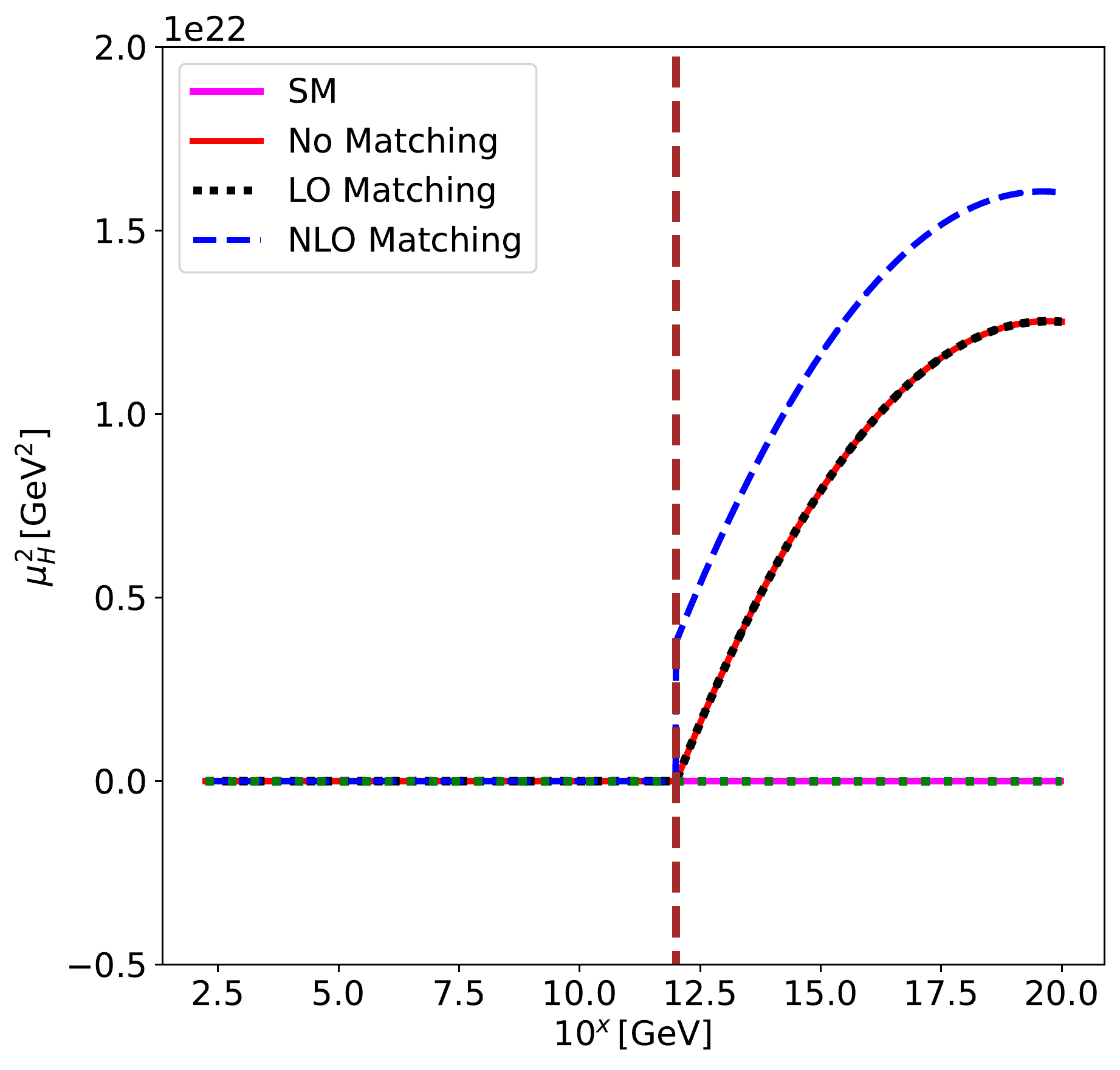} & 
	\includegraphics[width=0.78\textwidth]{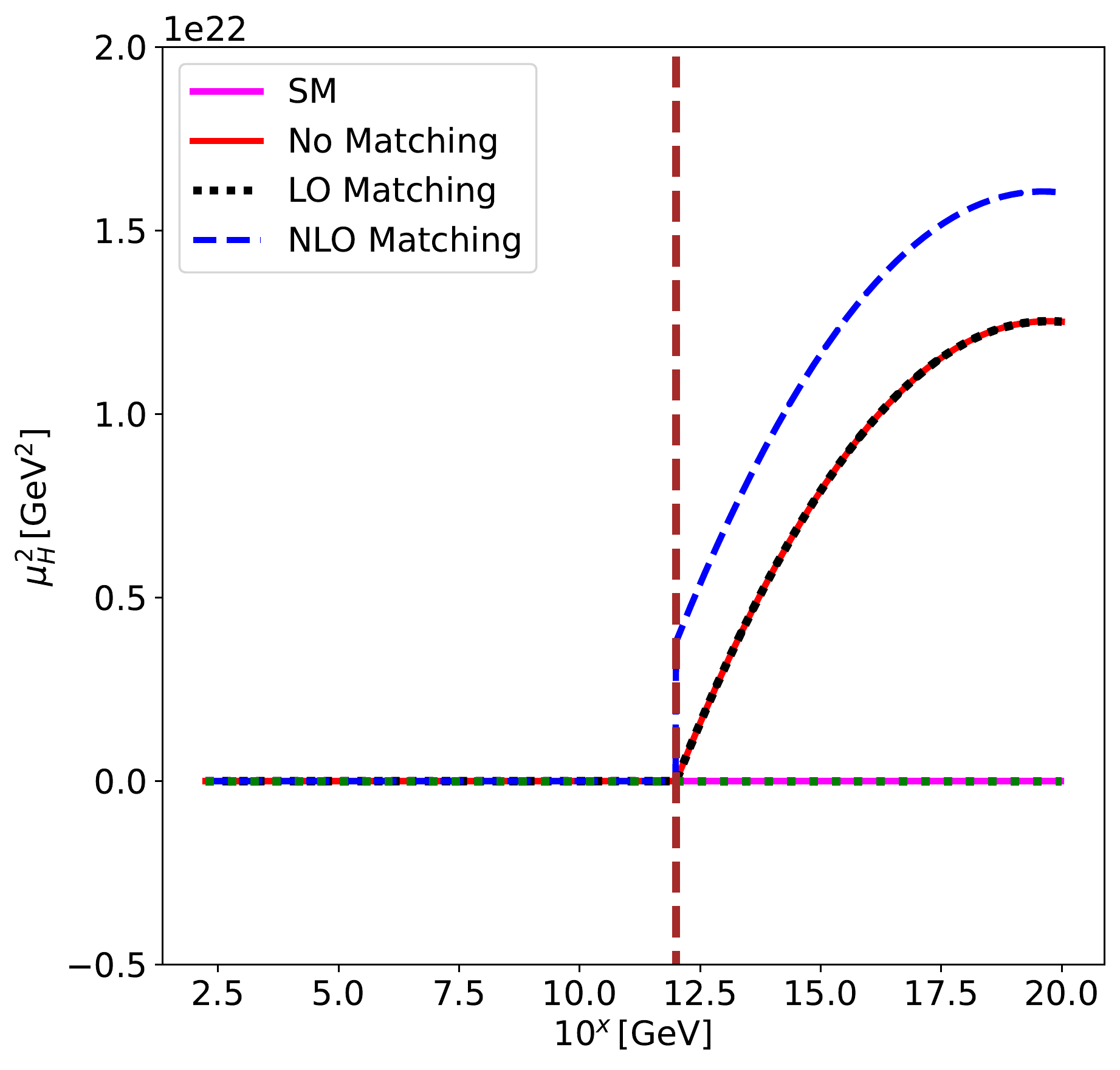} & 
	\includegraphics[width=0.78\textwidth]{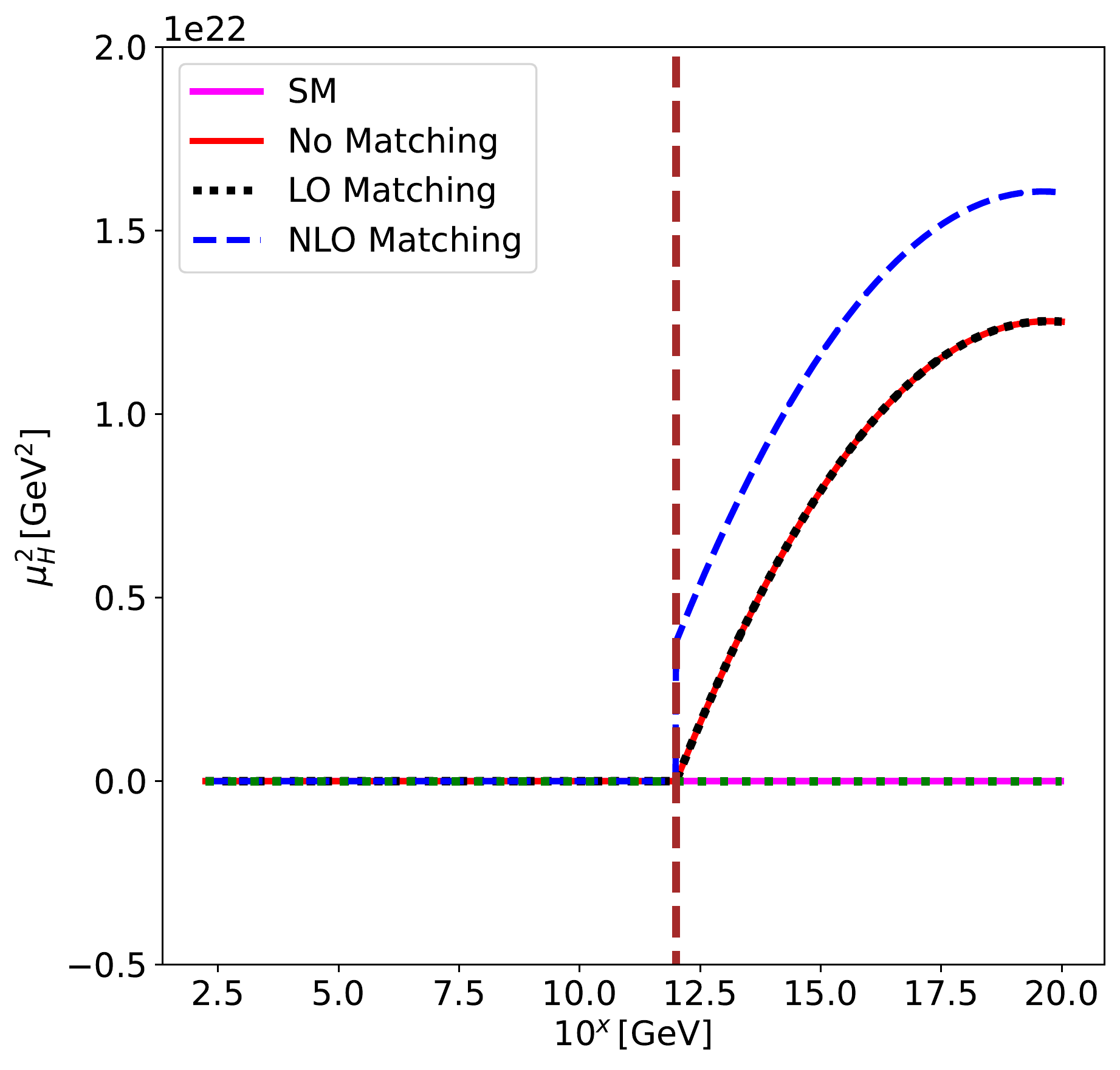}\\
	& \Huge{(Large Yukawa with $M=10^{12}$\,GeV)} & \\ 
\end{tabular}
  \end{adjustbox}
  }\caption{Same as figure\,\ref{fig:seesaw2mu} but for the large Yukawa case in table\,\ref{tab:seesaw2BM}.}\label{fig:seesaw2mu2}
\end{figure}

\begin{figure}[t]
\centering{
  \begin{adjustbox}{max width = \textwidth}
\begin{tabular}{ccc}
  \Huge{~~~~~$\lambda_4=-0.1$, $\lambda_5=-0.1$}	& \Huge{~~~~~$\lambda_4=0.1$, $\lambda_5=0.1$} & \Huge{~~~~~$\lambda_4=0.5$, $\lambda_5=0.5$} \\ 
\includegraphics[width=0.8\textwidth]{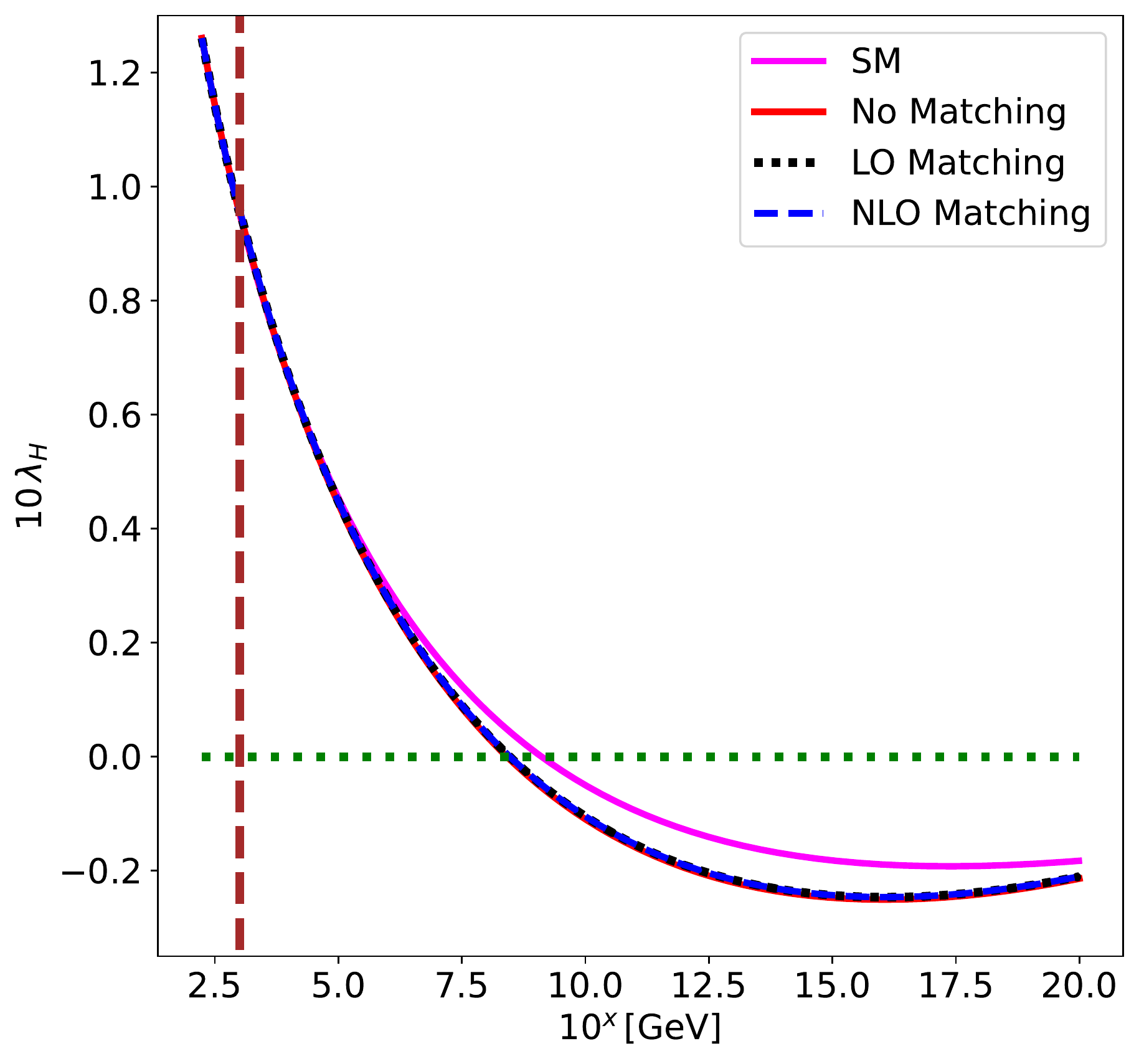} & 
	\includegraphics[width=0.8\textwidth]{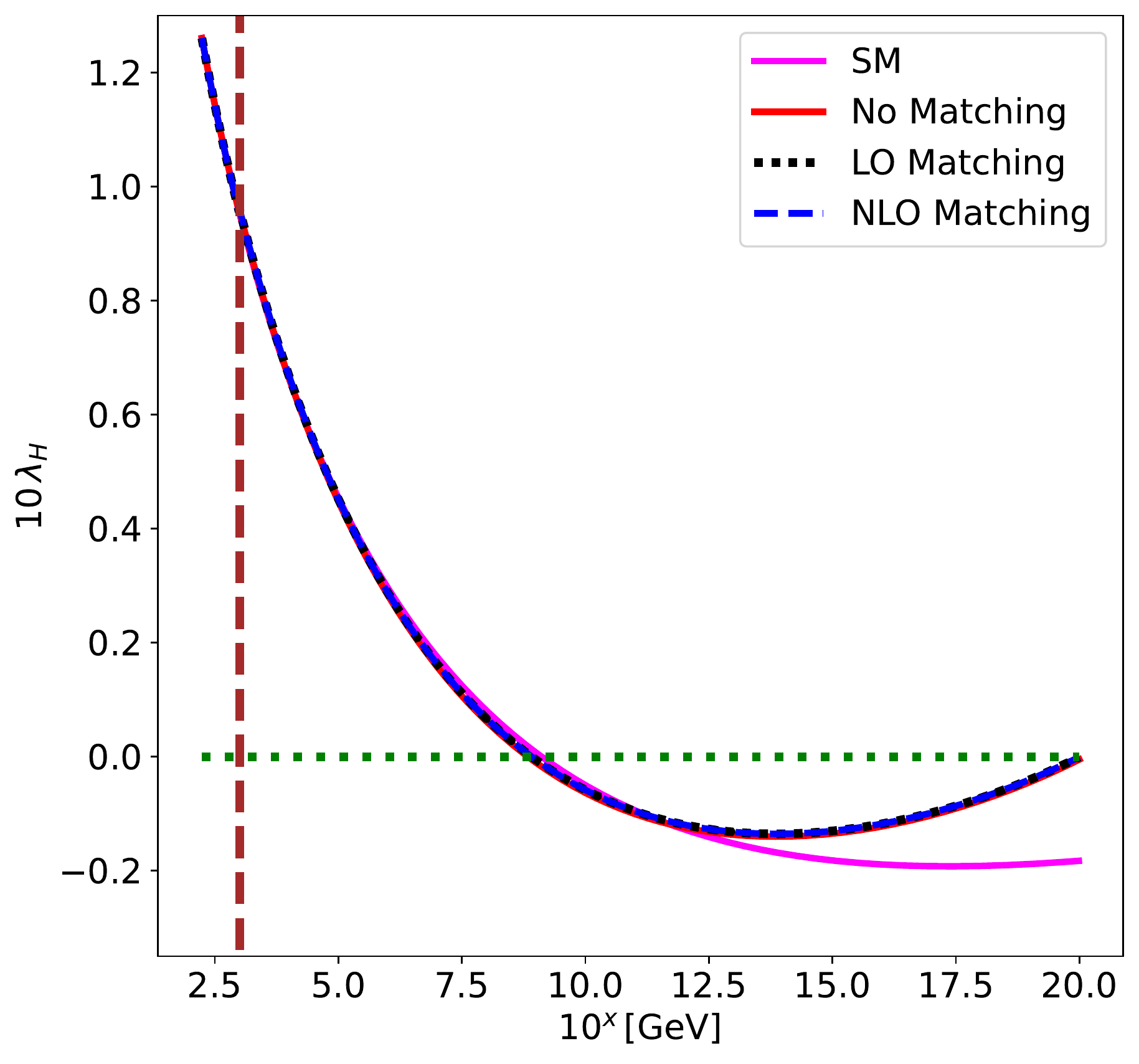} & 
	\includegraphics[width=0.76\textwidth]{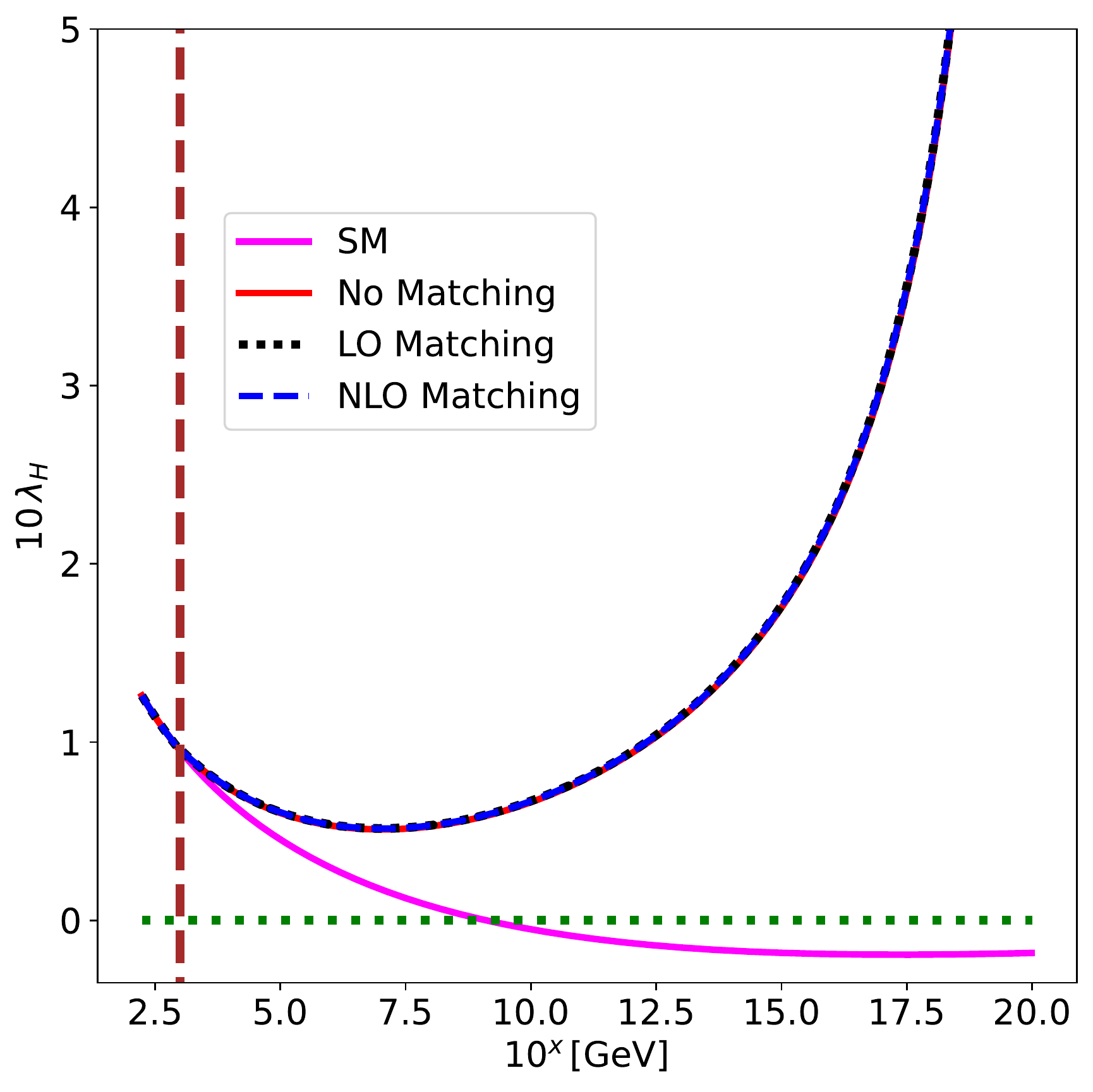}\\
	& \Huge{(Tiny Yukawa with $M=10^3$\,GeV)} & \\ 
\includegraphics[width=0.8\textwidth]{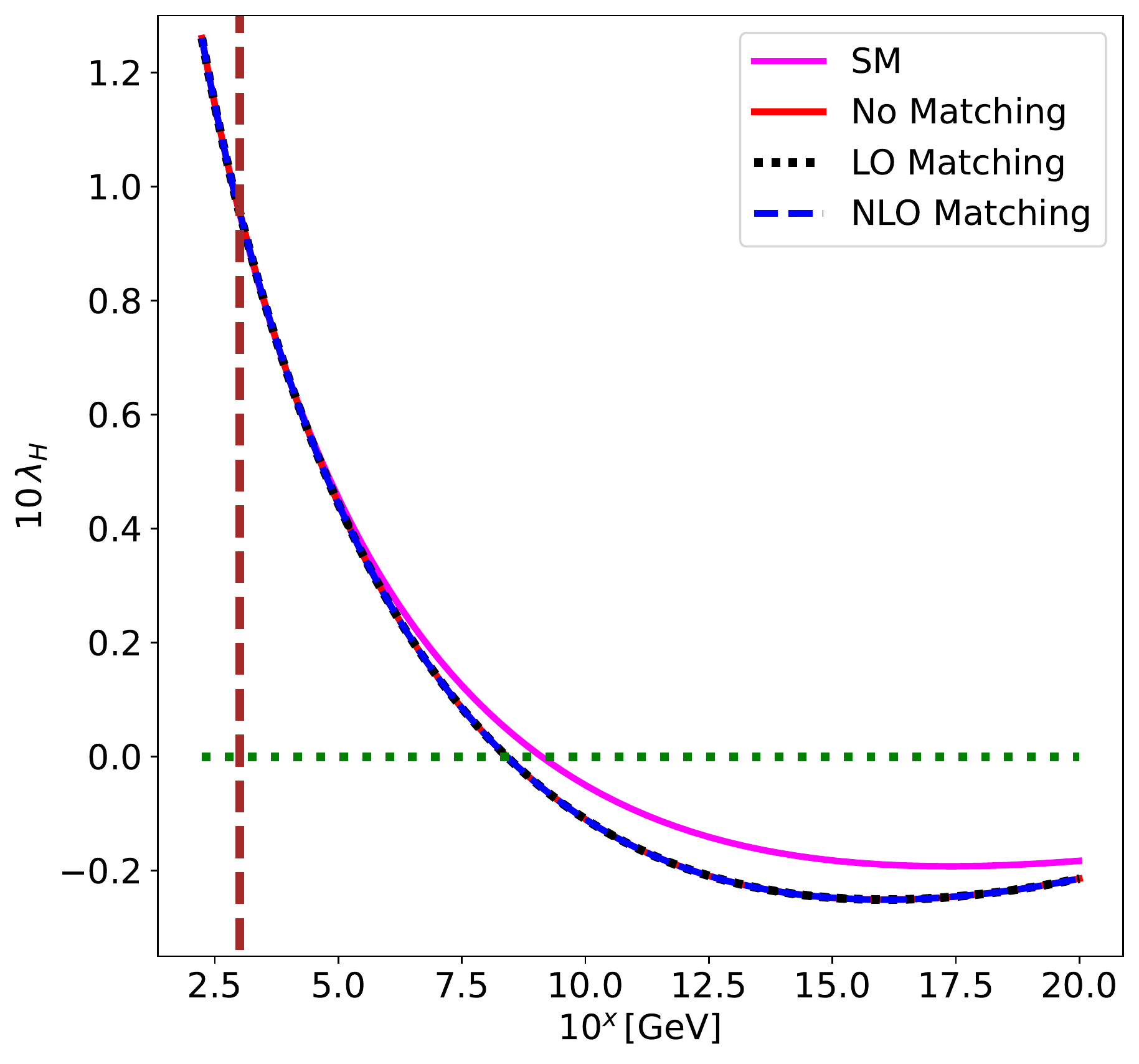} & 
	\includegraphics[width=0.8\textwidth]{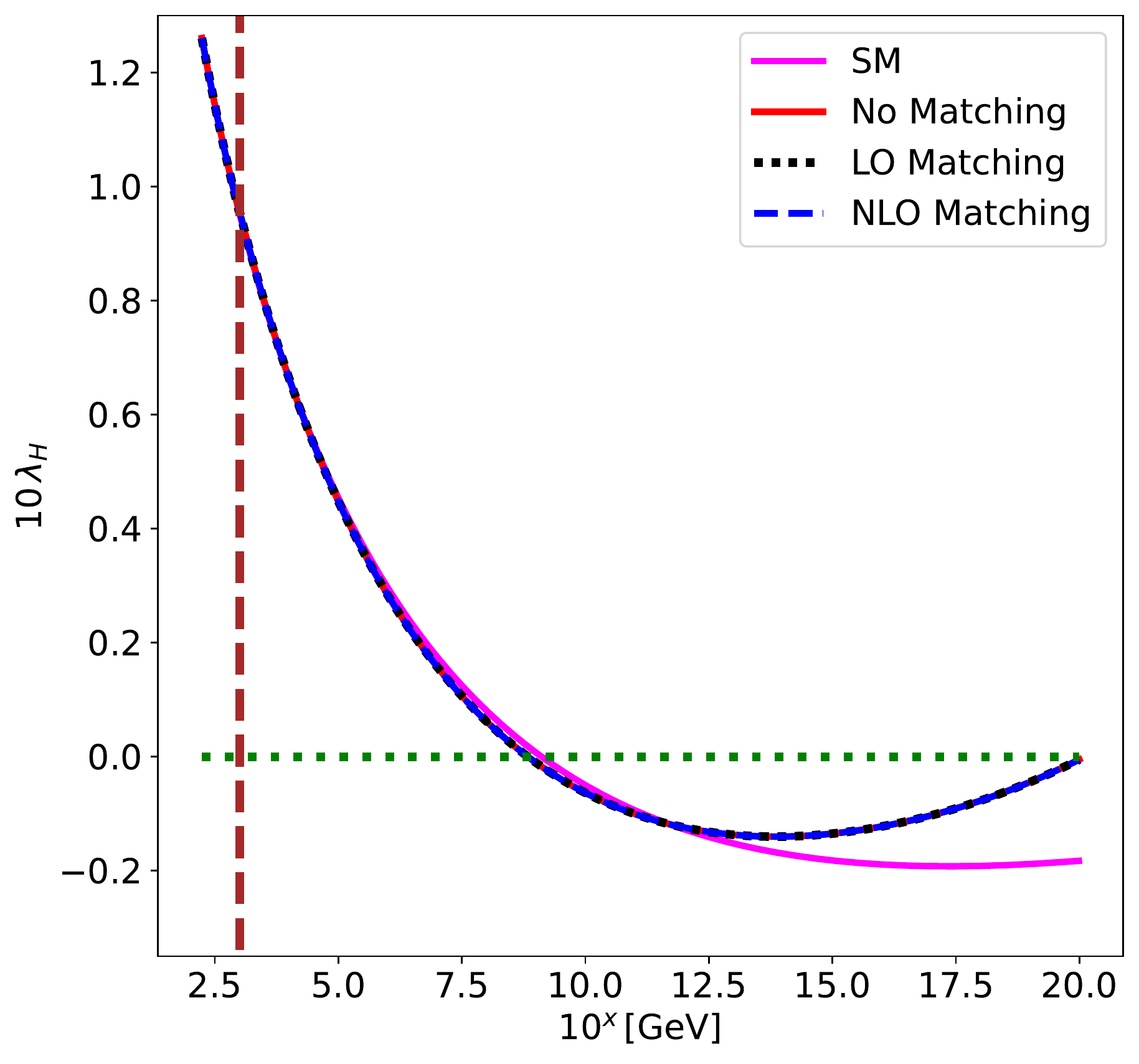} & 
	\includegraphics[width=0.76\textwidth]{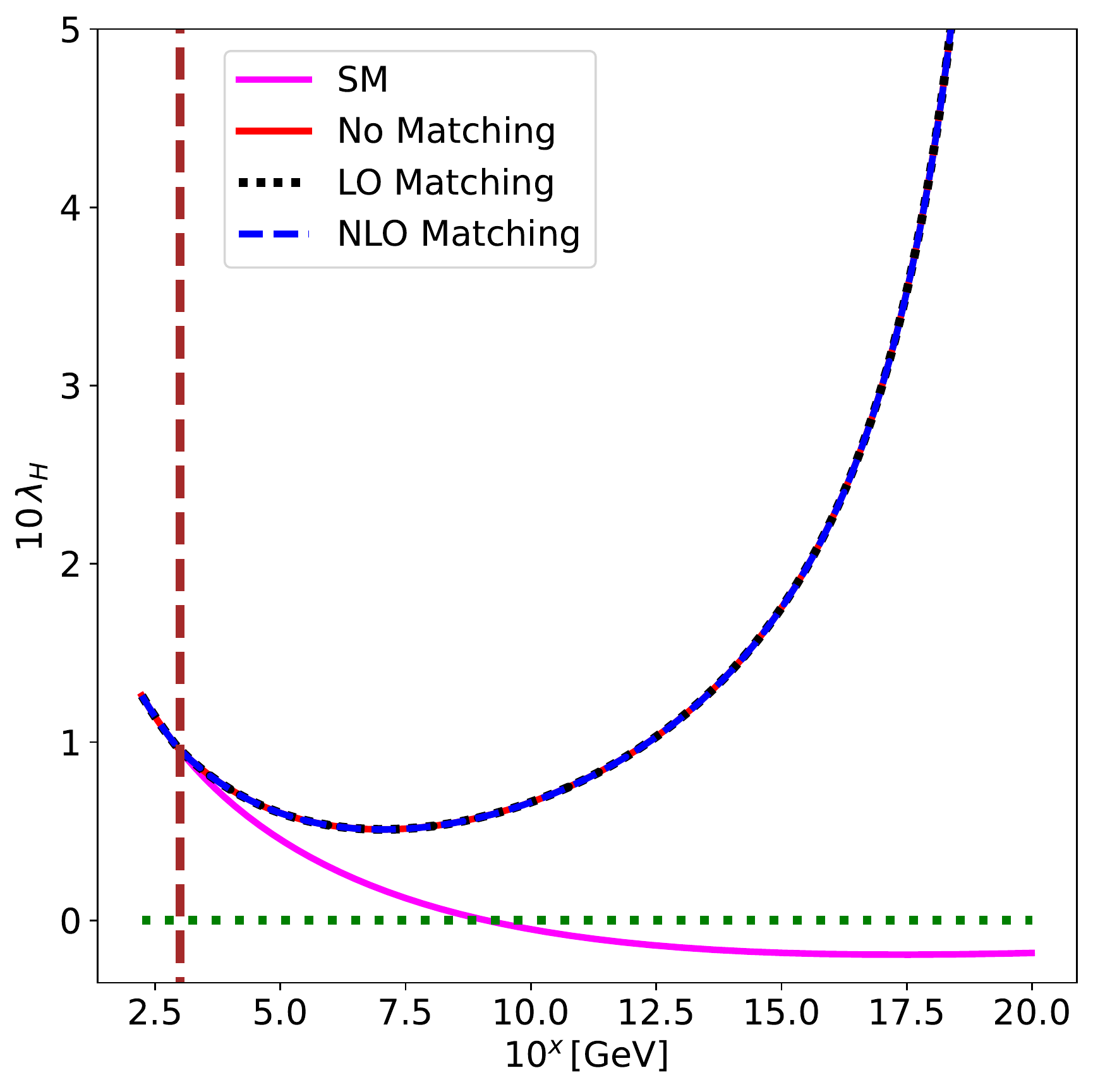}\\
	& \Huge{(Modest Yukawa with $M=10^3$\,GeV)} & \\ 
\includegraphics[width=0.8\textwidth]{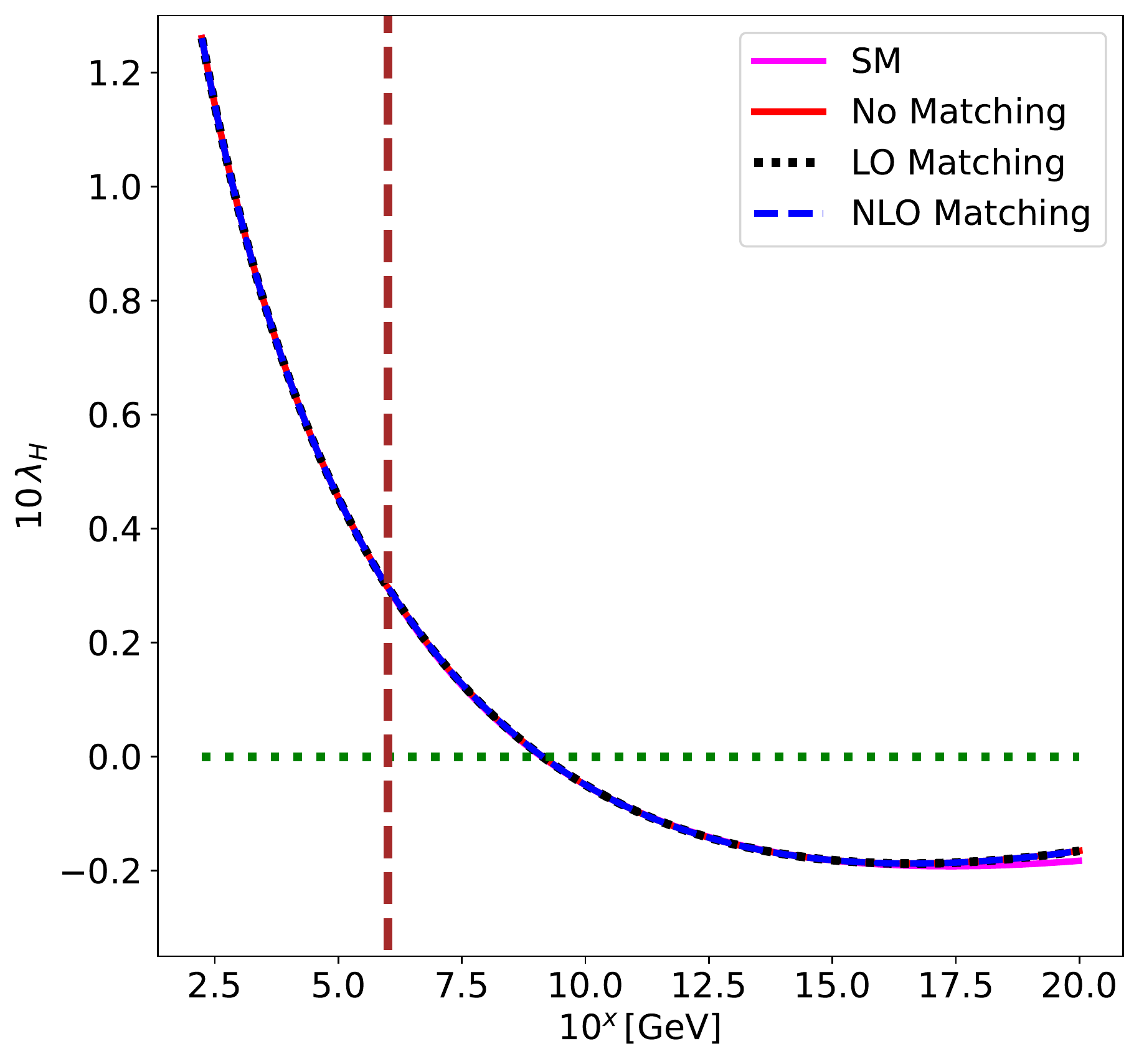} & 
	\includegraphics[width=0.8\textwidth]{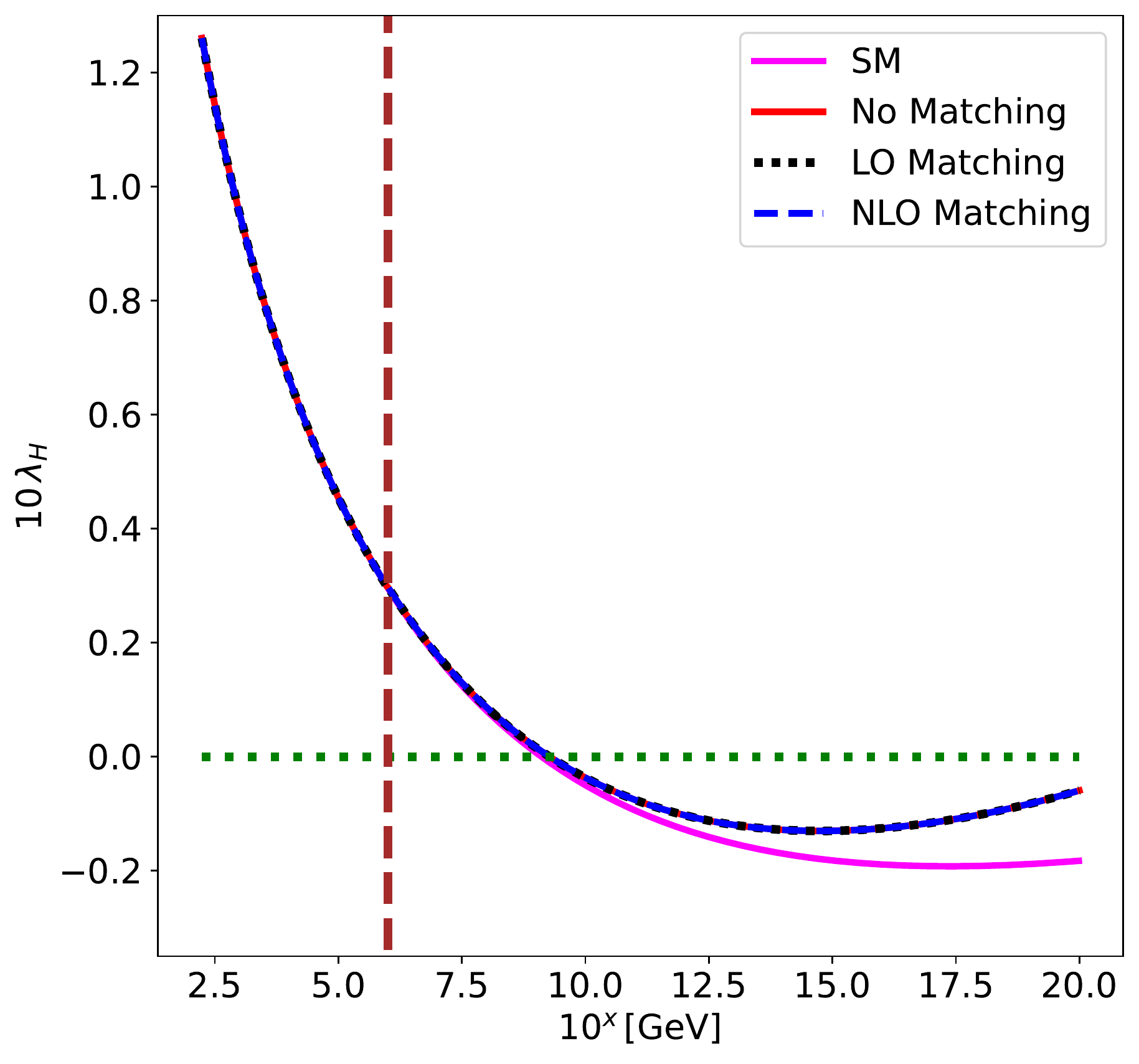} & 
	\includegraphics[width=0.8\textwidth]{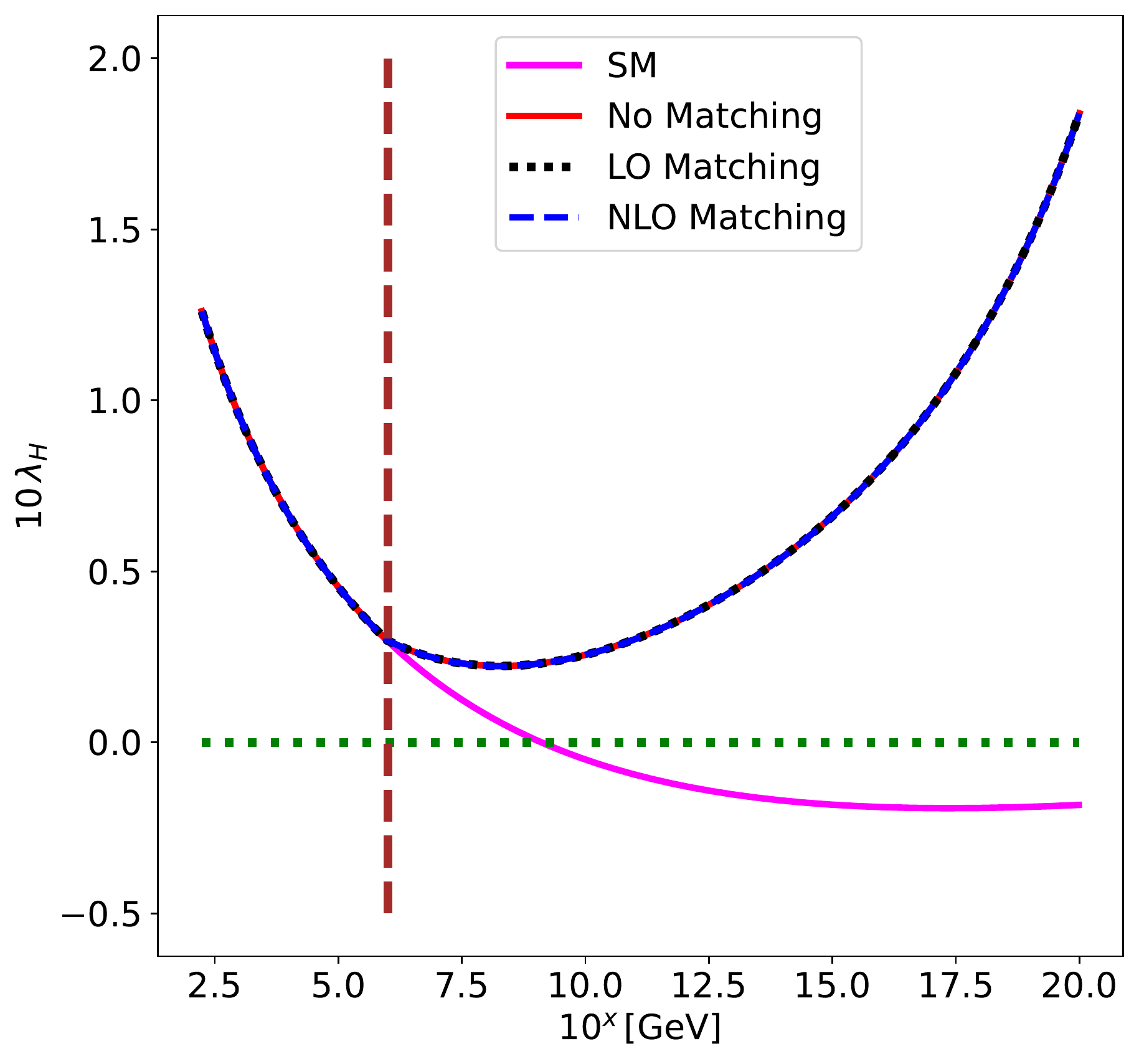}\\
	& \Huge{(Modest Yukawa with $M=10^6$\,GeV)} & \\ 
\end{tabular}
  \end{adjustbox}
  }\caption{Same as figure\,\ref{fig:seesaw2mu} but for the running of $\lambda_H$.}\label{fig:seesaw2lambda}
\end{figure}

\begin{figure}[t]
\centering{
  \begin{adjustbox}{max width = \textwidth}
\begin{tabular}{ccc}
  \Huge{~~~~~$\lambda_4=-0.1$, $\lambda_5=-0.1$}	& \Huge{~~~~~$\lambda_4=0.1$, $\lambda_5=0.1$} & \Huge{~~~~~$\lambda_4=0.5$, $\lambda_5=0.5$} \\ 
\includegraphics[width=0.8\textwidth]{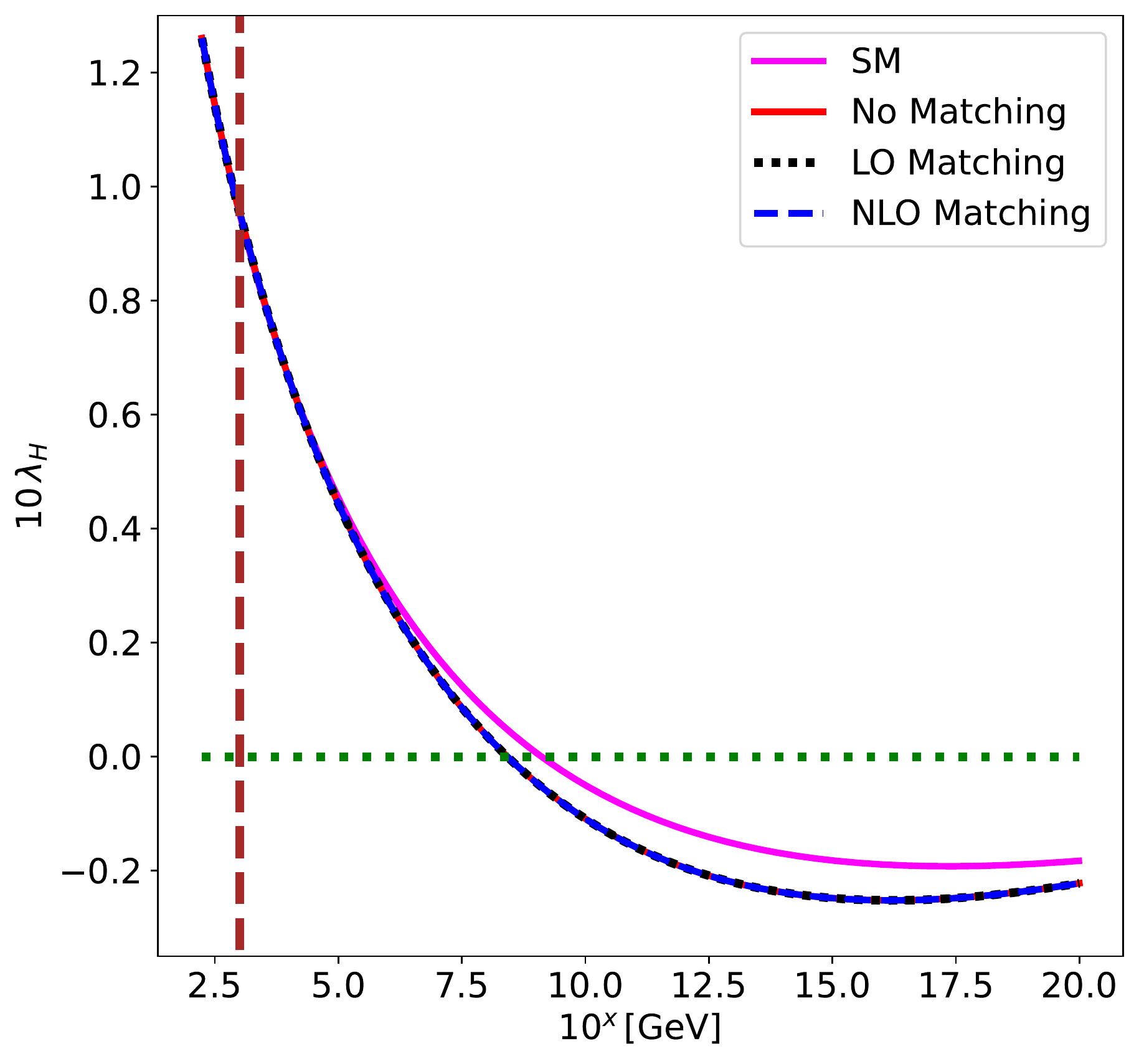} & 
	\includegraphics[width=0.8\textwidth]{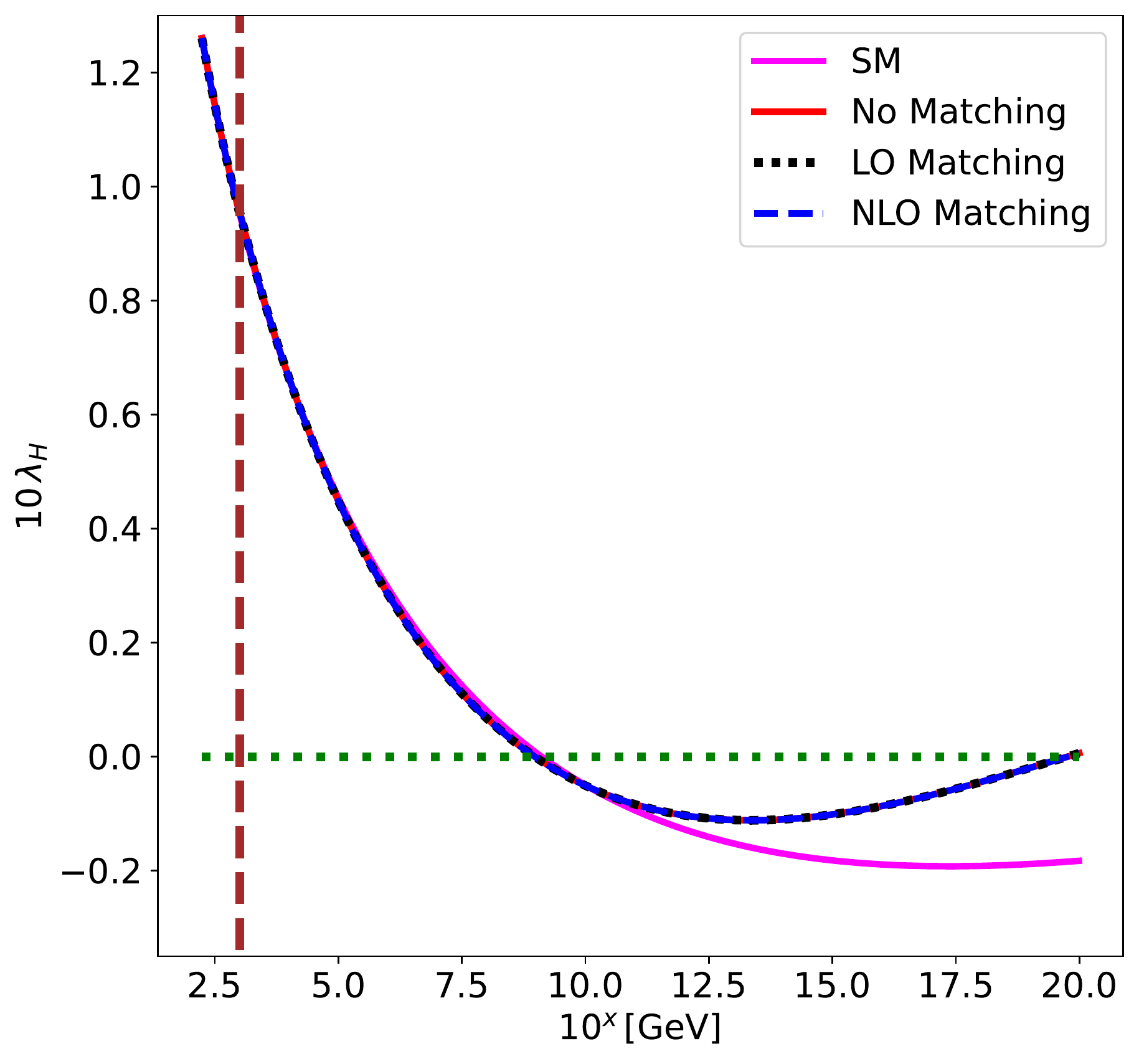} & 
	\includegraphics[width=0.76\textwidth]{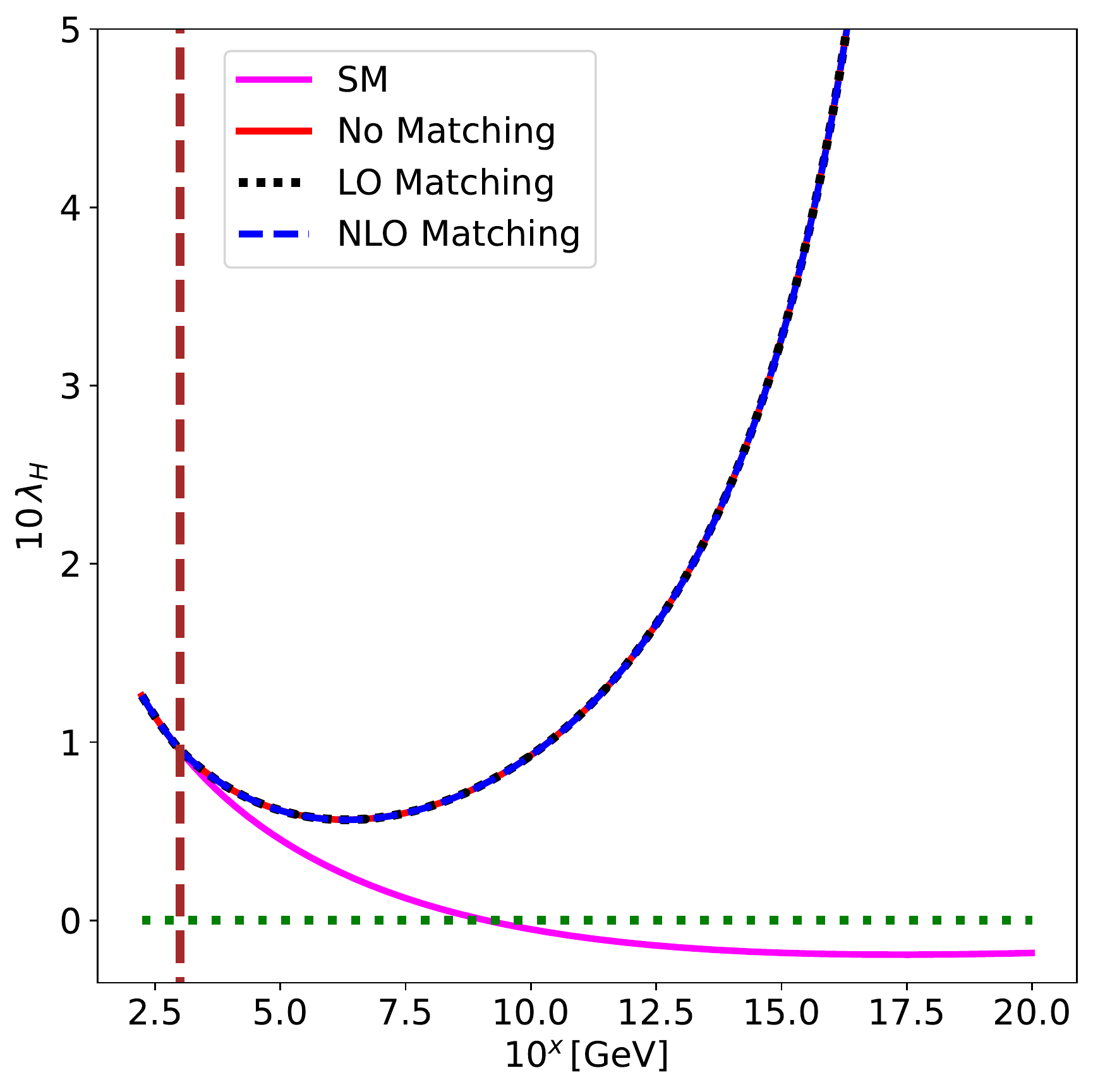}\\
	& \Huge{(Large Yukawa with $M=10^3$\,GeV)} & \\ 
\includegraphics[width=0.8\textwidth]{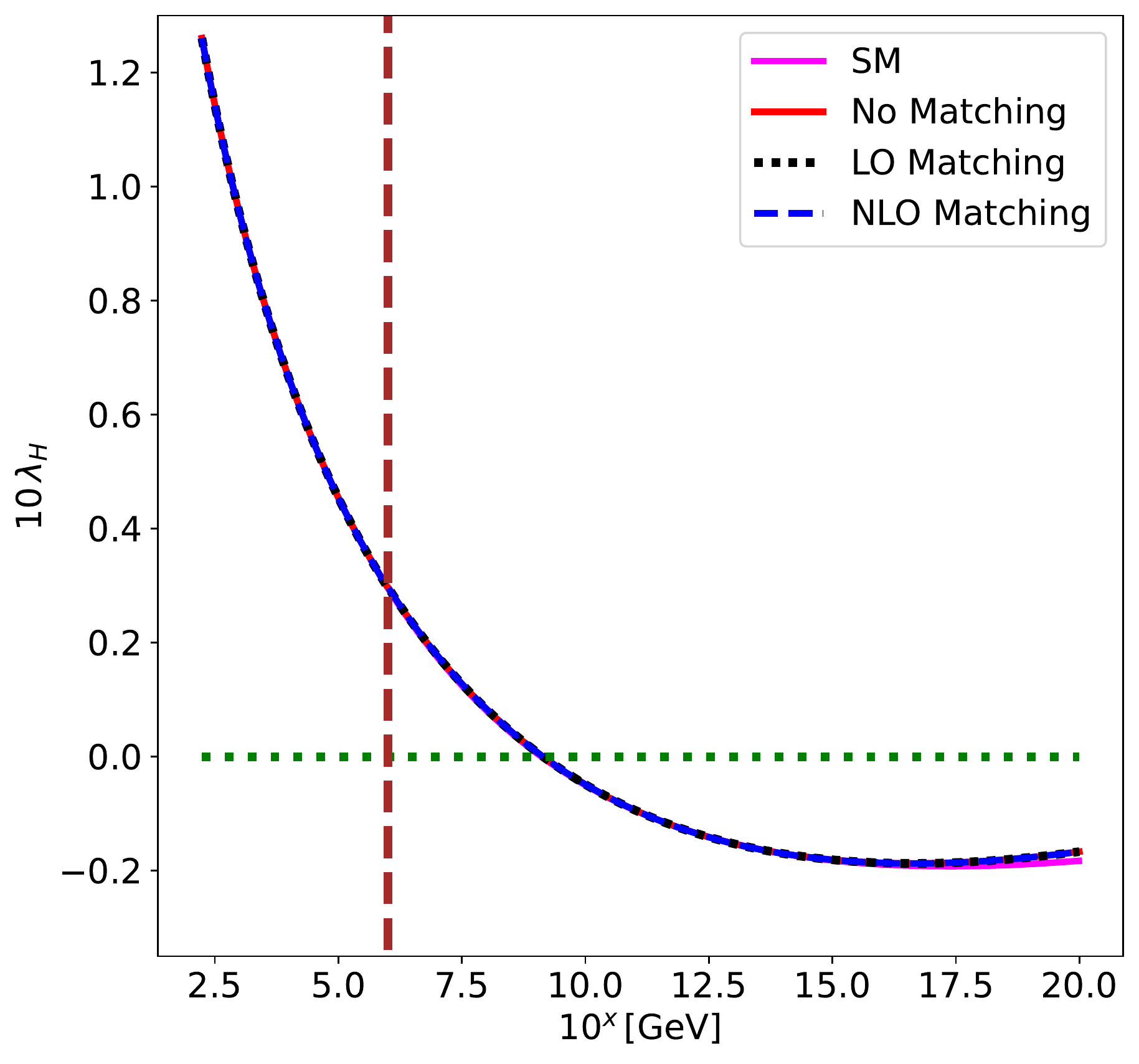} & 
	\includegraphics[width=0.8\textwidth]{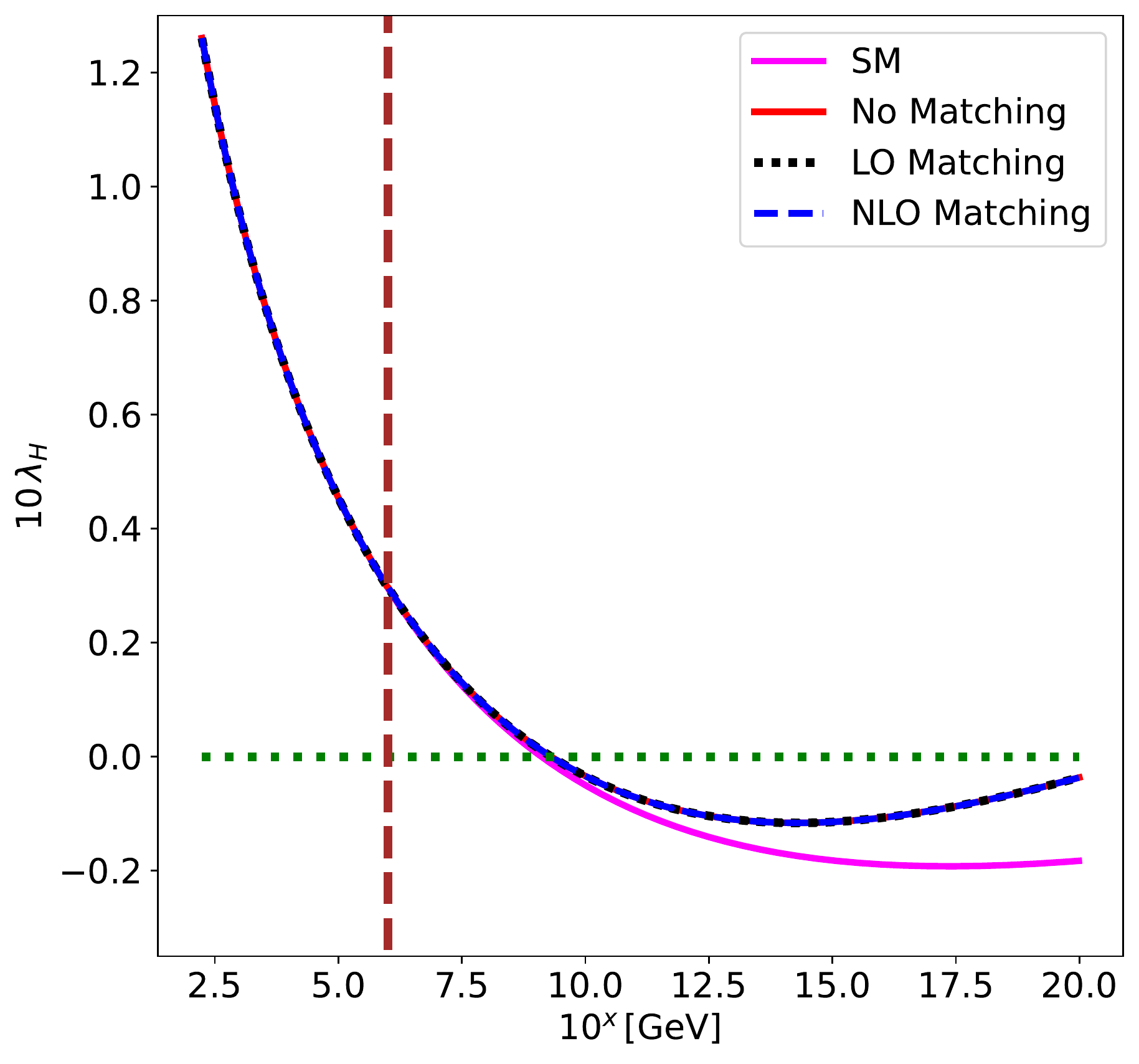} & 
	\includegraphics[width=0.78\textwidth]{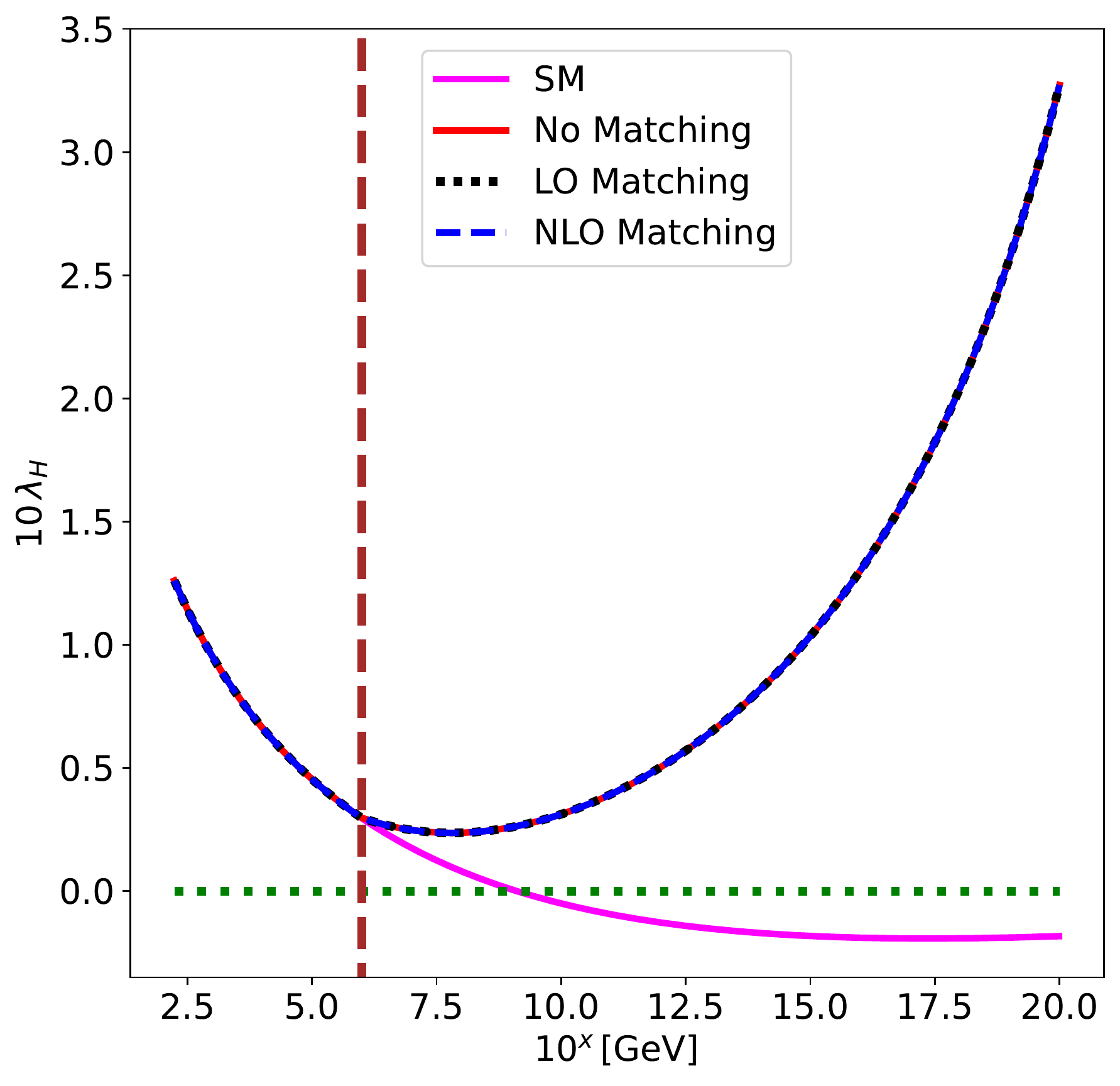}\\
	& \Huge{(Large Yukawa with $M=10^6$\,GeV)} & \\ 
\includegraphics[width=0.8\textwidth]{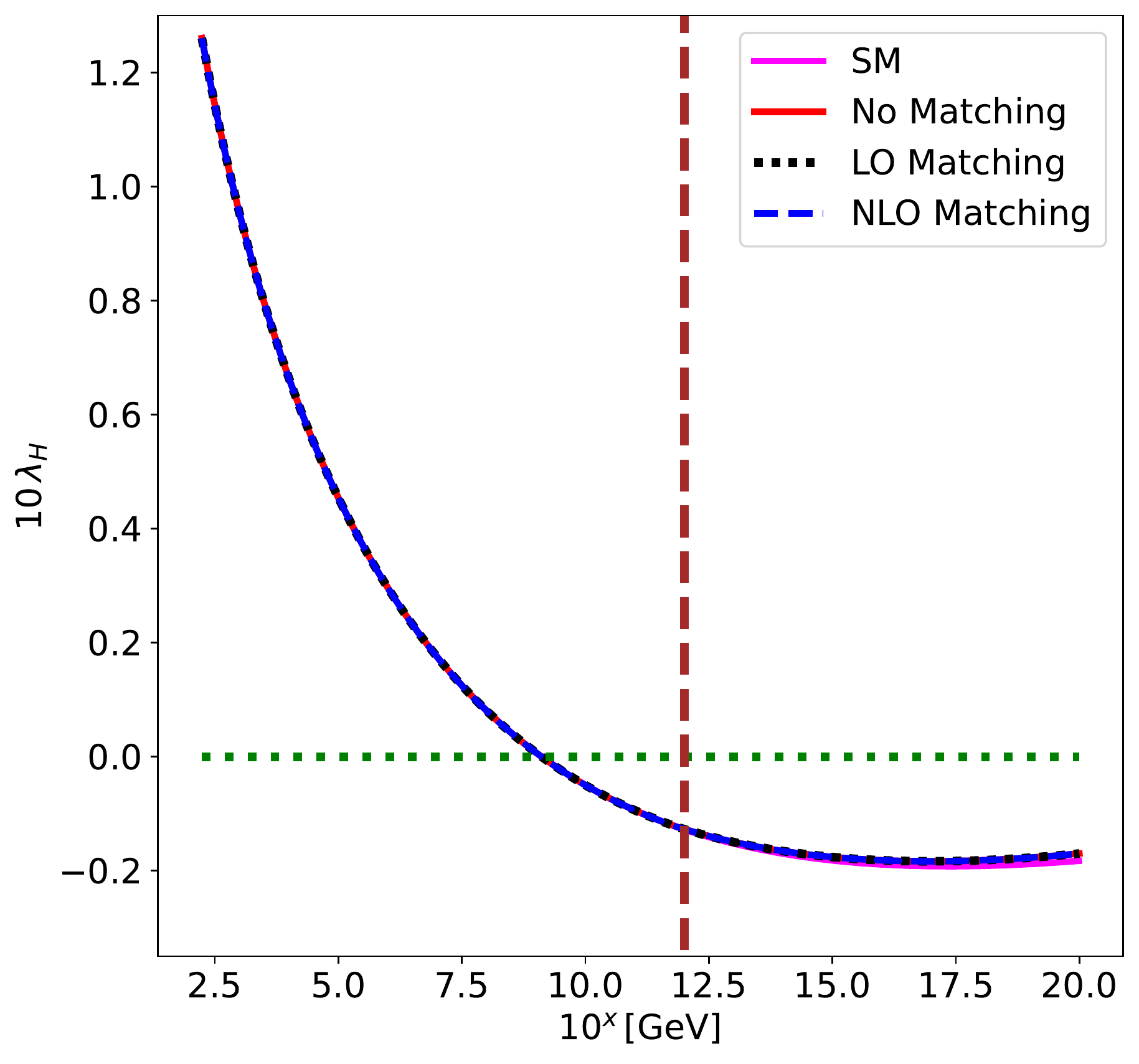} & 
	\includegraphics[width=0.8\textwidth]{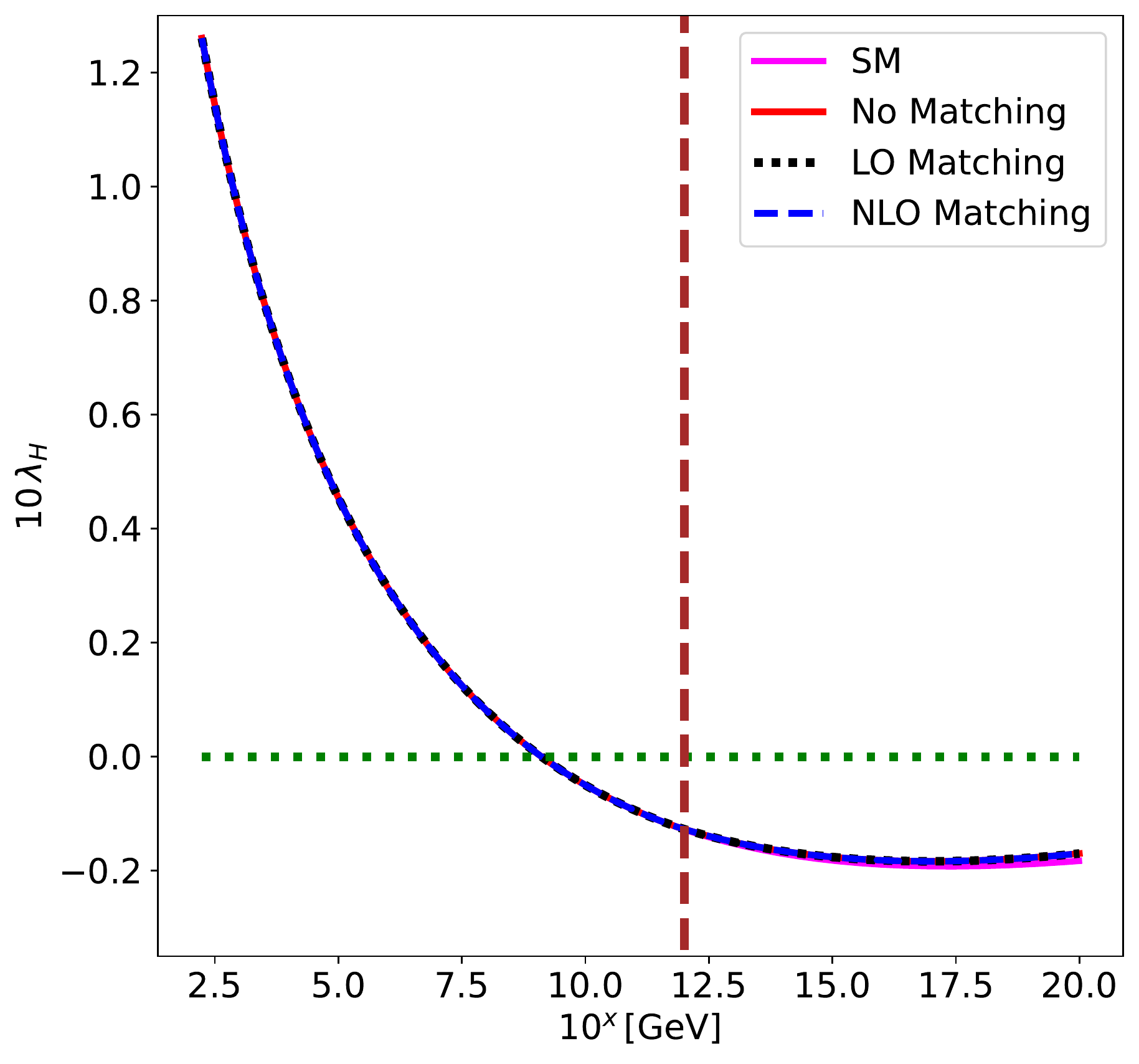} & 
	\includegraphics[width=0.8\textwidth]{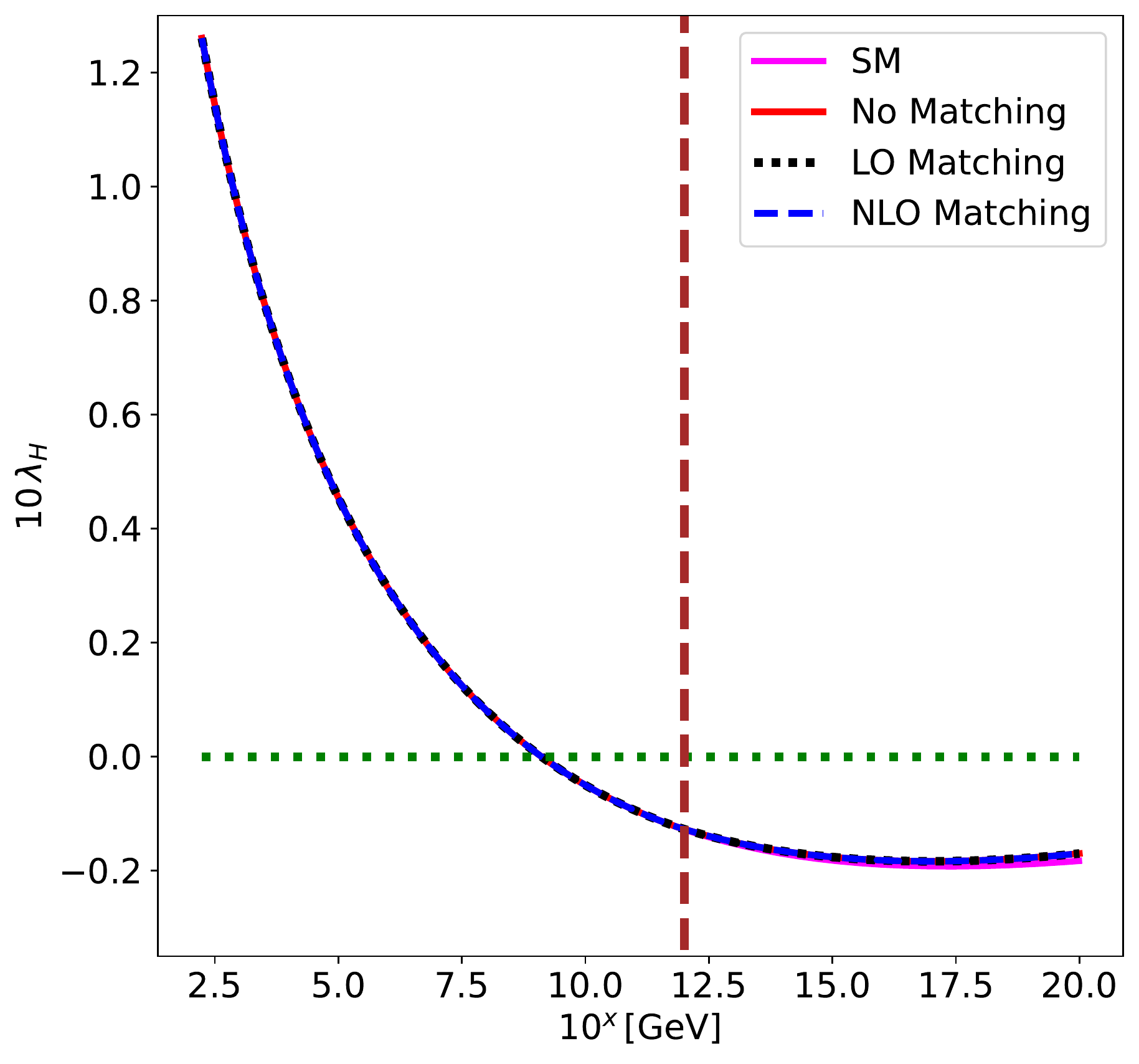}\\
	& \Huge{(Large Yukawa with $M=10^{12}$\,GeV)} & \\ 
\end{tabular}
  \end{adjustbox}
  }\caption{Same as figure\,\ref{fig:seesaw2mu2} but for the running of $\lambda_H$.}\label{fig:seesaw2lambda2}
\end{figure}

Solving the RGEs with the benchmark points in table\,\ref{tab:seesaw2BM}, we summarize our results in figures\,\ref{fig:seesaw2mu} and \ref{fig:seesaw2mu2} for the running of $\mu_H^2$, and figures\,\ref{fig:seesaw2lambda} and \ref{fig:seesaw2lambda2} for that of $\lambda_H$. A few points merit stressing:
\begin{itemize}
\item For tiny neutrino Yukawa couplings, the upper bound of $M$ is roughly fixed according to eq.\,\eqref{eq:vMbound}. For our representative benchmark $M=10^3$\,GeV, we find positive $\lambda_{4,5}$ are generically favored to trigger EWSB radiatively as implied in the first row of figure\,\ref{fig:seesaw2mu}. Especially, when $\lambda_{4,5}$ approaches unity, EWSB can even be invoked at a relatively low scale near $\sim10^4$\,GeV. Due to the lightness of the triplet in this case, we strongly recommend an investigation on this scenario at current/future colliders. Note also that the radiative breaking scale would be lifted to above $10^5$\,GeV if only tree-level matching is included, highlighting the importance of including the threshold effects from one-loop matching.
\item Increasing the neutrino Yukawa couplings while keeping $M=10^3$\,GeV does not change above observation, see the middle row of figure\,\ref{fig:seesaw2mu}. The reason is that the neutrino Yukawa only modifies the lepton Yukawa in the SM, and the latter only has a negligible impact on the running of $\mu_H^2$. However, increasing the neutrino Yukawa and the triplet mass simultaneously, the running of $\mu_H^2$ changes dramatically depending on the values of $\lambda_{4,5}$. For example, for negative $\lambda_{4,5}$ as shown in the first plot in the last row of figure\,\ref{fig:seesaw2mu}, tree-level matching may suggests radiative EWSB below the Planck scale. However, with the inclusion of one-loop matching, radiative EWSB becomes absent. Similarly, for positive $\lambda_{4,5}$ as presented in the last two plots in the last row of figure\,\ref{fig:seesaw2mu}, the negative shift due to the threshold effects from one loop matching implies the immediate presence of radiative EWSB at the matching scale, strongly suggesting the inclusion of one-loop matching for correctly interpreting/understanding the type-II seesaw model. On one hand, the difference between the first plot and the last two plots in the last row of figure\,\ref{fig:seesaw2mu} suggests the possibility of utilizing the Higgs potential stability, together with other phenomenological results, the Higgs to di-photon decay rate for example\,\cite{Du:2018eaw}, to determine the sign of $\lambda_{4,5}$; on the other hand, though our benchmark, i.e., $M=10^6$\,GeV, makes it difficult to explore this scenario at current/future colliders, our study suggests above observation could also occur below $10^6$\,GeV. Recall that the type-I and -III seesaw models can not radiatively trigger EWSB up to the Planck scale, one can thus expect to utilize this feature to distinguish the type-II seesaw model from the other two.
\item Discussion in last bullet remains the same qualitatively though differs quantitatively when further increasing the neutrino Yukawa couplings and/or the triplet mass. See figure\,\ref{fig:seesaw2mu2} for the results.
\item On the other hand, for the running of $\lambda_H$, since the shift from one-loop matching is always suppressed by $\mu_H^2/M^2$ (see our matching results in section\,\ref{sec:seesaw123}),\footnote{The suppression can either be understood directly from our numerical results in figures\,\ref{fig:seesaw2mu} and \ref{fig:seesaw2mu2}, or from the fact that the triplet is presumably heavier than the doublet from phenomenological considerations.} it shall be sufficient to include only tree-level matching. This is confirmed by our numerical results in figures\,\ref{fig:seesaw2lambda} and \ref{fig:seesaw2lambda2}.
\item Negative values of $\lambda_{4,5}$ only worsen the stability of the Higgs potential as seen from the first columns of figures\,\ref{fig:seesaw2lambda} and \ref{fig:seesaw2lambda2}. While positive $\lambda_{4,5}$ alleviate this instability of the Higgs potential as implied from the second and the third columns, their numerical values shall be large enough, for example, $\lambda_{4}=\lambda_{5}=0.5$, to ensure stability up to the Planck scale at loop level as suggested in the last columns of aforementioned two figures. We also want to point out that if one however goes beyond tree-level vacuum stability and perturbative unitarity, modest negative $\lambda_{4,5}$, for example $\lambda_{4}=\lambda_{5}=-0.5$ also help stabilize the Higgs potential even up to the Planck scale.
\end{itemize}

To understand our results analytically, we present in the following an analytical criterion for radiatively triggering EWSB in the type-II seesaw model. This can be achieved by requiring, at the matching scale,
\eqal{\mu_H^2 - \frac{1}{16\pi^2}\left( \frac{\mu ^2 \mu _H^4}{2 M^4} + 3\left(\mu ^2+\lambda _{45} M^2\right) \right) < 0.}
Since $v_\Delta\ll v_\Phi$ and $\mu_{H,0}\ll M$,\footnote{The latter condition can be seen either from our numerical result in figures\,\ref{fig:seesaw2mu}-\ref{fig:seesaw2mu2}, or from the validity of EFTs.}
\eqal{\frac{6v_\Delta^2}{v_\Phi^4}M^4 + 3\lambda_{45} M^2 > 16\pi^2\mu_{H,0}^2 - \frac{v_\Delta^2\mu_{H,0}^4}{v_\Phi^4} + \mathcal{O}\left( \frac{v_\Delta^2}{v_\Phi^2}, \frac{\mu_{H,0}^2}{M^2} \right).\label{eq:approxtype2}}
Further assuming an almost vanishing $v_\Delta$, one then has
\eqal{{\rm sgn}\left(\lambda_{45}\right) M^2 > \frac{16\pi^2 \mu_{H,0}^2}{3|\lambda_{45}|},}
forcing the sign of $\lambda_{45}$, i.e., ${\rm sgn}\left(\lambda_{45}\right)>0$, as already implied in the middle two plots of the first two rows of figure\,\ref{fig:seesaw2mu2}. The above inequality then simplifies to
\eqal{M > \frac{4\pi \mu_{H,0}}{\sqrt{3\lambda_{45}}}.}
We finally summarize this criterion for the type-II seesaw model as follows:
\begin{tcolorbox}
To radiatively trigger EWSB in the type-II seesaw model at the matching scale, it is equivalent to requiring, up to one-loop level,
\eqal{
\frac{6v_\Delta^2}{v_\Phi^4}M^4 + 3\lambda_{45} M^2 \gtrsim 16\pi^2\mu_{H,0}^2 - \frac{v_\Delta^2\mu_{H,0}^4}{v_\Phi^4},\quad M \lesssim |\mathcal{Y}|\cdot\frac{v_\Phi}{v_\Delta}\label{eq:criteria2EWSB1}}
with $v_{\Delta\,(\Phi)}$ the triplet (doublet) vev, $\mu_{H,0}$ the SM prediction of $\mu_H$ at the seesaw matching scale $M$, and $\mathcal{Y}$ of $\mathcal{O}(1)$ from the consideration of unitarity and perturbativity in eq.\,\eqref{eq:vMbound}. Especially, in the vanishing $v_\Delta$ limit, it simplifies to
\eqal{ M \gtrsim \frac{4\pi}{\sqrt{3\lambda_{45}}}\mu_{H,0} \quad,\label{eq:criteria2EWSB2}}
forcing positive $\lambda_{45}\equiv\lambda_4+\lambda_5$.
\end{tcolorbox}

While clear from our reasoning above, we comment here for the validity of the criterion above. The validity of the approximation in eq.\,\eqref{eq:approxtype2} resides in three facts: (1) the tininess of neutrino masses\,\cite{ParticleDataGroup:2020ssz}, (2) the precisely measured $\rho$ parameter\,\cite{ParticleDataGroup:2020ssz}, and (3) the absence of any new physics below $\sim$1\,TeV\,\cite{Baak:2014ora}. Though this work is not devoted to a detailed phenomenological study on these seesaw models, we want to stress here on the importance of one-loop matching on the Higgs potential, which might, for example, have a non-trivial role in baryogenesis\,\cite{Planck:2018vyg} and/or the generation of gravitational waves\,\cite{LIGOScientific:2016aoc,LIGOScientific:2018mvr,LIGOScientific:2020ibl}. At first sight, these corrections from one-loop matching might seem important only when tree-level matching is absent as for type-I and -II seesaw models, and are expected to be unimportant when tree-level matching is present. However, as we just see above for the type-II seesaw model, even when tree-level matching is non-vanishing, one-loop matching still plays a significant role in triggering EWSB radiatively.

There is one final remark that we would like to add to this criterion for the type-II seesaw model: These conditions, i.e., eqs.\eqref{eq:criteria2EWSB1} and \eqref{eq:criteria2EWSB2}, only guarantee radiative EWSB around the matching scale. Violation of these conditions does not necessarily forbid radiative EWSB above the matching scale. See, for example, the second column of figure\,\ref{fig:seesaw2mu}, where radiative EWSB is realized from the RG evolution.

To summarize, as seen above, the nature of EWSB can differ quite significantly in these three seesaw models, which can thus be used to distinguish them phenomenologically once new resonance(s) is (are) observed at colliders. On the other hand, we have seen the complementary role that has been played by the low-energy experiments along high-energy searches, the Higgs discovery for example. In next section, we will summarize observables present in these three models and qualitatively discuss how to use different experiments at different energy scales to distinguish them. We leave a detailed phenomenological study and quantitative results on this topic for a future project due to the scope of this work.

%% file: pheno.tex
\section{Seesaw model discrimination}\label{sec:pheno}
Besides the difference in triggering EWSB through radiative corrections for the three seesaw models discussed above, we also comment on how to distinguish these three models phenomenologically in this section. To that end, the most promising starting point would be to focus on operators that can only be induced by a specific seesaw model at tree-level. One example is the $\mathcal{O}_{eH}$ operator, which can only be induced by the type-II seesaw model at tree-level when ignoring the neutrino Yukawa couplings.\footnote{Beyond the tree level, all the three models can generate this operator. However, since their effects will be loop suppressed and it is the tree-level matching dominates corrections to SM predictions, we can safely perform the analysis by focusing on the dominant contributions at this point.} Since this operator will modify the Higgs production rate at lepton colliders such as CEPC\,\cite{CEPCStudyGroup:2018rmc,CEPCStudyGroup:2018ghi}, FCC-ee\,\cite{FCC:2018byv,FCC:2018evy}, ILC\,\cite{ILC:2007bjz} and CLIC\,\cite{CLICPhysicsWorkingGroup:2004qvu}, or even a muon collider, one can thus expect to make use of this channel as a promising candidate for identifying the type-II seesaw model from the other two seesaw models at colliders. 

However, due to the very limited number of operators from tree-level matching and given the upcoming precision era for particle physics in the next decades, it would be imperative to go beyond the leading order and to closely investigate effects originated from one-loop matching. Under this consideration, we gain many more operators that can only be induced by the type-II seesaw model as seen in table\,\ref{tab:pheno}. Due to the importance of lepto-quark operators in the development and precision tests of the SM, we would like to slightly extend our phenomenological discussion on these operators here. One example of such lepto-quark operators is $\mathcal{O}_{lequ}^{(1)}$, which can only be induced by the type-II seesaw model at one-loop level. This operator would introduce new interactions between leptons and quarks that could be detected by many precision experiments: (1) Parity-Violating Electron Scattering (PVES) experiments such as $Q_{\rm weak}$ at Jefferson Lab\,\cite{Qweak:2018tjf}, Atomic Parity Violation (APV)\,\cite{Bouchiat:1974kt,Macpherson:1991opp,Meekhof:1993zz,Edwards:1995zz,Vetter:1995vf,Wood:1997zq,Safronova:2017xyt}, polarized electron-deuteron Deep-Inelastic Scattering (eDIS) at SLAC\,\cite{Prescott:1979dh}, (2) neutrino-nucleon and neutrino-nucleus scattering experiments such as CHARM\,\cite{CHARM:1987pwr}, CCFR\,\cite{CCFR:1997zzq}, CDHS\,\cite{Blondel:1989ev}, NuTeV\,\cite{NuTeV:2001whx}, COHERENT\,\cite{COHERENT:2017ipa,COHERENT:2020iec}, FASER$\nu$\,\cite{Anchordoqui:2021ghd}, (3) neutrino oscillation experiments such as Daya Bay\,\cite{DayaBay:2012yjv,DayaBay:2018yms}, RENO\,\cite{RENO:2012mkc,RENO:2018dro}, Double Chooz\,\cite{DoubleChooz:2006vya,DoubleChooz:2019qbj}, T2K\,\cite{T2K:2019bcf}, NO$\nu$A\,\cite{NOvA:2019cyt}, JUNO\,\cite{JUNO:2015zny}, T2HK\,\cite{Hyper-KamiokandeProto-:2015xww}, DUNE\,\cite{DUNE:2016hlj} due to the presence of neutrino non-standard interactions.

While operators generated by a single seesaw model is perhaps the most promising starting point, there are secondary operators, induced simultaneously by two or all of the three seesaw models, that can also be used for the discussion here. This relies on the fact that the sign and/or the magnitude of corrections from these operators to SM predictions are different such that it would be possible to differentiate these models from precision measurements. Examples of this kind are the $\mathcal{O}_W$, $\mathcal{O}_{eB}$ and $\mathcal{O}_{ll}$ operators. We comment on how to utilize these three specific operators for model identification in the following:
\begin{itemize}
\item The $\mathcal{O}_W$ operator would contribute to the anomalous triple gauge couplings (aTGCs), which have be measured at the LHC\,\cite{CMS:2013ryd,CMS:2014xja,ATLAS:2016zwm,ATLAS:2018nci,CMS:2020gtj} and could also be measured at the  High-Luminosity LHC (HL-LHC) and future High-Energy (HE) colliders\,\cite{CEPCStudyGroup:2018rmc,CEPCStudyGroup:2018ghi,FCC:2018vvp,FCC:2018bvk,FCC:2018byv,FCC:2018evy,ILC:2007bjz,CLICPhysicsWorkingGroup:2004qvu}. This operator can be generated from both the type-II and -III seesaw models at one-loop level. However, the aTGCs will receive opposite sign shifts from these two models due to the fact that the Wilson coefficient from the type-III model is a factor of $-N_\Sigma$ larger than that from the type-II model as seen from our results in section\,\ref{sec:seesaw123}. This factor of $-N_\Sigma$ is included as a parameter in parentheses in table\,\ref{tab:pheno} and a similar convention is adopted for other operators therein as long as this factorization procedure can be done.
\item In comparison, the dipole operator $\mathcal{O}_{eB}$ can be induced by all the three seesaw models at one-loop level, whose Wilson coefficients in the Warsaw basis are given below for reference upon omitting the flavor indices:
\eqal{
C_{eB}^{\rm type-I} &= \frac{1}{16\pi^2}\frac{g_1}{24M^2}\left(Y_\nu Y_\nu^\dagger Y_e\right),\\
C_{eB}^{\rm type-II} &= \frac{1}{16\pi^2}\frac{g_1}{4M^2}\left(Y_\nu^\dagger Y_\nu Y_e\right),\\
C_{eB}^{\rm type-III} &= \frac{1}{16\pi^2}\frac{g_1}{8M^2}\left(Y_\Sigma^\dagger Y_\Sigma Y_e\right).
}
For simplicity, we use a common symbol $M$ to indicate the UV scale for the three seesaw models in above equalities, and similarly for $\mathcal{O}_{ll}$ below. Na\"ively, it seems that this operator would not be a smart choice for model discrimination since all the three models contribute non-vanishingly. However, since this operator contributes to lepton magnetic dipole moments that have been measured to an astonishing precision level\,\cite{Muong-2:2006rrc,Muong-2:2021ojo},\footnote{For the theoretical result from the SM or a result from the recent lattice calculation, see Refs.\,\cite{Borsanyi:2020mff,Aoyama:2020ynm}.} these three models are thus expected to be differentiable from precision measurements given their magnitude difference in contributing to lepton dipole moments.
\end{itemize}

\FloatBarrier
\begin{table}[H]
  \centering
  \resizebox{!}{0.43\textheight}{  
\begin{tabular}{|c|ccc|c|}
\hline
\multirow{2}{*}{Operators} & \multicolumn{3}{c|}{Seesaw models} & \multirow{2}{*}{$Y_{\nu,\Sigma}\stackrel{?}{=}0$}\\
\cline{2-4}
 & Type-I & Type-II & Type-III &  \\
 \hline
 \multicolumn{5}{|c|}{{\bf Tree-level matching}} \\
 \hline
 $(H^\dagger H)^2$ & \xmark & \cmark & \xmark & {No}\\
 $\mathcal{O}_{H}$ & \xmark & \cmark & \xmark &  {No}\\
 $\mathcal{O}_{H\Box}$ & \xmark & \cmark & \xmark & {No}\\
 $\mathcal{O}_{HD}$ & \xmark & \cmark & \xmark & {No}\\
 $\mathcal{O}_{eH}$ & \xmark & \cmark & \xmark & {\color{magenta}Yes}\\
 $\mathcal{O}_{uH}$ & \xmark & \cmark & \xmark & {No}\\
 $\mathcal{O}_{dH}$ & \xmark & \cmark & \xmark & {No}\\
 \hline
 \multicolumn{5}{|c|}{{\bf One-loop matching}} \\
 \hline
 $\mathcal{O}_{W}$ & \xmark & \cmark & \cmark $\left(-N_\Sigma\right)$ & No \\
 $\mathcal{O}_{HW}$ & \xmark & \cmark & \xmark &  {\color{magenta}Yes}\\ 
 $\mathcal{O}_{HB}$ & \xmark & \cmark & \xmark &  {\color{magenta}Yes}\\ 
 $\mathcal{O}_{HWB}$ & \xmark & \cmark & \xmark &  {\color{magenta}Yes}\\ 
 $\mathcal{O}_{eW}$ & \cmark & \cmark $\left(-\frac{3}{5}\right)$ & \cmark $\left(\frac{9}{5}\right)$ &  No\\ 
 $\mathcal{O}_{eB}$ & \cmark & \cmark (6) & \cmark (3) &  No\\ 
 $\mathcal{O}_{Hl}^{(1)}$ & \xmark & \cmark & \xmark &  {\color{magenta}Yes}\\ 
 $\mathcal{O}_{Hl}^{(3)}$ & \xmark & \cmark & \cmark &  {\color{magenta}Yes}\\ 
 $\mathcal{O}_{Hq}^{(1)}$ & \xmark & \cmark & \xmark &  {\color{magenta}Yes}\\ 
 $\mathcal{O}_{Hq}^{(3)}$ & \xmark & \cmark & \cmark &  {\color{magenta}Yes}\\ 
 $\mathcal{O}_{ll}$ & \xmark & \cmark & \cmark &  {\color{magenta}Yes}\\ 
 $\mathcal{O}_{qq}^{(1)}$ & \xmark & \cmark & \xmark &  No\\
 $\mathcal{O}_{qq}^{(3)}$ & \xmark & \cmark & \cmark $\left({4N_\Sigma}\right)$ &  No\\
 $\mathcal{O}_{lq}^{(1)}$ & \xmark & \cmark & \xmark &  {\color{magenta}Yes}\\ 
 $\mathcal{O}_{lq}^{(3)}$ & \xmark & \cmark & \cmark $\left({4N_\Sigma}\right)$ &  {\color{magenta}Yes}\\ 
 $\mathcal{O}_{ee}$ & \xmark & \cmark & \xmark &  {No}\\
 $\mathcal{O}_{uu}$ & \xmark & \cmark & \xmark &  {No}\\
 $\mathcal{O}_{dd}$ & \xmark & \cmark & \xmark &  {No}\\
 $\mathcal{O}_{eu}$ & \xmark & \cmark & \xmark &  {No}\\
 $\mathcal{O}_{ed}$ & \xmark & \cmark & \xmark &  {No}\\
 $\mathcal{O}_{ud}^{(1)}$ & \xmark & \cmark & \xmark &  {No}\\
 $\mathcal{O}_{le}$ & \xmark & \cmark & \xmark &  {\color{magenta}Yes}\\
 $\mathcal{O}_{lu}$ & \xmark & \cmark & \xmark &  {\color{magenta}Yes}\\
 $\mathcal{O}_{ld}$ & \xmark & \cmark & \xmark &  {\color{magenta}Yes}\\
 $\mathcal{O}_{qe}$ & \xmark & \cmark & \xmark &  {No}\\
 $\mathcal{O}_{qu}^{(1)}$ & \xmark & \cmark & \xmark &  {\color{magenta}Yes}\\
 $\mathcal{O}_{qu}^{(8)}$ & \xmark & \cmark & \xmark &  {\color{magenta}Yes}\\
 $\mathcal{O}_{qd}^{(1)}$ & \xmark & \cmark & \xmark &  {\color{magenta}Yes}\\
 $\mathcal{O}_{qd}^{(8)}$ & \xmark & \cmark & \xmark &  {\color{magenta}Yes}\\
 $\mathcal{O}_{ledq}$ & \xmark & \cmark & \xmark &  {\color{magenta}Yes}\\
 $\mathcal{O}_{quqd}^{(1)}$ & \xmark & \cmark & \xmark &  {\color{magenta}Yes}\\
 $\mathcal{O}_{lequ}^{(1)}$ & \xmark & \cmark & \xmark &  {\color{magenta}Yes}\\
\hline
\end{tabular}
}\caption{Check-marks (cross-marks) in blue (red) mean that operators in the first column can (cannot) be induced from the specific seesaw model. Flavor indices are implicit following the notations in\,\cite{Grzadkowski:2010es}. The last column specifies whether we assume vanishing neutrino Yukawa couplings.}\label{tab:pheno}
\end{table}

\begin{itemize}
\item Similarly, the 4-lepton operator $\mathcal{O}_{ll}$ can also be generated by all the three seesaw models, which will modify for example neutrino-electron scattering rate at CHARM-II\,\cite{CHARM-II:1994dzw}, neutrino trident production at CCFR\,\cite{CCFR:1997zzq} and electron parity-violating scattering at SLAC-E158\,\cite{SLACE158:2005uay}. Assuming tiny neutrino Yukawa couplings for phenomenological considerations,\footnote{Theoretically, this is equivalent to assuming the seesaw models to be light -- relative to the GUT scale -- for type-I and -III, and a large triplet vev for type-II.} the Wilson coefficients for the three models become
\eqal{
C_{ll,prst}^{\rm type-I} &= 0,\\
C_{ll,prst}^{\rm type-II} &= -\frac{1}{16\pi^2}\frac{1}{240M^2}\left[ 4 g_2^4\delta_{pt}\delta_{sr} + (3g_1^4 - 2 g_2^4)\delta_{pr}\delta_{st} \right],\\
C_{ll,prst}^{\rm type-III} &= -\frac{1}{16\pi^2}\frac{g_2^4}{30M^2}\left( 2\delta_{pt}\delta_{sr}N_\Sigma - \delta_{pr}\delta_{st} \right),
}
where $p,r,s,t$ are the flavor indices. Therefore, in the limit where neutrino Yukawa couplings vanish as indicated by the last column of table\,\ref{tab:pheno}, only the type-II and -III seesaw models would induce the $\mathcal{O}_{ll}$ operator. Recall that the MOLLER experiment\,\cite{MOLLER:2014iki} at the Jefferson Lab, and similarly the P2 experiment at MESA\,\cite{Berger:2015aaa}, would measure the change of electron weak charge, i.e., $\delta Q_e^W$, at the per-mil level and that its SM prediction $\left(\delta Q_e^W\right)^{\rm SM}$ for MOLLER is also known at the per-mil level from a recent two-loop calculation\,\cite{Du:2019evk}, one can thus utilize this precision observable to distinguish the type-II model from the type-III one from the fact that in the presence of $\mathcal{O}_{ll}$, one has $\left(\delta Q_e^W\right)^{\rm full}=\left(\delta Q_e^W\right)^{\rm SM} - 2v^2C_{ll,1111}$\footnote{For the EFT contribution, here we only keep the dominant terms at $\mathcal{O}(1/\Lambda^2)$ from the interference between tree-level SM contributions and the $\mathcal{O}_{ll}$ operator.} with $v$ the vev of the Higgs doublet.
\end{itemize}

Based on above criteria, we summarize all operators that can be used for model identification in table\,\ref{tab:pheno}. While already explained in different places above, we summarize our conventions used in table\,\ref{tab:pheno} below for reference:
\begin{itemize}
\item The first column lists the names of dimension-6 operators in Warsaw basis following the notations established in\,\cite{Grzadkowski:2010es}. All flavor indices are ignored. 
\item Cross-marks in red in the second column mean that the operators can not be generated in that specific model, while check-marks in blue imply the opposite.
\item The last column specifies whether we take vanishing neutrino Yukawa couplings in order to obtain results in the second column.
\end{itemize}

%% file: conclusions.tex
\section{Conclusions}\label{sec:conclusions}
Neutrino oscillations definitely suggest new physics beyond the SM since the latter predicts vanishing mixing among neutrinos and thus forbids the observed phenomena. In this work, we assume neutrino masses to be generated at tree level from the type-I, -II and -III seesaw models. These three models naturally induce the dimension-5 Weinberg operators after integrating out the heavy new degrees of freedom, which is done up to one loop with the functional method in this article.

However, due to the indistinguishability of the three seesaw models from the dimension-5 operators, we retain the dimension-6 ones from the functional matching. Even though the functional approach renders calculations of one-loop matching systematical and straightforward, the resulting operators commonly inherit redundancy from, for example, Fierz identities and equations of motion that has to be eliminated manually to obtain an independent set of operators. We explain in detail how to remove this redundancy in section\,\ref{sec:functionalmethod} and then summarize our results in tables\,(\ref{tab:treematching}-\ref{tab:1loopWarsaw3}) of section\,\ref{sec:seesaw123} for all the operators induced by the three seesaw models. Our results are presented in both the intermediate Green's basis and the Warsaw basis up to one-loop level for reference.

Na\"ively, one may expect that tree-level matching would possibly be enough to capture all interesting phenomena from a specific model based on simple arguments of perturbativity. However, this turns out not to capture all the interesting physics as can be seen from the last column of figure\,\ref{fig:seesaw1mu} for the type-I and -III seesaw models, as well as figures\,\ref{fig:seesaw2mu} and \ref{fig:seesaw2mu2} for the type-II seesaw model. We find that, on top of tree-level results, the threshold effects from one-loop matching could change the symmetry breaking pattern significantly. To be concrete, on one hand, while it is known that the type-I and -III models can not trigger radiative electroweak symmetry breaking even at the Planck scale, we find the inclusion of one-loop threshold effects introduces an extra positive shift to the Higgs mass, making it even more impossible to radiatively trigger EWSB. On the other hand, for the type-II seesaw model, we find radiative EWSB could be triggered at a much lower scale even below $10^4$\,GeV, making the type-II seesaw model rather interesting for a phenomenological study from this aspect. Based on an analytical analysis, we also present the criterion for radiative EWSB in the type-II seesaw model around the matching scale, see our eqs.\,\eqref{eq:criteria2EWSB1} and \eqref{eq:criteria2EWSB2} for the results.

Synergy of experiments performed at different energy scales in distinguishing the three seesaw models are also qualitatively discussed in section\,\ref{sec:pheno}. In particular, we extend our discussion for the representative $\mathcal{O}_{lequ}^{(1)}$, $\mathcal{O}_{W}$, $\mathcal{O}_{eB}$ and $\mathcal{O}_{ll}$ operators in table\,\ref{tab:pheno} due to their impact on various low- and high-energy experiments. However, due to the scope of this work, we leave a detailed phenomenological study on these three seesaw models for a future project.

%% file: appendix.tex
\appendix

\section{The Green's and Warsaw bases}

\begin{table}[H]
  \centering
  \resizebox{0.965\textwidth}{!}{  
\begin{tabular}{|c|c|c|c|c|c|}
\hline \multicolumn{2}{|c|}{\color{blue} Type: $X^3$} & \multicolumn{2}{|c|}{\color{blue} Type: $H^2D^4$} & \multicolumn{2}{|c|}{\color{blue} Type: $\psi^2HD^2$}  \\

\hline ${\cal O}_{W}$ & $\epsilon^{IJK} W^{I \nu}_{\mu} W^{J\rho}_{\nu} W^{K \mu}_{\rho}$ & ${\cal R}_{DH}$ & $(D_\mu D^\mu H)^\dagger (D_\nu D^\nu H)$ & ${\cal R}_{eHD1}$ & $\left(\bar{\ell} e\right)D_\mu D^\mu H$ \\

\cline{1-4} \multicolumn{2}{|c|}{\color{blue} Type: $X^2D^2$} & \multicolumn{2}{|c|}{\color{blue} Type: $H^4D^2$} & ${\cal R}_{eHD2}$ & $\left(\bar{\ell} i\sigma_{\mu\nu} D^\mu e\right)D^\nu H$ \\

\cline{1-4} ${\cal R}_{2W}$ & $-\frac{1}{2}\left(D_\mu W^{I\mu\nu}\right) \left(D^\rho W^I_{\rho\nu}\right)$ & ${\cal O}_{H\square}$ & $\left(H^\dagger H\right)\square \left(H^\dagger H\right) $ & ${\cal R}_{eHD3}$ & $\left(\bar{\ell} D_\mu D^\mu e\right) H$ \\
${\cal R}_{2B}$ & $-\frac{1}{2}\left(D_\mu B^{\mu\nu}\right) \left(D^\rho B_{\rho\nu}\right)$ & ${\cal O}_{HD}$ & $\left(H^\dagger D^\mu H\right)^* \left(H^\dagger D^\mu H\right) $ & ${\cal R}_{eHD4}$ & $\left(\bar{\ell} D_\mu e\right) D^\mu H$ \\

\cline{1-2}\cline{5-6} \multicolumn{2}{|c|}{\color{blue} Type: $X^2H^2$} & ${\cal R}_{HD}'$ & $\left(H^\dagger H\right) \left(D_\mu H\right)^\dagger \left(D^\mu H\right)$ & \multicolumn{2}{|c|}{\color{blue} Type: $\psi^2 X D$} \\

\hline ${\cal O}_{HW}$ & $H^\dagger H W^I_{\mu\nu} W^{I\mu\nu}$ & \multicolumn{2}{|c|}{\color{blue} Type: $H^6$} & ${\cal R}_{W\ell}$ & $\left(\bar{\ell} \sigma^I \gamma^\mu \ell\right) D^\nu W^I_{\mu\nu}$ \\

\cline{3-4} ${\cal O}_{HB}$ & $H^\dagger H B_{\mu\nu} B^{\mu\nu}$ & ${\cal O}_H$ & $\left(H^\dagger H\right)^3$ & ${\cal R}_{\tilde{W}\ell}$ & $\frac{1}{2}(\bar{\ell} \sigma^I \gamma^\mu i \overset{\leftrightarrow}{D}{}^\nu\ell) \tilde{W}^I_{\mu\nu}$ \\

\cline{3-4} ${\cal O}_{HWB}$ & $H^\dagger \sigma^I H W^I_{\mu\nu} B^{\mu\nu}$ & \multicolumn{2}{|c|}{Type: $\psi^2D^3$} & ${\cal R}_{B\ell}$ & $\left(\bar{\ell} \gamma^\mu \ell\right) D^\nu B_{\mu\nu}$ \\

\cline{1-4} \multicolumn{2}{|c|}{\color{blue} Type: $H^2 XD^2$} & ${\cal R}_{\ell D}$ & $\frac{1}{2}i\bar{\ell}\{D_\mu D^\mu, D\!\!\!\!/\, \}\ell$ & ${\cal R}_{\tilde{B}\ell}$ & $\frac{1}{2}(\bar{\ell} \gamma^\mu i \overset{\leftrightarrow}{D}{}^\nu\ell) \tilde{B}_{\mu\nu}$ \\

\hline ${\cal R}_{WDH}$ & $D_\nu W^{I\mu\nu} (H^\dagger i \overset{\leftrightarrow}{D}{}_\mu^I H)$ & \multicolumn{2}{|c|}{\color{blue} Type: $\psi^2XH$} & \multicolumn{2}{|c|}{\color{blue} Type: $\psi^2H^3$} \\

\cline{3-6} ${\cal R}_{BDH}$ & $\partial_\nu B^{\mu\nu} (H^\dagger i \overset{\leftrightarrow}{D}{}_\mu H)$ & ${\cal O}_{eW}$ & $(\bar{\ell}_L\sigma^{\mu\nu} e_R )\sigma^I H W^I_{\mu\nu}$ & ${\cal O}_{eH}$ & $ (H^\dagger H) (\bar{\ell} e {H})$\\

&  & ${\cal O}_{eB}$ & $(\bar{\ell}_L\sigma^{\mu\nu} e_R ) H B_{\mu\nu}$ &  ${\cal O}_{uH}$ & $ (H^\dagger H) (\bar{q} u \tilde{H})$ \\

&  &  &  &  ${\cal O}_{dH}$ & $ (H^\dagger H) (\bar{q} d {H})$  \\


\hline \multicolumn{2}{|c|}{\color{blue} Type: $\psi^2DH^2$} & \multicolumn{2}{|c|}{\color{blue} Type: $\psi_L\psi_L\psi_L\psi_L$} & \multicolumn{2}{|c|}{\color{blue} Type: $\psi_L\psi_L\psi_R\psi_R$}  \\

\hline ${\cal O}_{H\ell}^{(1)}$ & $(\bar{\ell}\gamma^\mu \ell)(H^\dagger i \overset{\leftrightarrow}{D}{}_\mu H)$ & ${\cal O}_{\ell\ell}$ & $(\bar{\ell} \gamma^\mu \ell)(\bar{\ell} \gamma_\mu \ell)$ & ${\cal O}_{\ell e}$ & $(\bar{\ell} \gamma^\mu \ell)(\bar{e} \gamma_\mu e)$ \\

${\cal R}_{H\ell}'^{(1)}$ & $(\bar{\ell}i \overset{\leftrightarrow}{D\!\!\!\!/\,} \ell)(H^\dagger H)$ & ${\cal O}_{qq}^{(1)}$ & $(\bar{q} \gamma^\mu q)(\bar{q} \gamma_\mu q)$ & ${\cal O}_{\ell u}$ & $(\bar{\ell} \gamma^\mu \ell)(\bar{u} \gamma_\mu u)$ \\

${\cal O}_{H\ell}^{(3)}$ & $(\bar{\ell}\sigma^I\gamma^\mu \ell)(H^\dagger i \overset{\leftrightarrow}{D}{}_\mu^I H)$ & ${\cal O}_{qq}^{(3)}$ & $(\bar{q} \gamma^\mu \sigma^I q)(\bar{q} \gamma_\mu \sigma^I q)$ & ${\cal O}_{\ell d}$ & $(\bar{\ell} \gamma^\mu \ell)(\bar{d} \gamma_\mu d)$ \\

${\cal R}_{H\ell}'^{(3)}$ & $(\bar{\ell}i \overset{\leftrightarrow}{D\!\!\!\!/\,}{}^I \ell)(H^\dagger \sigma^I H)$ & ${\cal O}_{\ell q}^{(1)}$ & $(\bar{\ell} \gamma^\mu \ell)(\bar{q} \gamma_\mu q)$ & ${\cal O}_{qe}$ & $(\bar{q} \gamma^\mu q)(\bar{e} \gamma_\mu e)$ \\

${\cal O}_{Hq}^{(1)}$ & $(\bar{q}\gamma^\mu q)(H^\dagger i \overset{\leftrightarrow}{D}{}_\mu H)$ & ${\cal O}_{\ell q}^{(3)}$ & $(\bar{\ell} \gamma^\mu \sigma^I \ell)(\bar{q} \gamma_\mu \sigma^I q)$ & ${\cal O}_{qu}^{(1)}$ & $(\bar{q} \gamma^\mu q)(\bar{u} \gamma_\mu u)$ \\

\cline{3-4} ${\cal R}_{Hq}'^{(1)}$ & $(\bar{q}i \overset{\leftrightarrow}{D\!\!\!\!/\,} q)(H^\dagger H)$ & \multicolumn{2}{|c|}{\color{blue} Type: $\psi_R\psi_R\psi_R\psi_R$} & ${\cal O}_{qu}^{(8)}$ & $(\bar{q} \gamma^\mu T^A q)(\bar{u} \gamma_\mu T^A u)$ \\

\cline{3-4} ${\cal O}_{Hq}^{(3)}$ & $(\bar{q}\sigma^I\gamma^\mu q)(H^\dagger i \overset{\leftrightarrow}{D}{}_\mu^I H)$ & ${\cal O}_{ee}$ & $(\bar{e} \gamma^\mu e)(\bar{e} \gamma_\mu e)$ & ${\cal O}_{qd}^{(1)}$ & $(\bar{q} \gamma^\mu q)(\bar{d} \gamma_\mu d)$ \\

${\cal R}_{Hq}'^{(3)}$ & $(\bar{q}i \overset{\leftrightarrow}{D\!\!\!\!/\,}{}^I q)(H^\dagger \sigma^I H)$ & ${\cal O}_{uu}$ & $(\bar{u} \gamma^\mu u)(\bar{u} \gamma_\mu u)$ & ${\cal O}_{qd}^{(8)}$ & $(\bar{q} \gamma^\mu T^A q)(\bar{d} \gamma_\mu T^A d)$ \\

\cline{5-6} ${\cal O}_{He}$ & $(\bar{e}\gamma^\mu e)(H^\dagger i \overset{\leftrightarrow}{D}{}_\mu H)$ & ${\cal O}_{dd}$ & $(\bar{d} \gamma^\mu d)(\bar{d} \gamma_\mu d)$ & \multicolumn{2}{|c|}{\color{blue} Type: $\psi_L\psi_R\psi_R\psi_L$ and $\psi_L\psi_R\psi_L\psi_R$} \\

\cline{5-6} ${\cal R}_{He}'$ & $(\bar{e}i \overset{\leftrightarrow}{D\!\!\!\!/\,} e)(H^\dagger H)$ & ${\cal O}_{eu}$ & $(\bar{e} \gamma^\mu e)(\bar{u} \gamma_\mu u)$ & ${\cal O}_{\ell edq}$ & $(\bar{\ell} e)(\bar{d} q)$ \\

${\cal O}_{Hu}$ & $(\bar{u}\gamma^\mu u)(H^\dagger i \overset{\leftrightarrow}{D}{}_\mu H)$ & ${\cal O}_{ed}$ & $(\bar{e} \gamma^\mu e)(\bar{d} \gamma_\mu d)$ & ${\cal O}_{quqd}^{(1)}$ & $(\bar{q}^j u)\epsilon_{jk}(\bar{q}^k d)$ \\

${\cal R}_{Hu}'$ & $(\bar{u}i \overset{\leftrightarrow}{D\!\!\!\!/\,} u)(H^\dagger H)$ & ${\cal O}_{ud}$ & $(\bar{u} \gamma^\mu u)(\bar{d} \gamma_\mu d)$ & ${\cal O}_{\ell equ}^{(1)}$ & $(\bar{\ell}^j e)\epsilon_{jk}(\bar{q}^k u)$ \\

${\cal O}_{Hd}$ & $(\bar{d}\gamma^\mu d)(H^\dagger i \overset{\leftrightarrow}{D}{}_\mu H)$ & & & & \\

${\cal R}_{Hd}'$ & $(\bar{d}i \overset{\leftrightarrow}{D\!\!\!\!/\,} d)(H^\dagger H)$  & & & & \\

\hline
\end{tabular}
}\caption{Operators induced by seesaw models in the Green's and Warsaw bases. Operators in both the Warsaw basis and the Green's basis are denoted by ${\cal O}_i$, while redundant operators in the Green's basis are denoted by ${\cal R}_i$. We classify the operators by their types as specified in blue. Flavor indices of the fermions are labeled by their order, which is not shown in the operators.}\label{tab:operatorbasis}
\end{table}

In this appendix we present the operators induced by seesaw models in the Green's and Warsaw basis. Operators in both the Warsaw basis and the Green's basis are denoted by ${\cal O}_i$, while redundant operators in the Green's basis are denoted by ${\cal R}_i$. Flavor indices of the fermions, which is omitted, are labeled by their orders in the operators. For example,
\eqal{{\cal O}_{lequ,prst}^{(1)} = (\bar{\ell}^j_p e_r)\epsilon_{jk}(\bar{q}^k_s u_t).}
The Hermitian derivatives read as
\eqal{H^\dagger i \overset{\leftrightarrow}{D}{}_\mu H = i H^\dagger D_\mu H - i (D_\mu H)^\dagger H \text{ and } H^\dagger i \overset{\leftrightarrow}{D}{}_\mu^I H = i H^\dagger \sigma^I D_\mu H - i (D_\mu H)^\dagger \sigma^I H.}
Dual tensors are defined by $\tilde{X}_{\mu\nu} = \frac{1}{2}\epsilon_{\mu\nu\rho\sigma}X^{\rho\sigma}$ $(\epsilon_{0123}=+1)$ as in Ref. \cite{Grzadkowski:2010es}. 